\documentclass[11pt,a4paper]{article}
\pdfoutput=1
\usepackage{jcappub}
\usepackage{graphicx}
\usepackage{aas_macros_JCAP}
\usepackage{blindtext}
\usepackage{bm}
\usepackage{verbatim}

\newcommand{\hMpc}{\textrm{Mpc}/h}

\newcommand{\Msun}{M_{\odot}/h}
\newcommand{\void}{\mathrm{v}}
\newcommand{\halo}{\mathrm{h}}

\newcommand{\tracer}{\mathrm{t}}
\newcommand{\matter}{\mathrm{m}}

\newcommand{\coreDens}{\hat{n}_\mathrm{min}}
\newcommand{\ellipticity}{\varepsilon}
\newcommand{\mr}{\texttt{mr}}
\newcommand{\hr}{\texttt{hr}}
\newcommand{\MR}{\texttt{midres}}
\newcommand{\HR}{\texttt{highres}}
\newcommand{\UHR}{\texttt{ultra-hr}}
\newcommand{\uhr}{\texttt{uhr}}
\newcommand{\Mag}{\texttt{Magneticum}}

\title{Why Cosmic Voids Matter: Nonlinear~Structure~\&~Linear~Dynamics}

\author[a]{Nico Schuster,}
\author[a,b]{Nico Hamaus,}
\author[a,c]{Klaus Dolag,}
\author[a,b,d]{Jochen Weller}

\affiliation[a]{Universit\"ats-Sternwarte M\"unchen, Fakult\"at f\"ur Physik, Ludwig-Maximilians-Universit\"at, Scheinerstr. 1, 81679 M\"unchen, Germany}
\affiliation[b]{Excellence Cluster ORIGINS, Bolzmannstr. 2, 85748 Garching, Germany}
\affiliation[c]{Max-Planck-Institut f\"ur Astrophysik, Karl-Schwarzschild-Str. 1, 85748 Garching, Germany }
\affiliation[d]{Max-Planck-Institut f\"ur extraterrestrische Physik, Giessenbachstr. 1, 85748 Garching, Germany}

\emailAdd{nico.schuster@physik.lmu.de}
\emailAdd{n.hamaus@physik.lmu.de}

\abstract{
We use the Magneticum suite of state-of-the-art hydrodynamical simulations to identify cosmic voids based on the watershed technique and investigate their most fundamental properties across different resolutions in mass and scale. This encompasses the distributions of void sizes, shapes, and content, as well as their radial density and velocity profiles traced by the distribution of cold dark matter particles and halos. We also study the impact of various tracer properties, such as their sparsity and mass, and the influence of void merging on these summary statistics. Our results reveal that all of the analyzed void properties are physically related to each other and describe universal characteristics that are largely independent of tracer type and resolution. Most notably, we find that the motion of tracers around void centers is perfectly consistent with linear dynamics, both for individual, as well as stacked voids. Despite the large range of scales accessible in our simulations, we are unable to identify the occurrence of nonlinear dynamics even inside voids of only a few Mpc in size. This suggests voids to be among the most pristine probes of cosmology down to scales that are commonly referred to as highly nonlinear in the field of large-scale structure.
}

\date{\today}


\begin{document}
\maketitle

\newpage

\section{Introduction\label{sec:intro}}
Observations of the cosmic microwave background (CMB) radiation have revealed the presence of tiny fluctuations in temperature and density when the Universe was only $370\,000$ years old~\cite{Bennett2003,Planck2020}. Since then, gravity amplified these fluctuations by many orders of magnitude, building up what is known as the \emph{cosmic web} in the process. What starts off as a Gaussian random field that is fully specified by its two-point function, evolves into an intricate system of gravitationally bound structures. These are composed of a multi-scale network of sheets, connected by filaments, which in turn are connected by dense nodes~\cite{Zeldovich1970}. At the same time the remaining space is evacuated and expands at an increasing pace, creating enormous voids~\cite{Gregory1978,Joeveer1978,Kirshner1981,Zeldovich1982,Bertschinger1985,vdWeygaert1993}.

Today we can only infer this web of structures indirectly via large surveys that map out its luminous tracers, such as galaxies. The dominant fraction of the cosmic web's matter content is assumed to be composed of some form of \emph{cold dark matter} (CDM), whose nature is yet to be determined. On top of that, the observed late-time acceleration in the expansion of space has revealed the presence of \emph{dark energy}, so far consistent with a cosmological constant $\Lambda$~\cite{Riess1998,Perlmutter1999}. Despite these big unknowns, simulations have allowed us to investigate structure formation in a $\Lambda$CDM universe in great detail~\cite[e.g.,][]{Springel2001a,Dolag2016}.

In recent years the study of cosmic voids therein has received particular attention, primarily triggered by the discovery that they can be utilized as very efficient probes of cosmology~\cite{Biswas2010,Lavaux2012,Sutter2012b,Hamaus2014a,Pisani2015a,Hamaus2015,Pisani2019,Moresco2022}. This benefit originates from the fact that void interiors can be seen as small independent pocket universes of low background density featuring super-Hubble expansion rates, which distinguishes them from collapsed structures that have decoupled from the Hubble flow, like dark matter halos. Nevertheless, in analogy to halos, cosmic voids exhibit universal characteristics, such as the shape of their radial density profile~\cite{Hamaus2014b,Ricciardelli2014}. The latter has become a focus of interest in itself with numerous recent follow-up studies, both in the context of $\Lambda$CDM~\cite[e.g.,][]{Chan2014,Leclercq2015,Cautun2016,SanchezC2017,Pollina2017,Chantavat2017,Fang2019,Stopyra2021,Shim2021,Tavasoli2021}, as well as with the aim of exploring modifications to general relativity (GR)~\cite[e.g.,][]{Zivick2015,Cai2015,Barreira2015,Falck2018,Baker2018,Paillas2019,Davies2019,Perico2019,Wilson2021,Tamosiunas2022,Fiorini2022}, the impact of massive neutrinos~\cite[e.g.,][]{Massara2015,Banerjee2016,Kreisch2019,Schuster2019,Zhang2020,Contarini2021,Bayer2021,Kreisch2022} and models of particle dark matter~\cite[e.g.,][]{Yang2015,Reed2015,Baldi2018,Lester2021,Arcari2022}.

In addition, the study of radial velocity profiles has become important in the context of modeling redshift-space distortions (RSD) for the observable shapes of voids in redshift space~\cite[e.g.,][]{Paz2013,Hamaus2014c,Hamaus2016,Cai2016,Hawken2017,Hamaus2017,Massara2018,Achitouv2019,Correa2021b,Massara2022}. However, tracer velocities can hardly be measured directly, so these models make use of local mass conservation to relate velocities to densities. Although voids represent nonlinear objects with significant density contrasts with respect to the mean background value, the relationship between density and velocity revealed itself to be remarkably well described by the continuity equation at merely \emph{linear} order in perturbation theory~\cite{Hamaus2014b}. This insight has been key for the success of RSD models for voids. In contrast, such models are challenged by complex nonlinearities and non-Gaussian statistics appearing in the context of galaxy clustering, which considerably limit the extraction of cosmological information to relatively large scales~\cite{Scoccimarro2004}.

The aim of this paper is to study the fundamental properties of voids in more detail with the help of hydrodynamical simulations. We compare different estimators of the density and velocity profiles of stacked voids and explore the use of weighting schemes. Furthermore, we examine how mass cuts and subsamplings of the underlying tracer distributions affect these profiles and test how well linear mass conservation holds up around voids. The outline of the paper is as follows: section~\ref{sec:Magneticum} provides details on the simulation suite and section~\ref{sec:methods} introduces the void finder applied to our tracer catalogs, as well as estimators for void profiles. Section~\ref{sec:catalogs} describes both halo and void catalogs obtained from the simulations and in section~\ref{sec:profiletypes} we investigate the different types of void profiles and the impact of void merging. Section~\ref{sec:mass_conservation} applies linear mass conservation to the individual and stacked density profiles around voids and compares the results with their velocity profiles. Finally, we summarize our conclusions in section~\ref{sec:conclusion}.

\section{The Magneticum simulations}
\label{sec:Magneticum}

This work makes use of simulations from the \Mag\footnote{\url{http://www.magneticum.org}} suite, a series of state-of-the-art hydrodynamical simulations that cover a variety of cosmological volumes at different mass resolutions. We will describe the \Mag{} suite briefly below, for more details we refer to previous work using these simulations~\cite[e.g.,][]{Hirschmann2014,Dolag2015,Steinborn2015,Teklu2015,Bocquet2016,Dolag2016,Remus2017,Castro2018,Angelinelli2022}. All the \Mag{} runs used in this work adopt a flat $\Lambda$CDM cosmology according to the best fitting values of WMAP7~\cite{Komatsu2011}, with $h = 0.704$, $\Omega_{\Lambda} = 0.728$, $\Omega_\mathrm{m} = 0.272$, $\Omega_\mathrm{b} = 0.0456$, $\sigma_8 = 0.809$, and $n_s = 0.963$. The simulations have been performed using an advanced version of the tree particle mesh-smoothed particle hydrodynamics (TreePM-SPH) code \textsc{P-Gadget3}~\cite{Springel2005}, including an improved SPH solver~\cite{Beck2016}. This code also implements a large variety of processes describing the evolution of baryons, such as the distribution of multiple metal species~\cite{Dolag2017}. Additionally, there are prescriptions describing black hole growth and the feedback from active galactic nuclei (AGN), based on the work in references~\cite{Springel2005a,DiMatteo2005,Fabjan2010}.

Previous studies using the \Mag{} simulations include the reproduction of various observables on large scales, such as the observed thermal history of the Universe~\cite{Young2021} and the observed SZ-Power spectrum~\cite{Dolag2016}. At the cluster scale, these simulations were used to successfully reproduce the observable luminosity-relation in X-ray~\cite{Biffi2013}, the pressure profile and chemical composition of the intra-cluster medium~\cite{Gupta2017,Dolag2017,Biffi2018b}, as well as the higher concentration of halos in fossil groups~\cite{Ragagnin2019} and properties of gas between galaxy clusters~\cite{Biffi2022}. Due to the aforementioned implementation of baryonic physics, various galaxy properties, such as realistic stellar masses~\cite{Naab2017,Lustig2022}, the impact of cluster environments on galaxies~\cite{Lotz2019}, the evolution of post-starburst galaxies~\cite{Lotz2021}, and the population of AGNs~\cite{Hirschmann2014,Steinborn2016,Biffi2018b} were successfully reproduced.

Our analysis mostly focuses on the \Mag{} boxes 0 and 2b, since these comprise the largest volume of medium (box 0) and high (box 2b) mass resolution. Their volumes are large enough to contain a sufficient number of cosmic voids (see table~\ref{table_Magneticum_voids}), with typical sizes in the range of $5-80\;\hMpc$ (see figure~\ref{fig_void_function_halo_CDM_merging_rad_ell_core}). Therefore, from now on we will refer to box 0 as \MR{} (\mr{}) and box 2b as \HR{} (\hr{}). Previous work on the tracer bias around voids already made use of the \MR{} simulation~\cite{Pollina2017}. For additional tests on the linear mass conservation, we further use box 4 with even higher resolution, albeit fewer voids, referred to as \UHR{} (\uhr{}). In this paper we restrict ourselves to snapshots at redshift $z=0.29$. For more details on the number of particles, box size $L_\mathrm{Box}$, and mass resolution of these \Mag{} boxes we refer the reader to table~\ref{table_Magneticum_sims}.

\begin{figure}[t]
\centering
\resizebox{\hsize}{!}{
    \includegraphics[trim = 0 15 0 0]{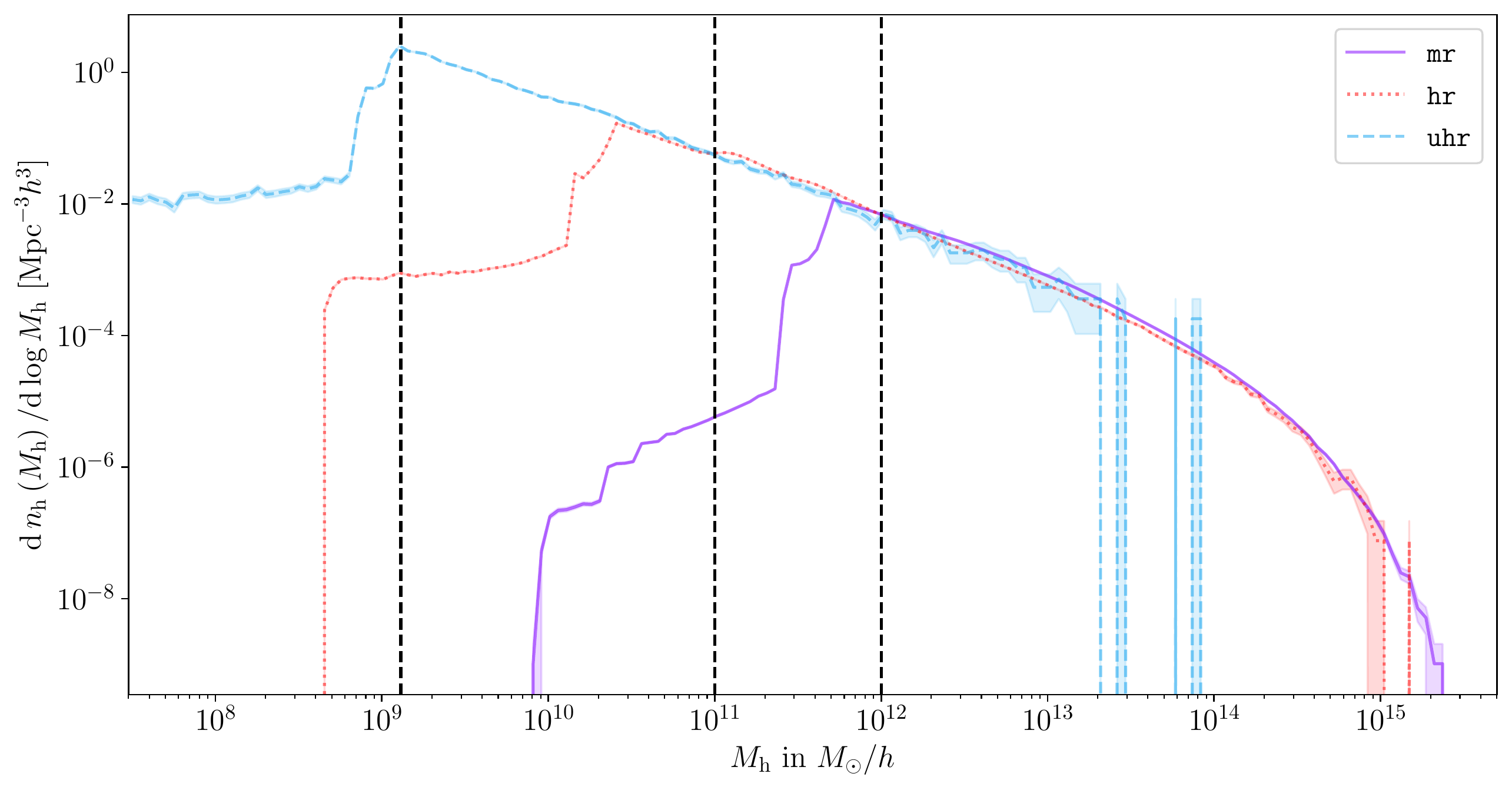}}
\caption{Halo mass function in the \MR{}, \HR{} and \UHR{} simulations at $z = 0.29$ (see table~\ref{table_Magneticum_sims}). Vertical lines indicate the chosen halo mass cuts for void finding at $1.3 \times 10^{9} \Msun$ (\uhr{}), $1.0 \times 10^{11} \Msun$ (\hr{}) and $1.0 \times 10^{12} \Msun$~(\mr{}). }
\label{fig_halo_mass_function}
\end{figure}

\begin{table}[ht]
\centering
\begin{tabular}{|c | c c c c c c|} 
 \hline
 Name & Box & $L_\mathrm{Box}$ & $N_\mathrm{particles}$ & $m_\mathrm{CDM}$ & $m_\mathrm{baryon}$ & $z$ \\ [0.5ex] 
 \hline
 \rule{0pt}{3ex}
 \MR{} (\mr{}) & 0 & 2688 & $2 \times 4536^3$ & $1.3 \times 10^{10}$ & $2.6 \times 10^9$ & 0.29  \\
 \HR{} (\hr{}) & 2b & 640 & $ 2 \times 2880^3$ & $6.9 \times 10^8$ & $1.4 \times 10^8$ & 0.29 \\
 \UHR{} (\uhr{}) & 4 & 48 & $ 2 \times 576^3$ & $3.6 \times 10^7$ & $7.3 \times 10^6$ & 0.29 \\ [1ex]
 \hline
\end{tabular}
\caption{Properties of the \Mag{} simulation boxes used in this work. $L_\mathrm{Box}$ is in units of $\hMpc$ and masses ($m_\mathrm{CDM}$ \& $m_\mathrm{baryon}$) are in units of $\Msun$.}
\label{table_Magneticum_sims}
\end{table}

We identify subhalos and their properties via the \textsc{SubFind} algorithm~\cite{Springel2001b}, modified to take baryonic components into account~\cite{Dolag2009}. Henceforth, we will refer to these subhalos simply as `halos'. The center of each halo is defined as the location of the member particle with the minimal gravitational potential and halo masses are defined by the total mass of their baryonic and CDM particles, according to the subhalo mass definition of \textsc{SubFind}. Due to the different mass resolutions used in the simulations, we chose a minimum halo mass of $1.0 \times 10^{12} \Msun$ in \MR{}, $1.0 \times 10^{11} \Msun$ in \HR{} and $1.3 \times 10^{9}  \Msun$ in \UHR{} to select halos for void finding. These mass cuts are above the resolution limit of each corresponding box at redshift $z = 0.29$, as can be seen in figure~\ref{fig_halo_mass_function}. Halos with masses below these mass cuts will not be considered in any further analysis in this work. In all the aforementioned boxes we analyze voids found in halos from the hydrodynamical simulations and voids found in subsamplings of the underlying CDM that formed these halos.

\section{Methodology \label{sec:methods}}

\subsection{VIDE void finding \label{subsec:void_finding}}
To identify voids in both the halo, as well as the underlying CDM distribution, we make use of the Void IDentification and Examination toolkit {\textsc{vide}}\footnote{\url{https://bitbucket.org/cosmicvoids/vide_public/wiki}} \cite{Sutter2015}. \textsc{vide} implements an enhanced version of the ZOnes Bordering On Voidness algorithm \textsc{zobov}~\cite{Neyrinck2008}, which is a watershed algorithm~\cite{Platen2007} that identifies local basins in the three-dimensional density field estimated from the positions of tracer particles. This density field is constructed via Voronoi tessellation, where each tracer particle \textit{j} is assigned a unique Voronoi cell of volume $\textit{V}_j$. Voronoi cells are defined as the volume of space that is closer to its associated tracer particle than to any other particle in the entire simulation box. Hence, the volumes of all Voronoi cells combined make up the whole simulation box. The density anywhere inside the cell of particle \textit{j} is then simply given by $n_j = 1 / \textit{V}_j$. Starting from the local minima in the density field, the watershed algorithm searches for neighbouring cells with monotonically increasing density to find extended basins of density depressions, our cosmic voids. 

Additionally, \textsc{vide} allows for a density-based merging threshold within \textsc{zobov}, which is a free parameter. Adjacent basins will be added to a void only if the lowest density along their common ridge line is below this threshold in units of the mean tracer density $\bar{n}$. A low threshold prevents voids from extending deeply into overdense structures, thereby limiting the depth of the void hierarchy~\cite{Neyrinck2008,Sutter2015}. In case two adjacent basins are merged, there will be two voids with one encompassing the other, creating a `parent' and a `child' void, hence the original number of voids is maintained even after merging. The difference is that the child-level void is now considered as a sub-void of the parent-level void, which encompasses the volume of both voids. In this way, every void can only have one parent, but multiple children (sub-voids). The default value is set to a very low number to prevent any merging of adjacent basins, which results in a sample of parent voids without any sub-voids. A higher value approaching values of order one and above creates a void hierarchy with potentially many levels of sub-voids, but the total number of voids remains independent of the merging threshold. Unless stated otherwise, in this paper we use a default merging threshold of $10^{-9}\bar{n}$ (no merging). 

In the literature it is also common to use a threshold value of $0.2\bar{n}$ for merging voids, which has a special physical significance. It derives from the spherical expansion model for an inverted top-hat perturbation in an Einstein-de Sitter universe, where the value of $0.2\bar{n}$ marks the characteristic density inside the top hat when shell crossing occurs in its boundaries~\cite{Blumenthal1992,Sheth2004,Neyrinck2008}. However, this only applies when the threshold is defined in the full matter density field and when spherical symmetry is assumed. For voids defined in the number density field of tracers one additionally has to account for the tracer bias~\cite{Pollina2017,Ronconi2019,Contarini2019,Verza2022}. Moreover, the sparsity of tracer particles effectively smooths the density field on scales below their typical separation, which can affect density ridges between voids~\cite{Sutter2014a}.

When investigating the effects of merging, we therefore restrict ourselves to the two extreme thresholds for merging, namely a value close to zero for no merging (our default) and a value of infinity for a fully merged void hierarchy. The former case will be referred to as an \emph{isolated}, the latter as a \emph{merged} void catalog, respectively. The resulting catalogs encompass all cases with a finite merging threshold in between. The \emph{merged} catalog consists of one parent void with a deep hierarchy of children, whereas the \emph{isolated} catalog only contains separate, non-overlapping voids.

Performing the void finding with \textsc{vide} results in catalogs of non-spherical voids with various properties. The center of each void is defined as the volume-weighted barycenter of all of its constituent Voronoi cells at the comoving tracer locations $\bm{x}_j$:

\begin{equation}
\label{eq:void_barycenter}
\mathbf{X}_\mathrm{v} = \frac{\sum_j \, \bm{x}_j \, \textit{V}_j}{\sum_j \, \textit{V}_j}.
\end{equation}
We can think of this barycenter as the geometric center of a void, since its position is constrained by its boundary, which contains most of the void's tracer particles. This definition makes the position of the center very robust against Poisson fluctuations and preserves information about the void topology. Note that the barycenter does not necessarily coincide with the position of the lowest density Voronoi cell due to the lack of spherical symmetry in voids. For the same reason we can only define an \emph{effective radius} $r_\void$, which corresponds to the radius of a sphere of identical volume as the void. It is calculated as the sum over all its associated Voronoi cell volumes:
\begin{equation}
\label{eq:void_radius}
r_\void = \left( \frac{3}{4 \pi} \sum_j  \, \textit{V}_j \right)^{1/3} .
\end{equation}
For quantifying the shapes of voids, \textsc{vide} calculates their inertia tensor, defined via:
\begin{equation}
\label{eq:void_inertia_tensor}
\begin{split}
M_{xx} &= \sum_j \left( y_j^2 + z_j^2 \right)  \, , \\
M_{xy} &= - \sum_j x_j \, y_j \,,
\end{split}
\end{equation}
where we sum over all member particles with comoving coordinates $x_j$, $y_j$, and $z_j$ relative to the void's center. The other components of the inertia tensor are defined accordingly. The void ellipticity is then given in terms of the smallest ($J_1$) and largest ($J_3$) eigenvalues of the inertia tensor~\cite{Sutter2015}:
\begin{equation}
\label{eq:void_ellipticity}
\varepsilon = 1 - \left( \frac{J_1}{J_3} \right)^{1/4} \,.
\end{equation}

Further void properties of interest include the core density $\coreDens$ and the compensation $\Delta_\tracer$~\cite{Schuster2019}. The index `t' for tracers is either `h' in case of finding voids in the halo distribution or `m' for CDM, respectively. The core density is the density inside the largest Voronoi cell of a void, thus the cell of minimal density, expressed in units of the mean tracer density:
\begin{equation}
\label{eq:void_coreDensity}
\coreDens =   \frac{ n_{\mathrm{min}}   }{  \bar{n} }\,.
\end{equation}
The compensation is a measure of whether a void contains more or less tracer particles $N_\tracer$ than an average patch of the Universe of same volume $V$. It conveys information about the environment the void is located in, whether it be one of higher or lower local average density $\hat{n}_{\mathrm{avg}}$ in units of the mean $\bar{n}$. Voids with $\Delta_\tracer > 0$ are referred to as being overcompensated and voids with $\Delta_\tracer < 0$ as being undercompensated~\cite{Hamaus2014a,Hamaus2014b}. The compensation is defined as:
\begin{equation}
\label{eq:void_compensation}
\Delta_\tracer \equiv \frac{N_\tracer/V}{\bar{n}} - 1 = \hat{n}_{\mathrm{avg}} - 1\,.
\end{equation}

\subsection{Void profiles \label{subsec:void_profiles}}
The density profile $n_\void^{(i)}(r) / \bar{n} - 1$ of an individual void $i$ is defined as the spherically averaged density contrast around the center of a void from its mean value $\bar{n}$ in the Universe. When using tracer particles, the density in radial shells of thickness $2 \delta r$ at a comoving distance $r$ from the void center at the origin can be written down in its general form as:
\begin{equation}
n_\void^{(i)}(r) = \frac{3}{4\pi \, \bar{w}} \sum_{j } \frac{w_j \, \, \Theta(r_j)}{ \left( r + \delta r \right)^3 - \left( r - \delta r \right)^3 } \, \, ,
\label{eq:general_density_profile} 
\end{equation}
where $\Theta( r_j) \equiv \vartheta \left[ r_j - \left( r - \delta r \right) \right] \, \vartheta \left[- r_j + \left( r + \delta r \right) \right] $ uses two Heaviside step functions $\vartheta$ to define the radial bins of the profile, with $r_j$ being the distance of tracer $j$ from the void center. Here we sum over all tracers $j$ in the vicinity of the void, up to our desired maximal distance. The $w_j$ in equation~(\ref{eq:general_density_profile}) are a placeholder for optional weights and $\bar{w} = \frac{1}{N_\tracer} \sum_k w_k$ is the average of weights over all $N_\tracer$ tracers. For the usual (unweighted) density profile we simply set $w_j=1$ for every tracer. Another option is to include mass weighting for halos as tracers, which we will explore in section \ref{subsec:mass_weight_profiles}.

Besides the individual density profiles of voids, we are also interested in stacked profiles. These stacks are simply an average over the individual profiles in ranges of different void properties, typically their radius $r_\void$. In order to maintain characteristic void features, such as their compensation wall, at a constant location from the void center in a given stack, and in order to better compare different stacks with each other, we calculate the profile of every individual void using constant radial bin sizes when expressed in units of their void radius, i.e. $\delta r/r_\void = \mathrm{const}$. In this way every void's compensation wall is centered around $r = r_\void$, and not scattered among different distances when using constant units in physical scale. In this manner the stacked density profile is given by an average over equation~(\ref{eq:general_density_profile}):
\begin{equation}
n_\void(r) = \frac{1}{N_\void} \sum\limits_{i} n_\void^{(i)}(r) \, .
\label{eq:stacked_density_profile} 
\end{equation}
When calculating density profiles of voids using CDM tracers, we will denote these profiles by $\rho_\void (r) / \bar{\rho} -1$ instead of $n_\void (r) / \bar{n } -1  $ for halo tracers, and $n_\void^{(\mathrm{w})}(r) / \bar{n}^{(\mathrm{w})} - 1$ for mass-weighted density profiles based on halo tracers.

To investigate the radial movement of tracer particles around a void, we calculate its velocity profile around the barycenter. In our definition, positive velocities correspond to an outflow of tracers from the void, whereas velocities are negative when tracers move towards the void center. The velocity profile of each individual void $i$ can be estimated by calculating:
\begin{equation}
\label{eq:individual_velocity_profile} 
u_\void^{(i)} (r) = \frac{  \sum_j \bm{u}_j \cdot \hat{\bm{r}}_j \,\, V_j \,\, \Theta(r_j)  }
{  \sum_j  V_j \,\, \Theta(r_j) } \, .
\end{equation}
Here $\bm{u}_j$ is the peculiar velocity vector of a given tracer $j$, $\hat{\bm{r}}_j = \bm{r}_j / r_j$ the unit vector connecting the void center with the tracer particle $j$, and $V_j$ its Voronoi cell volume. By weighting the individual particle velocities with their Voronoi volumes $V_j$, we ensure that the velocity profile is a volumetric representation of the true underlying velocity field~\cite{Hamaus2014b}. This takes into account that a uniform sampling of velocity fields from an uneven tracer distribution is biased high in denser and biased low in emptier regions~\cite{Massara2022}.

Moving on to stacked velocity profiles, there are two different ways of calculating them. The first option is equivalent to the way of stacking density profiles by simply averaging over the individual velocity profiles of each void in the sample:
\begin{equation}
\label{eq:individual_stack_velocity_profile} 
u_\void(r) = \frac{1}{N_\void} \sum\limits_{i} u_\void^{(i)}(r) \,.
\end{equation}
We will refer to stacks calculated with this method as \emph{individual stacks}. Alternatively, we can average the denominator and numerator of equation~(\ref{eq:individual_velocity_profile}) separately for all voids of the stack before taking their ratio:
 \begin{equation}
\label{eq:global_stack_velocity_profile} 
u_\void(r) = \frac{ \sum_i \left[ \sum_j \bm{u}_j \cdot \hat{\bm{r}}_j \,\, V_j \,\, \Theta(r_j) \right]^{(i)} }
{ \sum_i \left[ \sum_j  V_j \,\, \Theta(r_j) \right]^{(i)} } \,.
\end{equation}
This method will be referred to as \emph{global stacks}. There are typically many voids, in particular small ones, which contain no or very few tracer particles near their centers. Hence, as no velocity can be estimated there, it is set to zero by default. Even stacking many such individual profiles via equation~(\ref{eq:individual_stack_velocity_profile}) will result in no better velocity estimate. However, using global stacking one first gathers the tracers within the entire sample of voids and then divides by the normalization in equation~(\ref{eq:global_stack_velocity_profile}). This guarantees better statistics for tracer counts in shells, but statistically favours larger voids, whose shell volumes are bigger in physical scale. Both methods come with advantages and disadvantages. Depending on the applications they are used for, one may be preferred over the other. We will discuss this issue in more detail in the following sections.

Finally, local mass conservation allows us to relate the density profile with the velocity profile via the linearized continuity equation~\cite{Peebles1980,Peebles1993}:
\begin{equation}
u_\void(r,z) = - \frac{1}{3} \, \frac{f(z)}{b_\tracer} \, \frac{H(z)}{1+z} \, r  \, \Delta(r)  \,,
\label{eq:velocity_density}
\end{equation}
where $f(z) = \Omega_\matter^{\gamma}(z)$ is known as the linear growth rate of density perturbations. Here $\gamma \simeq 0.55$ is the growth index of matter perturbations in GR, $\Omega_\matter(z)$ the matter density parameter, $H(z)$ the Hubble parameter and $b_\tracer$ the bias of tracer particles, with $b_\tracer = 1$ for CDM. Lastly, $\Delta (r)$ is the integrated density contrast, defined as:
\begin{equation}
\Delta (r) = \frac{3}{r^3} \int^r_0 \left( \frac{n_\void(q) }{\bar{n}} - 1\right) q^2 \,\text{d}q .
\label{eq:integrated_density_contrast} 
\end{equation}
Combining equations~(\ref{eq:velocity_density}) and~(\ref{eq:integrated_density_contrast}) then yields:
\begin{equation}
    u_\void( r, z) = - \frac{\Omega_\matter^{\gamma}(z)}{b_\tracer} \, \frac{ H(z)  }{1 + z} \, \frac{1}{r^2} \,
    \int_0^r \left( \frac{n_\void(q) }{\bar{n}} - 1 \right) q^2 \,  \text{d}q \,,
\label{eq:velocity_relation} 
\end{equation}
from which we can see the direct relation between density and velocity profiles. In section~\ref{sec:mass_conservation} we will investigate how well this equation holds up in the environments around cosmic voids.

\section{Magneticum catalogs \label{sec:catalogs}}

\subsection{Tracers \label{subsec:tracer_catalogs}}
As void finding with \textsc{vide} only requires the positions of tracer particles of any kind, we will use both the distributions of halos and CDM in our simulations to identify voids in. We do not identify voids in the total distribution of matter (including baryons) in this work. From here on, we will refer to the voids identified in the CDM distribution simply as \emph{CDM voids} and correspondingly voids identified in the halo distribution will be referred to as \emph{halo voids}. The corresponding numbers and number densities of halos with masses above the mass cuts of $1.0 \times 10^{12} \Msun$ in \MR{}, $1.0 \times 10^{11} \Msun$ in \HR{} and $1.3 \times 10^{9} \Msun$ in \UHR{} are shown in table~\ref{table_Magneticum_voids}. The \MR{} simulations most closely match the expected tracer densities attainable with state-of-the-art galaxy surveys, such as Euclid~\cite{Hamaus2022}.

\setlength{\tabcolsep}{3.5pt}
\begin{table}[t]
\centering
\begin{tabular}{|c | c c c c c c|} 
 \hline
 Name & $M_\mathrm{cut}\,[\Msun]$ & $N_\halo\,[\times 10^6]$ & $\bar{n}_\halo \, [(\hMpc)^{-3}]$   & $\bar{r}_\tracer\,[\hMpc]$ & $N_\void$ in halos &  $N_\void$ in CDM \\ [0.5ex]
 \hline
  \rule{0pt}{3ex} 
 \MR{} &   $1.0 \times 10^{12}$ & $62.1$ & $3.2 \cdot 10^{-3}$ & $6.8$ & $356\,597$ & $600\,273$ \\
 \HR{}  &   $1.0 \times 10^{11}$ & $8.21$ & $3.1 \cdot 10^{-2}$ & $3.2$ & $33\,324$ & $52\,951$ \\ 
 \UHR{}  &   $1.3 \times 10^{9}$ & $0.136$ & $1.2 \cdot 10^{0}$ & $0.93$ & $346$ & $424$ \\ [1ex] 
 \hline
\end{tabular}
\caption{Number of halos $N_\halo$ with $M_\halo \geq M_\mathrm{cut}$, mean halo density $\bar{n}_\halo$ and tracer separation $\bar{r}_\tracer$, number of halo voids and number of CDM voids in the different \Mag{} runs, all at redshift $z = 0.29$. For the CDM voids, subsamples of the CDM tracers that match $N_\halo$ are used for void finding.}
\label{table_Magneticum_voids}
\end{table}

In both \MR\ and \HR\ boxes we additionally perform subsamplings of the CDM particles which closely match the total number of halos used for the void finding. This is done in order to have identical mean tracer separations $\bar{r}_\tracer$ and more specifically to find halo- and CDM-defined voids of roughly similar ranges in void radii. Moreover, these matched subsamplings eliminate differences in the statistics derived from CDM and halo voids that are merely caused by different tracer number densities. In \MR{} this equates to subsampling to around $0.066 \%$ of the total CDM particles, whereas in \HR\ we subsample to around $0.034 \%$ of all CDM particles. The CDM voids of \UHR{} will be of no further relevance. While voids found in the three-dimensional distribution of matter are not directly accessible to observations, we nevertheless study these CDM voids to compare them to the ones identified using halos.

\subsection{Voids \label{subsec:void_catalogs}}
Here we present the \emph{isolated} and \emph{merged} void catalogs extracted from the distribution of halos and CDM in the \MR{} and \HR{} boxes at redshift $z = 0.29$. For details on the number of tracers used for the void finding, we refer to table~\ref{table_Magneticum_voids} and section~\ref{subsec:tracer_catalogs}. Table~\ref{table_Magneticum_voids} additionally contains the numbers of voids that were found using mass cuts and subsamplings, with no difference in void numbers between \emph{isolated} and \emph{merged} catalogs. We will not investigate the void properties and profiles from the \UHR{} simulation further due to the extremely low number of voids, instead we only use \UHR{} for a resolution study of linear mass conservation in section~\ref{sec:mass_conservation}.

\begin{figure}[t]

               \centering

               \resizebox{\hsize}{!}{

                               \includegraphics[trim=12 0 -12 0]{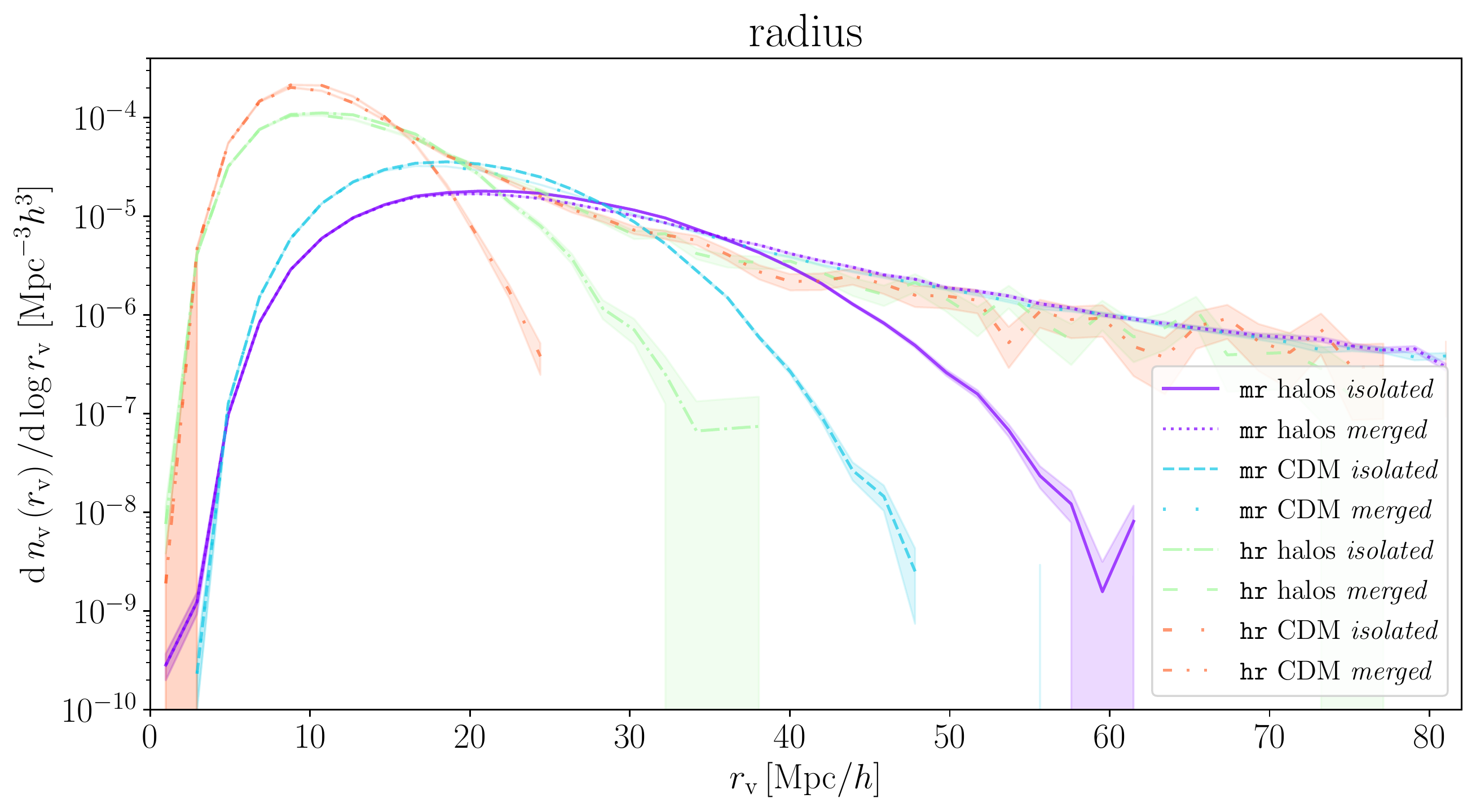}}

               \resizebox{\hsize}{!}{

                               \includegraphics[trim=0 5 0 7, clip]{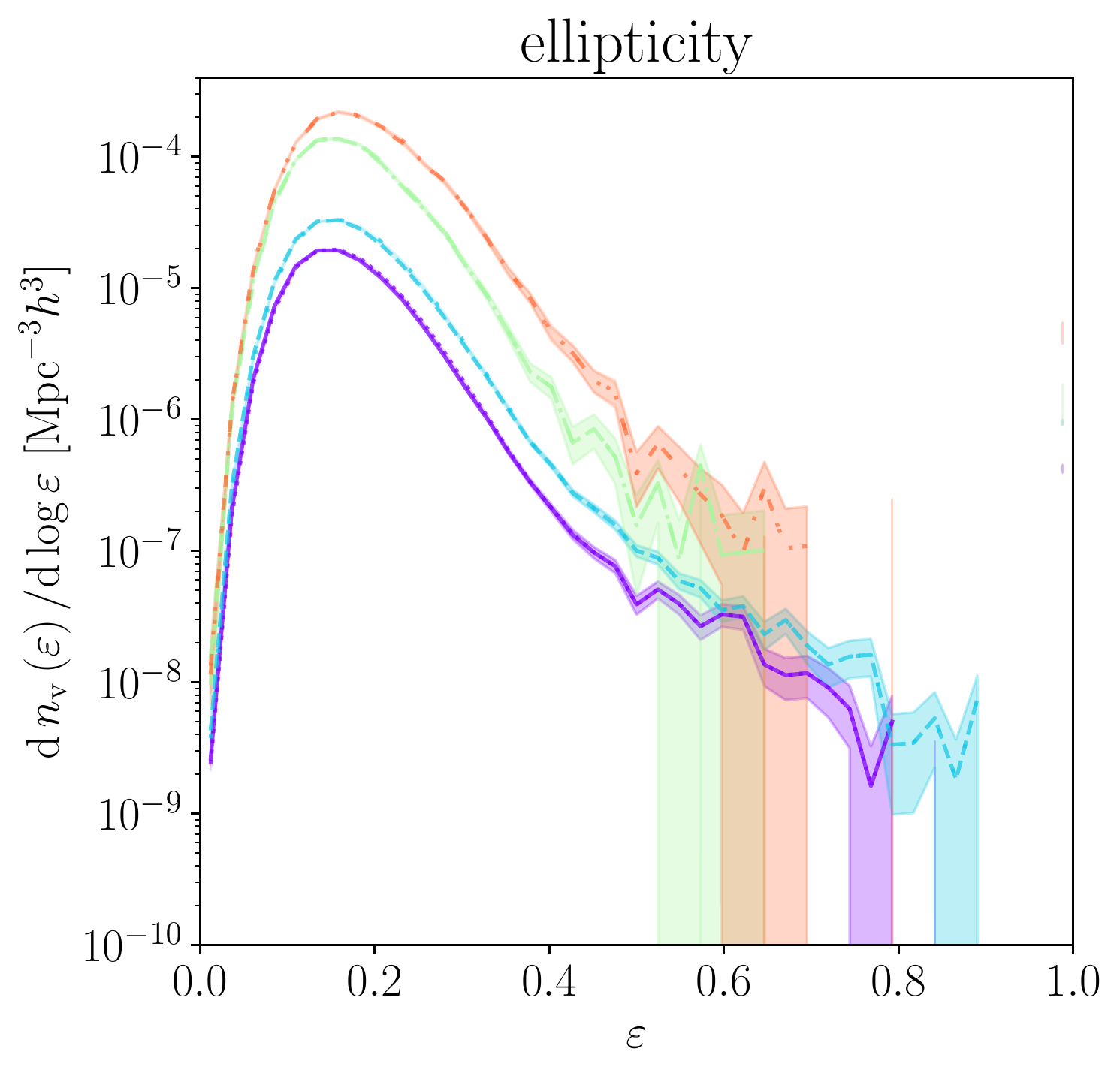}

                               \includegraphics[trim=0 5 0 7, clip]{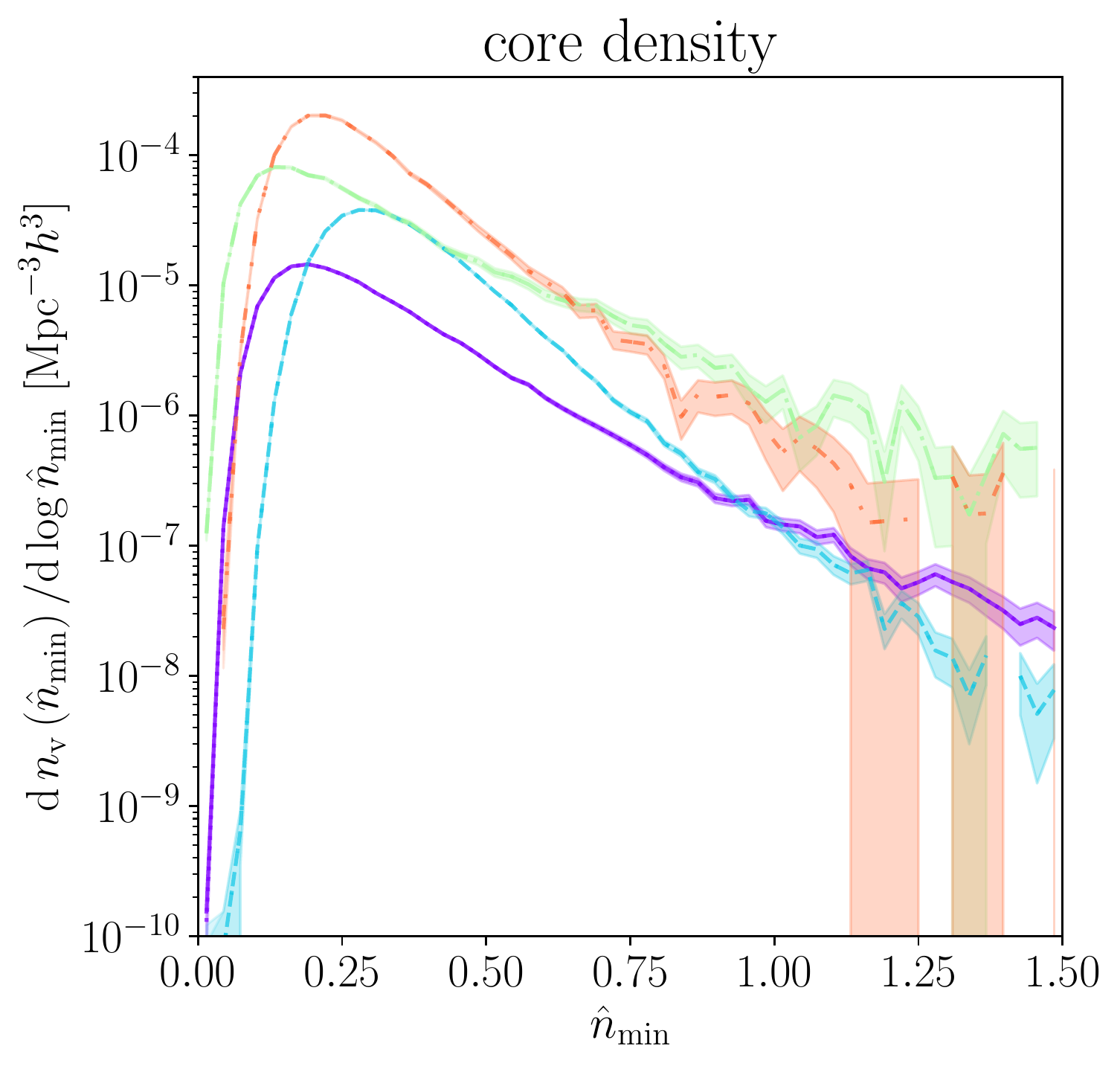}}

               \caption{Abundance of \emph{isolated} and \emph{merged} voids defined in the distribution of halos and CDM in the \MR{} and \HR{} simulations at redshift $z = 0.29$. Shown as a function of void radius (top), ellipticity (bottom left) and core density (bottom right).}

               \label{fig_void_function_halo_CDM_merging_rad_ell_core}

\end{figure}

\begin{figure}[t]

               \centering

               \resizebox{\hsize}{!}{

                               \includegraphics[trim=0 5 0 5, clip]{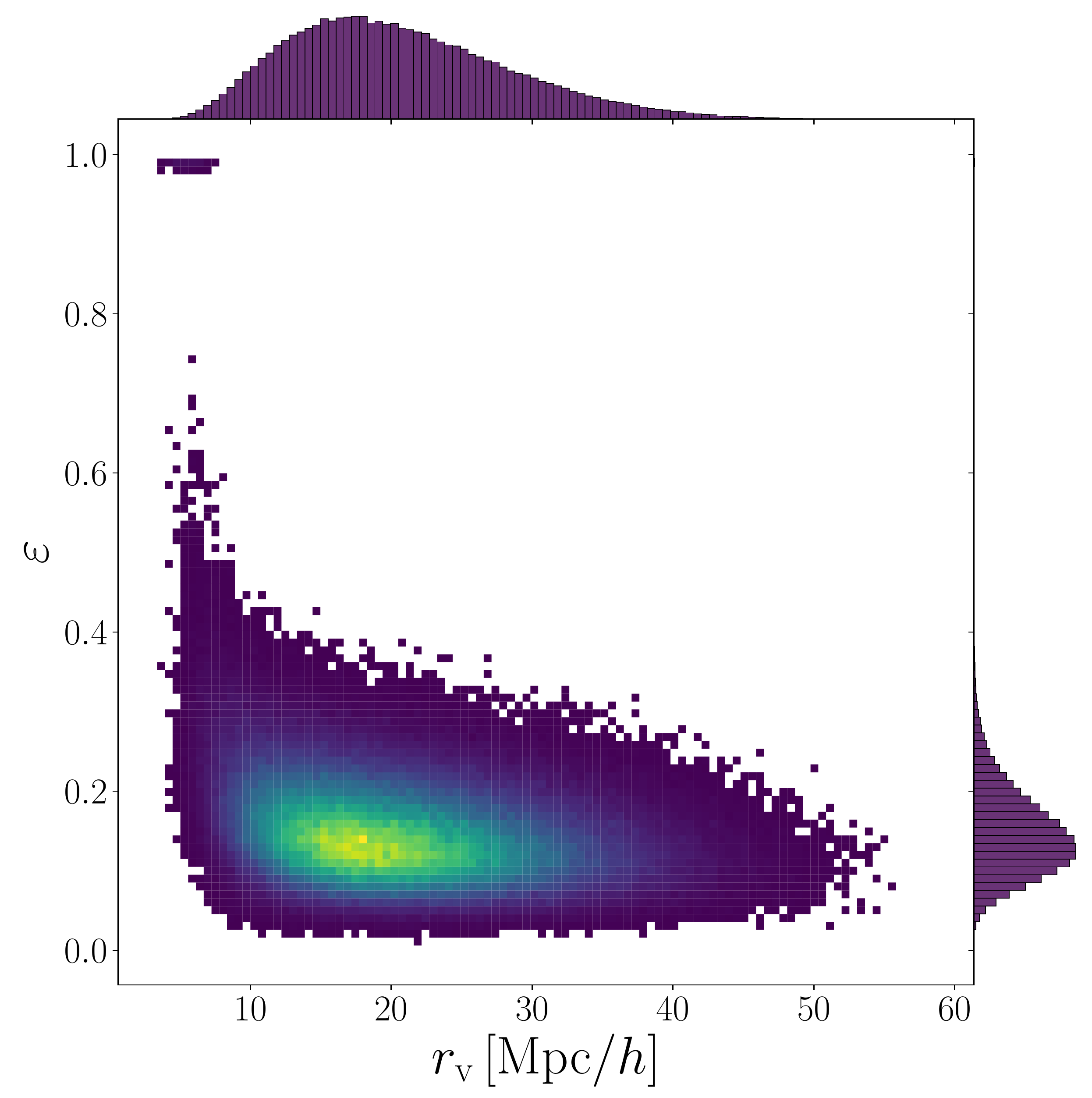}

                               \includegraphics[trim=0 5 0 5, clip]{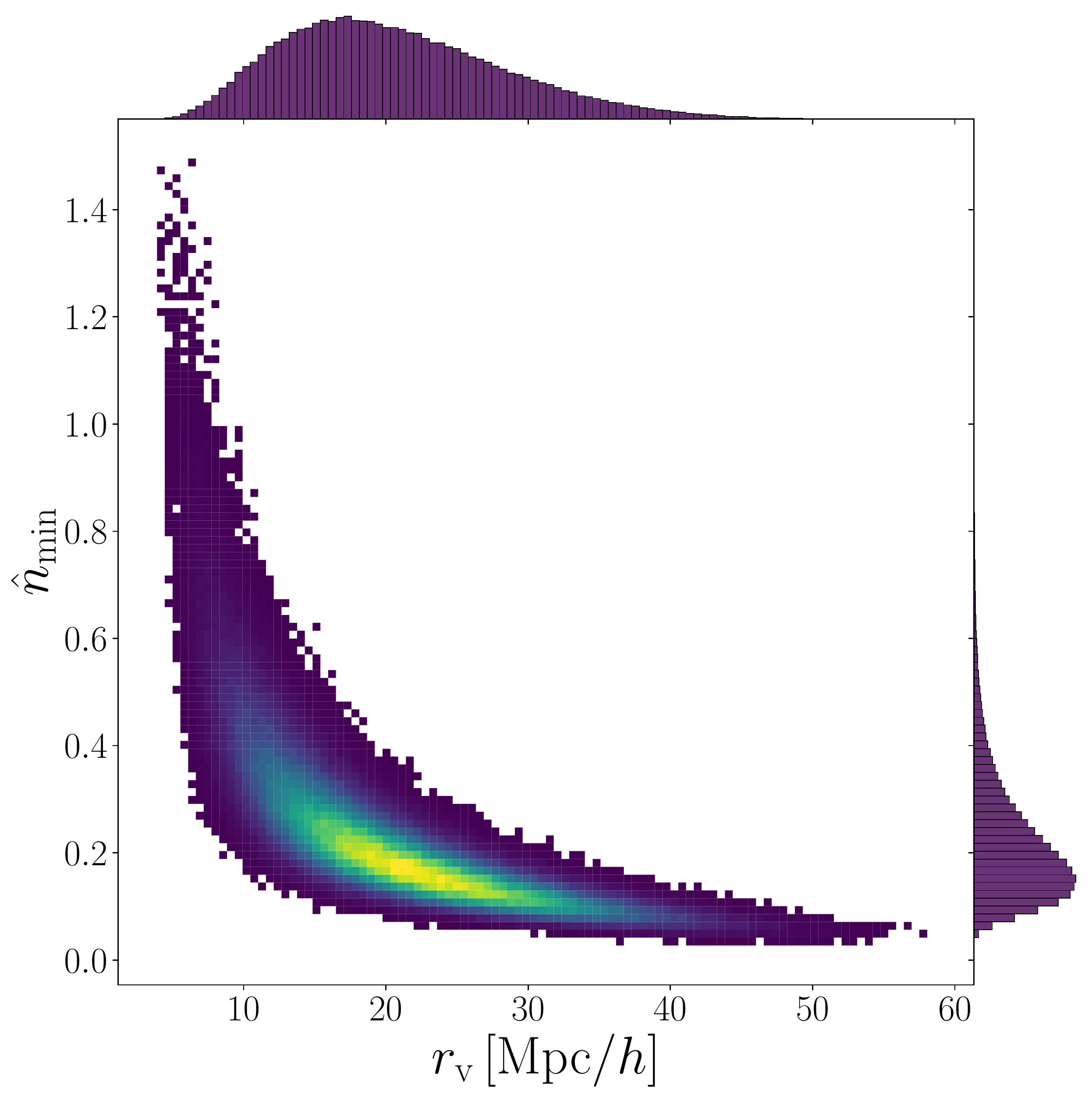}}

               \resizebox{0.5\hsize}{!}{

                               \includegraphics[trim=0 10 0 5, clip]{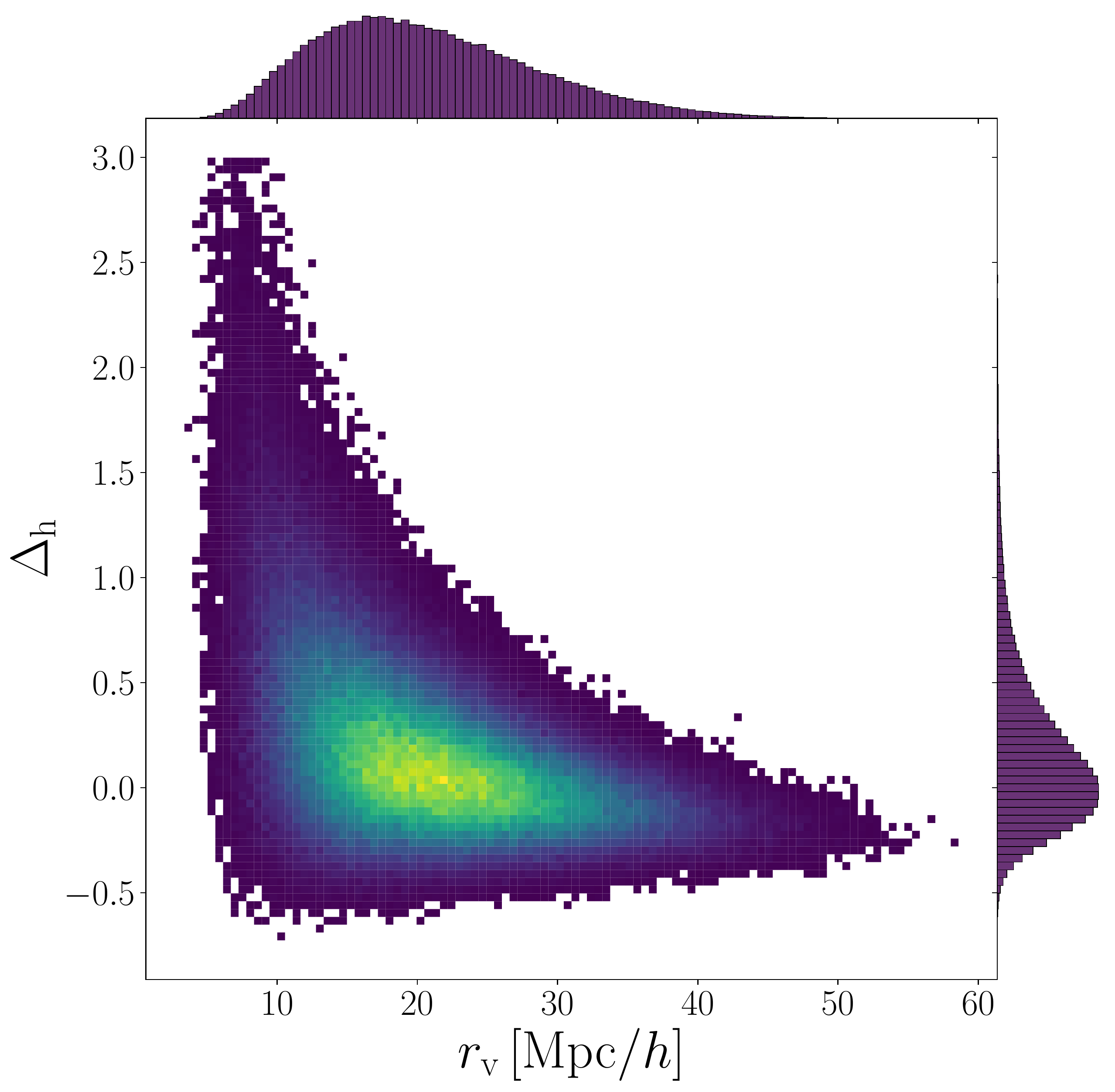}}

               \caption{Two-dimensional distributions of radius, ellipticity (top left), core density (top right) and compensation (bottom) from \emph{isolated} halo voids in the \MR{} simulation. Brighter colors correspond to higher numbers of voids per bin.}

               \label{fig_radius_ellipticity_coreDens_compensation_dependance}

\end{figure}

In figure~\ref{fig_void_function_halo_CDM_merging_rad_ell_core} we present the void abundance of \emph{isolated} and \emph{merged} voids as a function of their radius (up to around $80\,\hMpc$), ellipticity, and core density (up to a value of $1.5$). The former is also known as the void size function~\cite[e.g.,][]{Jennings2013,Pisani2015a,Contarini2019,Verza2019,Contarini2022b, Verza2022b}. We first note that whether or not voids are \emph{merged} or \emph{isolated} only significantly affects their distribution in radius, but their ellipticities and core densities are almost identical. Voids from the \HR{} simulation reach smaller radii than in \MR{}, because the better resolution provides more CDM particles and low-mass halos, and thus a higher overall tracer density.

Note that although the total number of halos and CDM particles are matched for each box, we observe almost twice as many CDM voids (see table~\ref{table_Magneticum_voids}), which implies smaller voids in the CDM for a fixed simulation volume. This difference is due to the bias of halos, which are preferentially located in regions of high matter density. CDM particles sample the density field without this bias, which enables the detection of voids in less dense regions as well. Therefore, establishing a one-to-one correspondence between halo voids and CDM voids from the same simulation is not feasible~\cite{Sutter2014a}.

However, the void size functions of \emph{merged} halo and CDM voids in both \MR{} and \HR{} all agree within their error bars for radii larger than about $45\,\hMpc$. Figure~\ref{fig_void_function_halo_CDM_merging_rad_ell_core} suggests that the abundance of \emph{merged} voids above a certain size is much less dependent on tracer bias, subsampling fraction, mass cuts, and resolution. We additionally ran \textsc{vide} on a subsample of 200 million CDM tracers in \MR{}, which confirmed this result for \emph{merged} voids. No such convergence can be found in the void size function of \emph{isolated} voids, which continuously fragment into smaller ones as the density of tracers increases. However, at the smallest radii the void size function is not affected by merging. These are voids in the \emph{isolated} catalog, which get relabelled to sub-voids in the \emph{merged} catalog. As noted before, the total number of voids in \emph{merged} and \emph{isolated} catalogs is identical, since no new voids are `created', only some voids are \emph{merged}, resulting in the hierarchical structure of parent and child voids.

From the lower left plot in figure~\ref{fig_void_function_halo_CDM_merging_rad_ell_core} it is evident that the ellipticity distribution is more or less identical for voids in all cases, the curves in \HR{} are simply shifted vertically towards higher values due to the larger number density of voids that can be found when resolving CDM particles at higher density, or halos of lower mass. All distributions consistently peak around a value of $\varepsilon\simeq 0.15$. There is a mild indication for merged voids to be slightly more elliptical, likely due to their increased amounts of substructure, which increases the possible complexity of void shapes.

In contrast, the distributions in core density on the lower right panel exhibit clear differences across tracer type and resolution. Halo voids typically exhibit deeper core densities than their CDM counterparts due to halo bias, which amplifies fluctuations in the density field~\cite{Sutter2014a,Pollina2017}. Moreover, the core densities show clear differences between \MR{} and \HR{} beyond a vertical shift of the distributions. Their maxima move towards lower density values as the resolution increases. This is because more nonlinear density fluctuations can be resolved on smaller scales. On the other hand, void merging does not affect the core densities, because the local density minima remain the same by construction.

The joint two-dimensional distributions of void properties are shown in figure~\ref{fig_radius_ellipticity_coreDens_compensation_dependance} for the case of \emph{isolated} halo voids of the \MR{} simulation. Voids from other tracer types, merging thresholds, and resolutions have qualitatively similar distributions. Their ellipticity only slightly depends on void radius, but the distribution becomes narrower towards larger voids. Among the smallest voids we find a few highly elliptical ones with $\varepsilon \approx 1$, which are most likely spurious due to the effects of tracer sparsity (see section~\ref{subsec:lin_theory_individual_voids}). The core density exhibits a stronger correlation with void radius, as evident from the upper right plot of figure~\ref{fig_radius_ellipticity_coreDens_compensation_dependance}. Larger voids tend to have lower core densities with a more narrow distribution~\cite{Nadathur2015}. Towards small voids the distribution widens considerably, featuring some voids with minimum densities even above the mean background value. This is an expected consequence of a parameter-free void definition based on the watershed technique, which does not impose any density threshold on the interior of voids. Because it is purely topological, this method is able to identify local basins above the mean density. Finally, the distribution of void compensation with radius is depicted in the bottom plot. Its shape falls in between the ones for ellipticity and core density, exhibiting an anti-correlation between compensation and void radius. Small voids can be either overcompensated when found in very dense environments, or undercompensated when found inside larger voids. Because their distribution is very skewed towards positive compensation, however, on average small voids are overcompensated. On the contrary, large voids are preferentially undercompensated~\cite{Hamaus2014a}.

\section{Profiles \label{sec:profiletypes}}
In this section we focus on the stacked density and velocity profiles of voids. The profiles are calculated individually for each void out to five times its effective radius in bins of width $0.1 \times r_\void$ and then stacked (averaged) in bins of different void properties. We use \emph{isolated} halo voids as our default, but investigate the impact of merging and CDM voids as well. Additionally, we calculate density profiles of halo voids using the underlying distribution of subsampled CDM. We then distinguish between their number density profiles and matter density profiles, depending on whether halos or CDM particles are used for the profile calculation, respectively. The latter case is relevant in weak lensing studies, where voids are identified via luminous tracers (e.g., galaxies) and the matter density field around them is probed by the gravitational shear of background objects~\cite{SanchezC2017,Fang2019}.

\subsection{Density profiles \label{subsec:density_profiles}}

\begin{figure}[t]

               \centering

               \resizebox{\hsize}{!}{

                               \includegraphics[trim=7 50 0 5, clip]{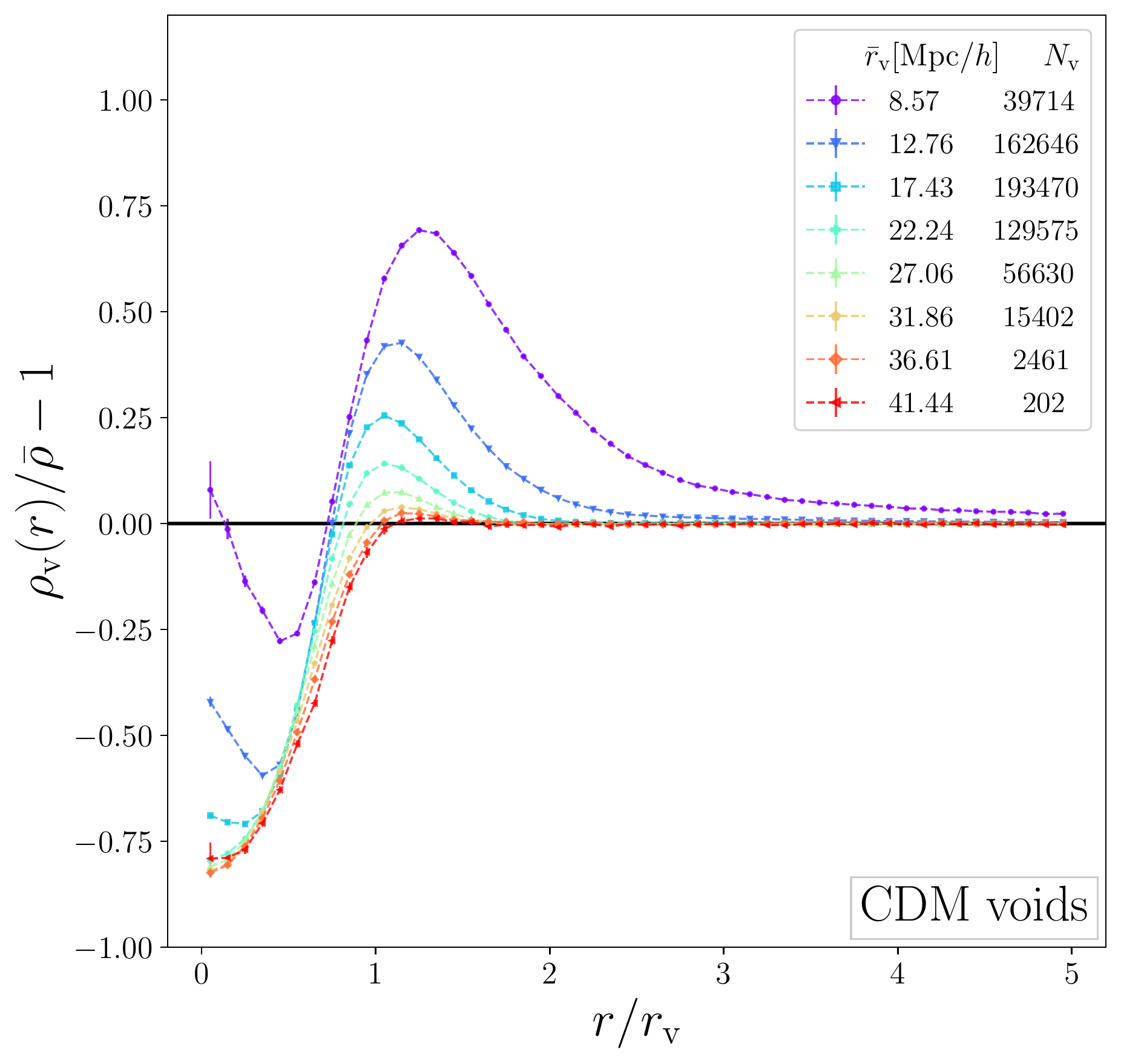}

                               \includegraphics[trim=7 50 20 5, clip]{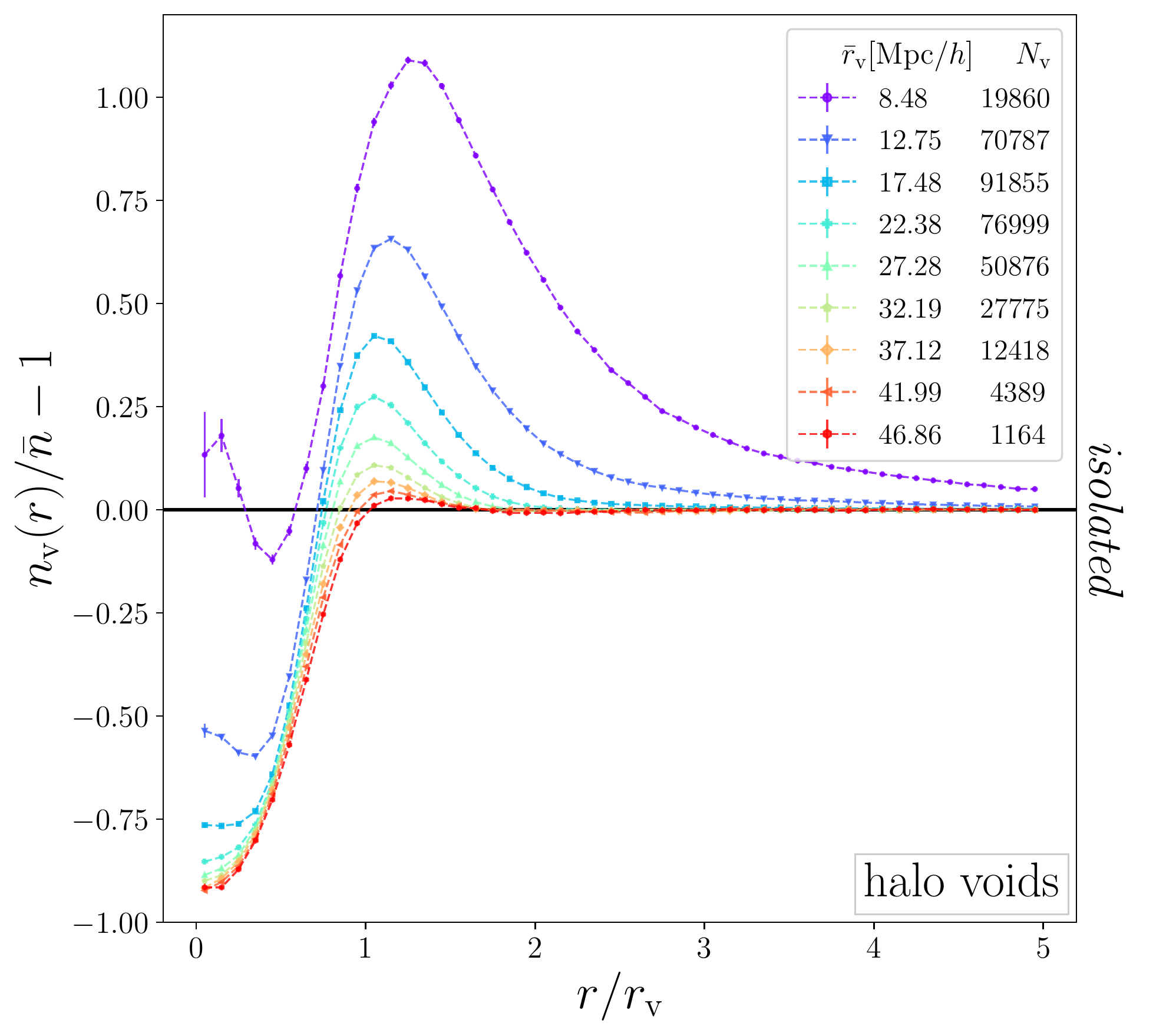}}

               \resizebox{\hsize}{!}{

                               \includegraphics[trim=0 10 0 0, clip]{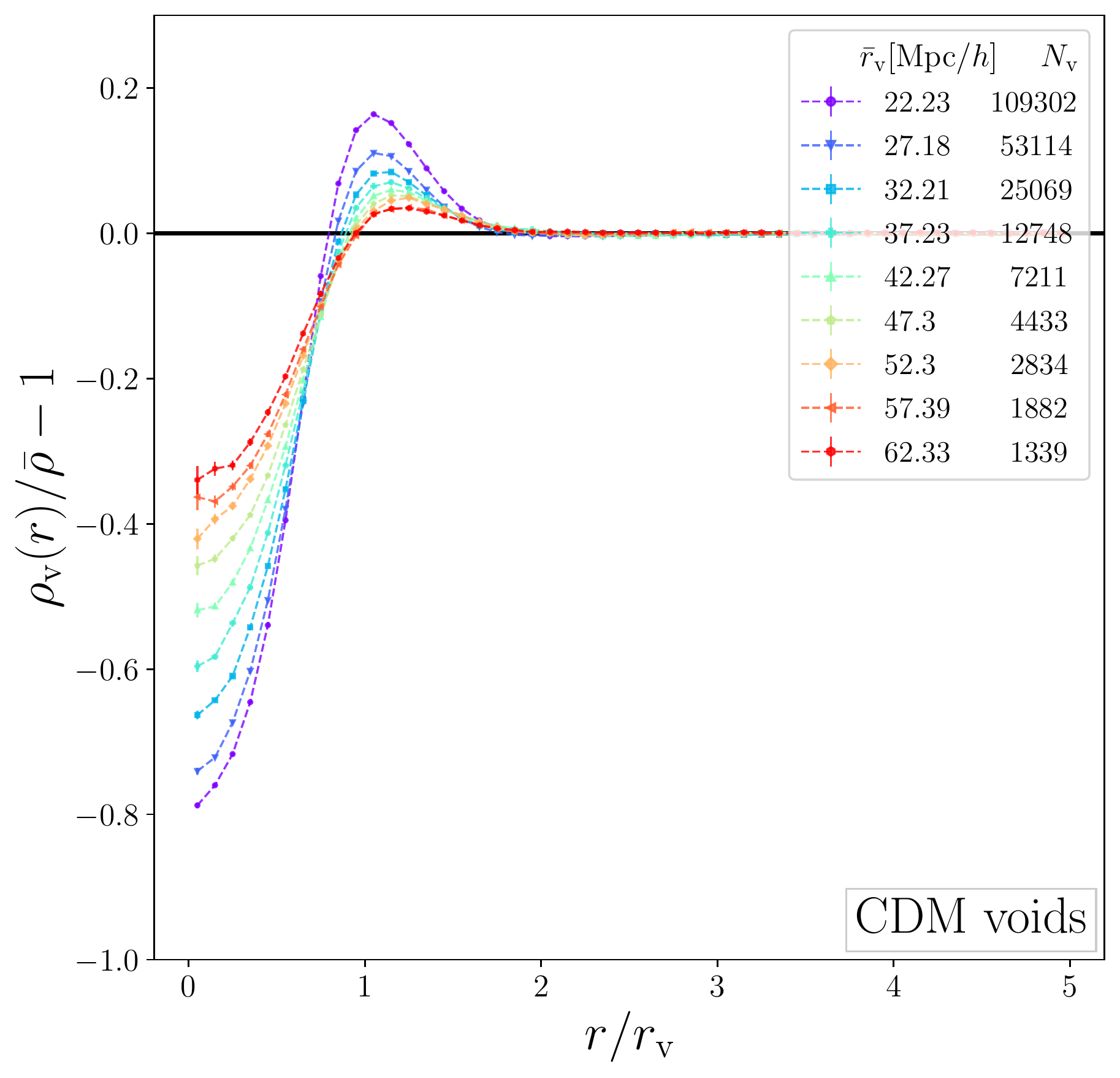}

                               \includegraphics[trim=0 10 20 0, clip]{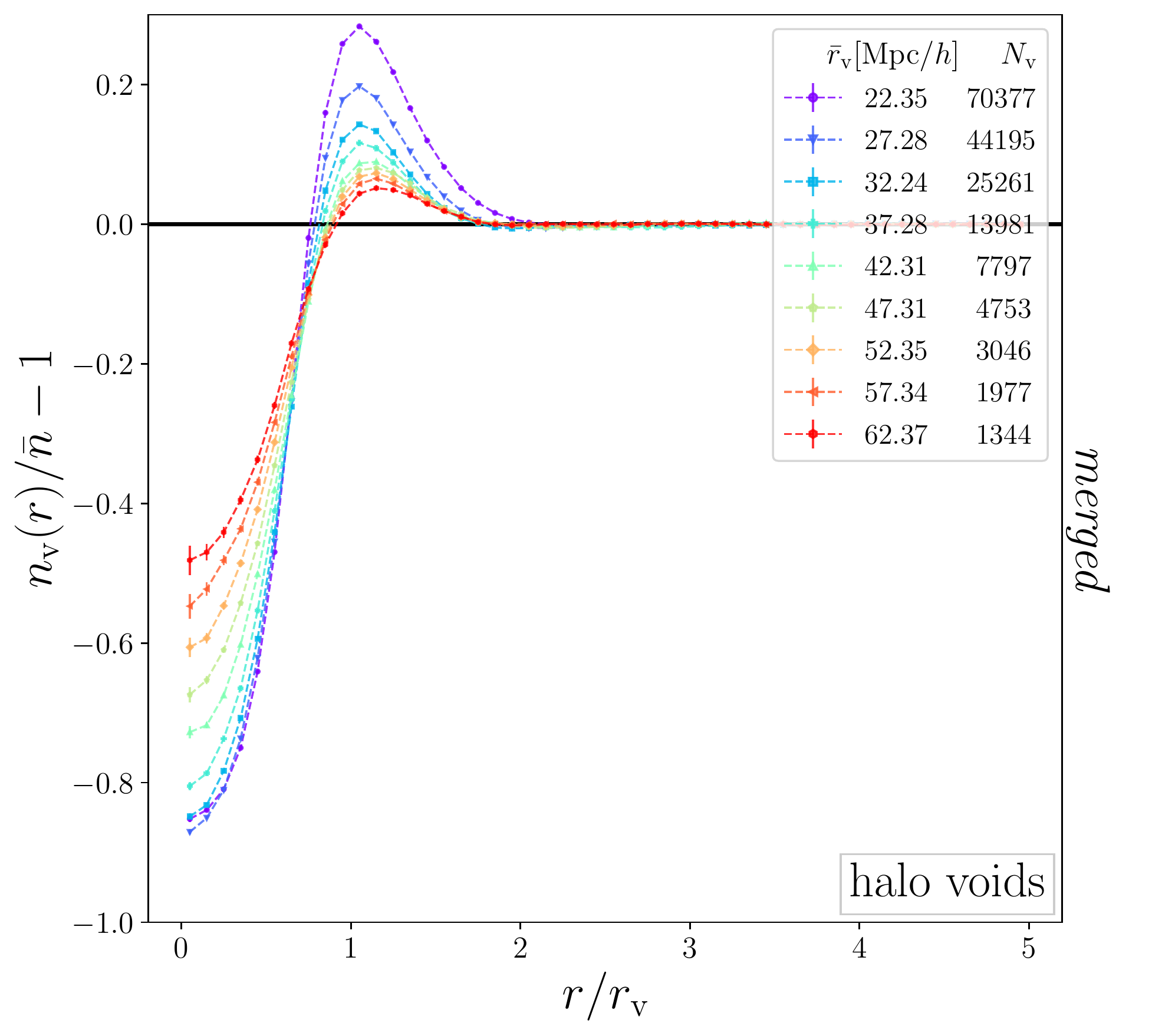}}

               \caption{Density profiles from the \MR{}{} simulation for \emph{isolated} (top) and \emph{merged} (bottom) voids identified in CDM (left) and halos (right). Profiles are stacked in contiguous void radius bins of width 5 $\hMpc$, starting at $5\,\hMpc$ for \emph{isolated} voids and at $20\,\hMpc$ for \emph{merged} voids. The mean void radius of each bin is indicated in the legend. Error bars show standard deviations on the mean profiles.}

               \label{fig_density_mr_CDM_halo_merging}

\end{figure}

\begin{figure}[t]

               \centering

               \resizebox{\hsize}{!}{

                               \includegraphics[trim=0 50 0 5, clip]{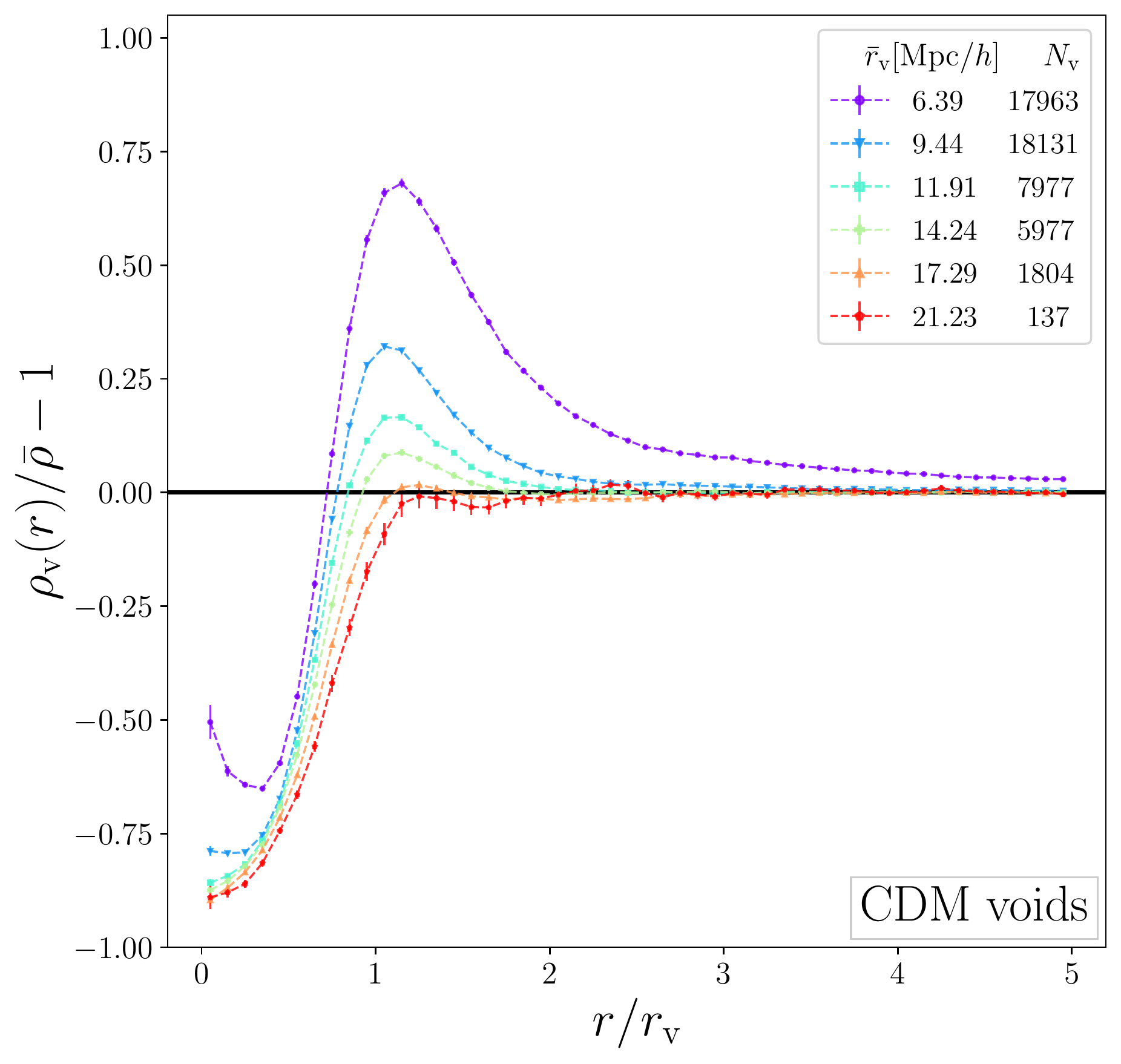}

                               \includegraphics[trim=0 50 20 5, clip]{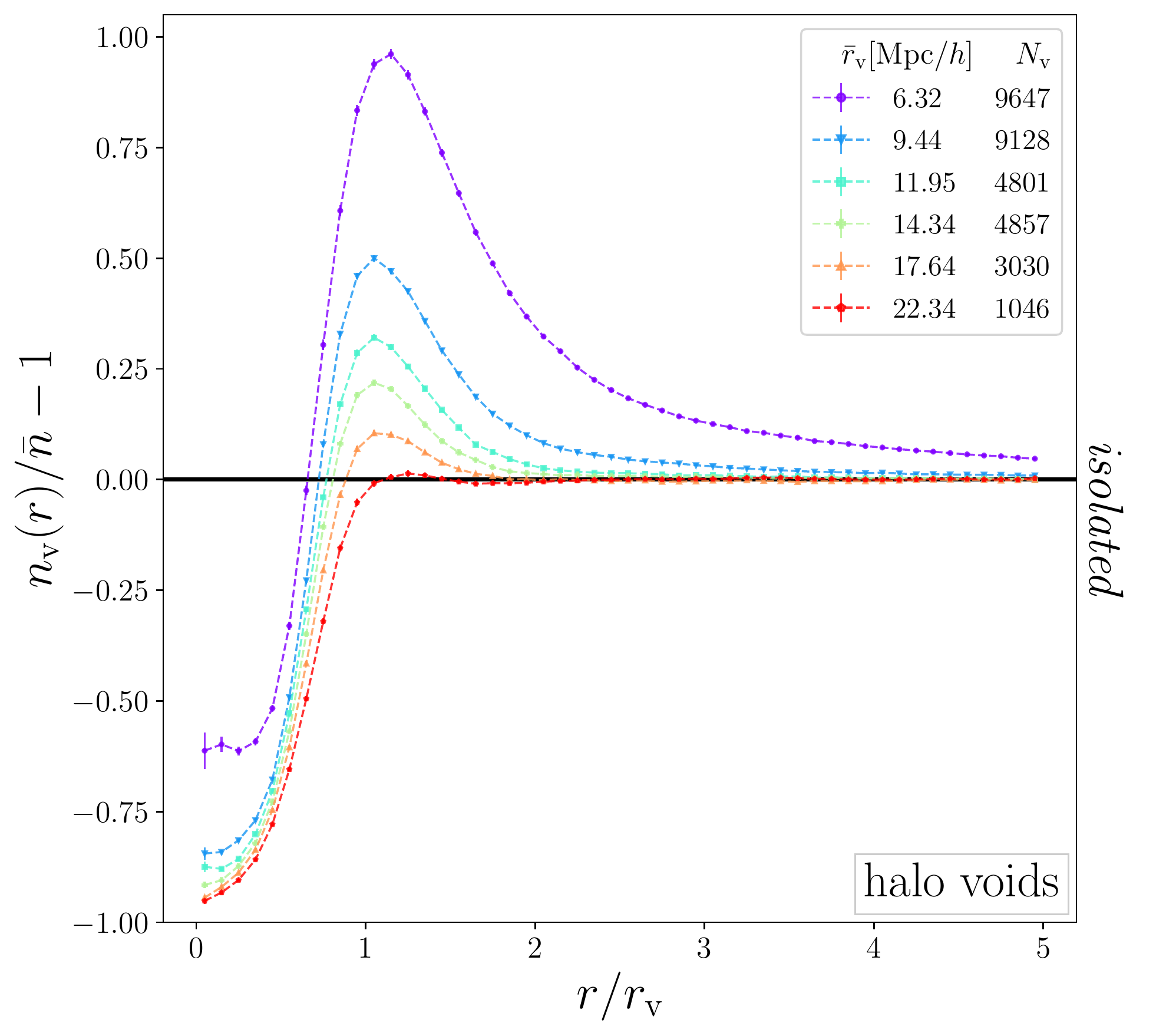}}

               \resizebox{\hsize}{!}{

                               \includegraphics[trim=0 10 0 5, clip]{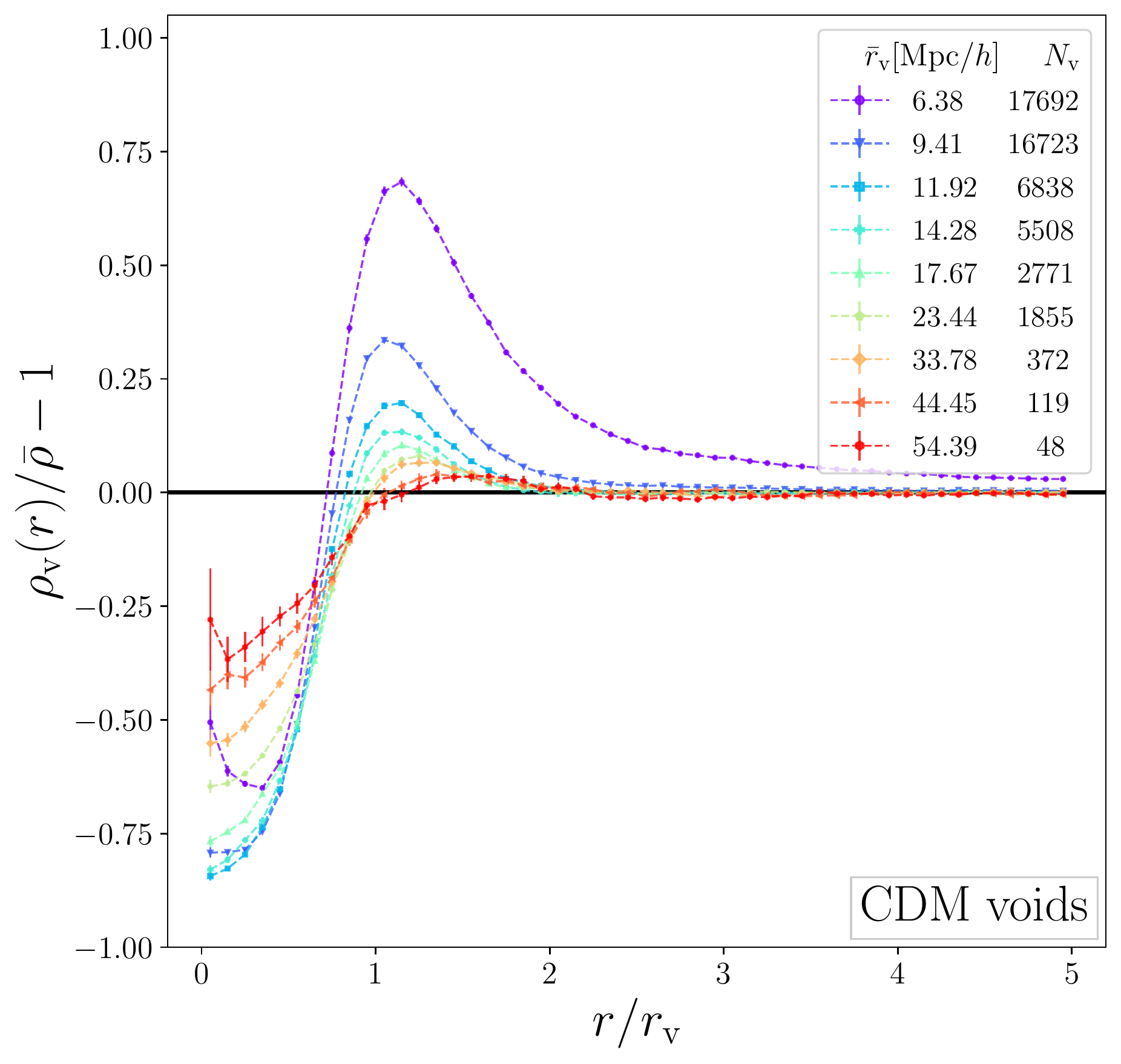}

                               \includegraphics[trim=0 10 20 5, clip]{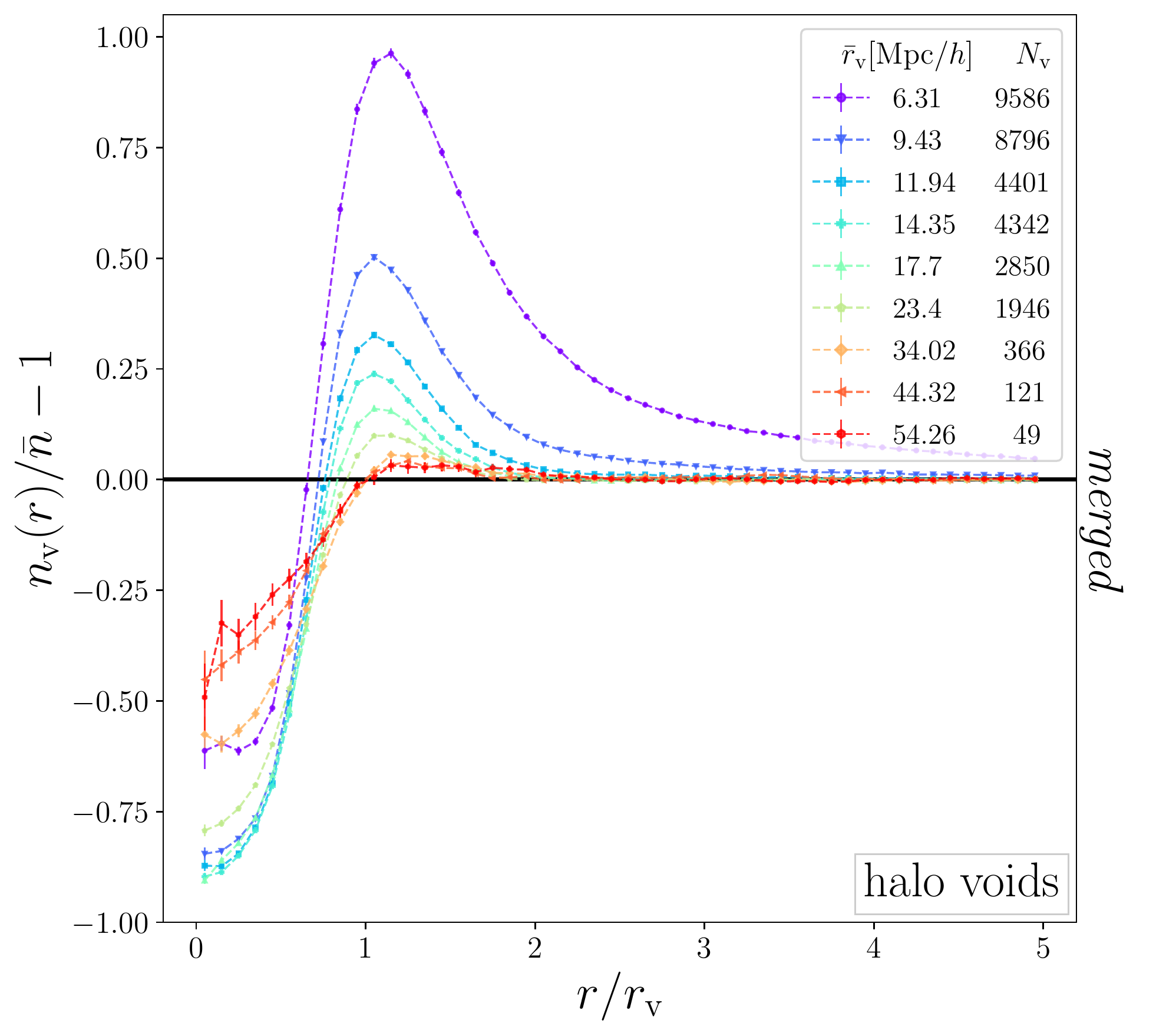}}

               \caption{Same as figure~\ref{fig_density_mr_CDM_halo_merging}, but in the \HR{} simulation. Void radius bins have been adapted to better cover their ranges, for \emph{isolated} voids between $4\,\hMpc$ and $30\,\hMpc$ (top) and for \emph{merged} voids between $4\,\hMpc$ and $60\,\hMpc$ (bottom).}

               \label{fig_density_hr_CDM_halo_merging}

\end{figure}

Figure~\ref{fig_density_mr_CDM_halo_merging} depicts the matter density profiles of CDM voids on the left and the number density profiles of halo voids on the right panels, both for \emph{isolated} (top) and \emph{merged} voids (bottom) from the \MR{} simulation. All profiles are stacked in contiguous void radius bins of $5\,\hMpc$ width. Legends indicate the number of voids and the mean void radius per bin. For \emph{isolated} voids the smallest bin starts at $5\,\hMpc$ and for \emph{merged} voids only at $20\,\hMpc$, since the stacked profiles of smaller \emph{merged} voids are nearly indistinguishable to the \emph{isolated} ones.

The density around \emph{isolated} voids gradually increases with smaller void size, reaching values above even the mean background density in the center of the smallest voids. Since their size is close to the mean tracer separation, one might classify these small voids as spurious. However, they are mostly embedded in environments of relatively high density~\cite{Hamaus2014a}, so the local mean tracer separation is smaller as well. They also exhibit clearly defined compensation walls around $r = r_\void$, which topologically defines them as voids. The slight increase in density towards the center of the smallest voids is caused by the sparse sampling statistics on scales below the mean tracer separation. As they are only defined via a few particles, the density estimate from equation~(\ref{eq:general_density_profile}) returns biased results. In particular, the density is biased high when the shell volume in the denominator is very small near the void center, but happens to contain a tracer particle.

When comparing CDM voids with halo voids, we observe that the latter have higher compensation walls and are slightly deeper near their center. This is consistent with the voids' core density distributions in figure~\ref{fig_void_function_halo_CDM_merging_rad_ell_core}, and the impact of halo bias amplifying the CDM density fluctuations~\cite{Sutter2014a,Pollina2017}. However, we emphasize that a comparison of CDM voids with halo voids of the same size is not necessarily meaningful, due to their significantly different void size functions. For \emph{merged} voids we find a somewhat different behavior, as shown in the lower panels of figure~\ref{fig_density_mr_CDM_halo_merging}. While the density profiles of the smallest voids remain virtually unchanged, merging gives rise to larger voids with slightly higher compensation walls and shallower cores than \emph{isolated} voids, an effect already seen in reference~\cite{Hamaus2014b}. This happens because \emph{merged} voids contain sub-voids and hence sub-structures defining those. When the density profiles of \emph{merged} voids are stacked, these sub-structures effectively get smoothed out, which leads to shallower cores and slightly higher compensation walls in the profiles.

The density profiles of voids from the \HR{} simulation are depicted in figure~\ref{fig_density_hr_CDM_halo_merging}.
In order to obtain a sufficient number of voids in each stack, we now adapt the bin widths in radius, covering a range from $4\,\hMpc$ to $30\,\hMpc$ for \emph{isolated} voids on the top, and from $4\,\hMpc$ to $60\,\hMpc$ for \emph{merged} voids on the bottom. We find consistent results in comparison to the \MR{} simulation shown in figure~\ref{fig_density_mr_CDM_halo_merging}, but resolve even smaller voids (as evident from figure~\ref{fig_void_function_halo_CDM_merging_rad_ell_core}). Again, we caution against directly comparing voids of a given size from different resolution simulations, similarly as for voids defined by different tracer types. Only in the regime where their void size functions converge, typically for larger \emph{merged} voids, this may be meaningful.

\begin{figure}[t]
\centering
\resizebox{\hsize}{!}{
    \includegraphics{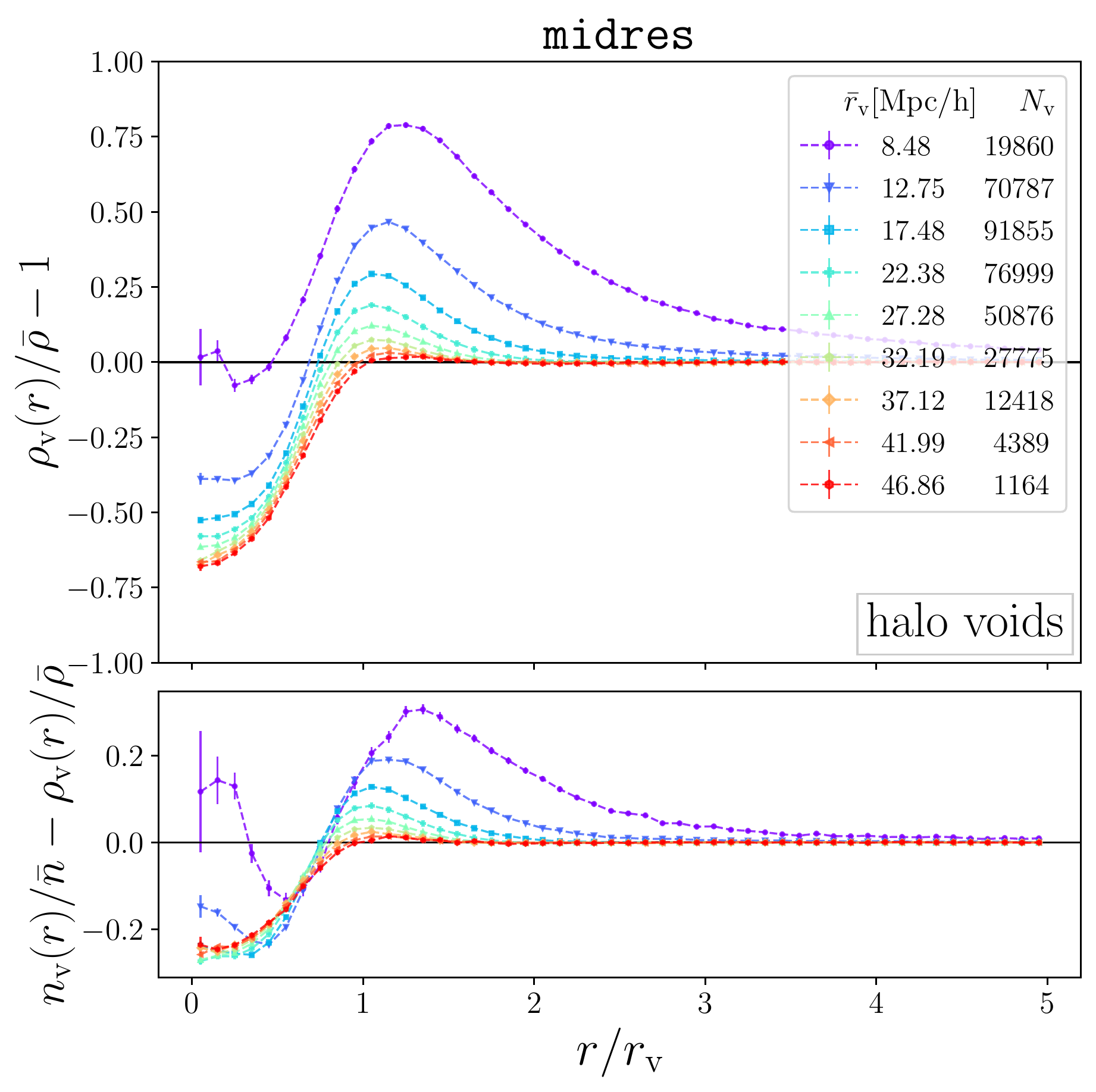}
    \includegraphics{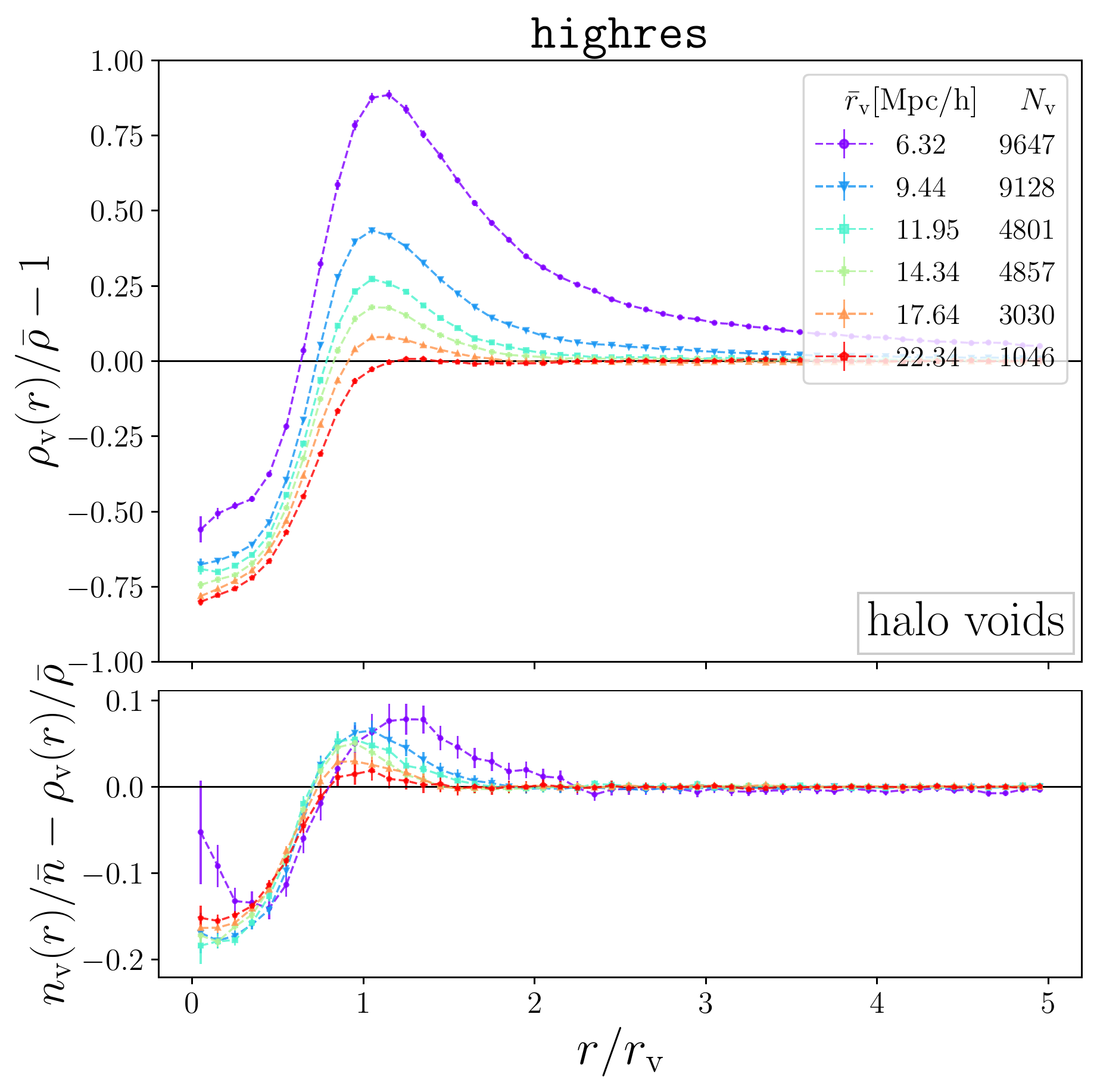}}
\caption{Same as figures~\ref{fig_density_mr_CDM_halo_merging} and~\ref{fig_density_hr_CDM_halo_merging}, but showing the matter density profiles based on CDM particles around \emph{isolated} halo voids in the \MR{} (left) and \HR{} simulation (right). Differences between halo number density and matter density profiles are shown on the bottom.}
\label{fig_matter_density_mr_hr_haloCDM}
\end{figure}

\begin{figure}[t]

               \centering

               \resizebox{\hsize}{!}{

                               \includegraphics[trim=7 50 0 5, clip]{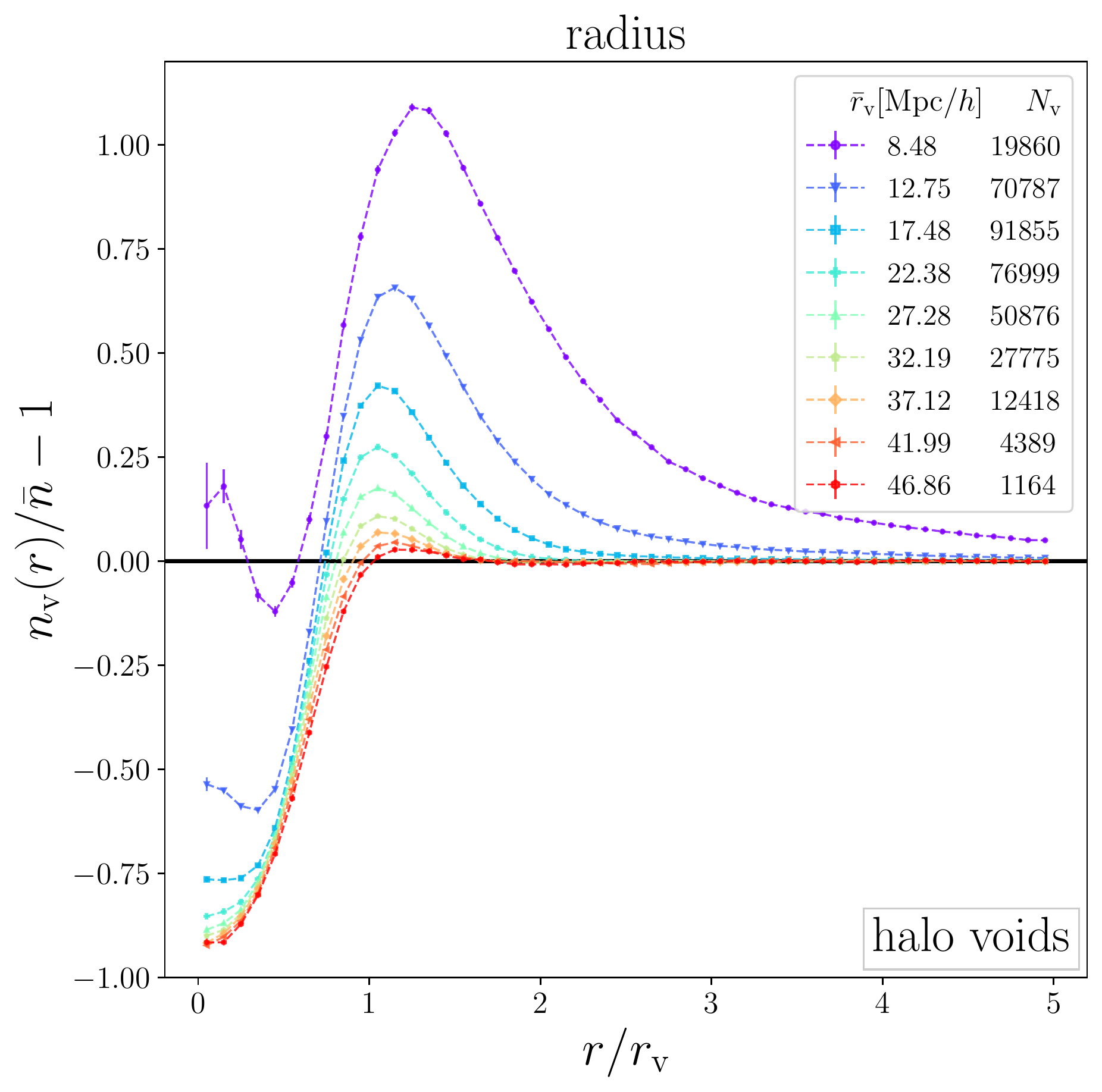}

                               \includegraphics[trim=7 50 0 5, clip]{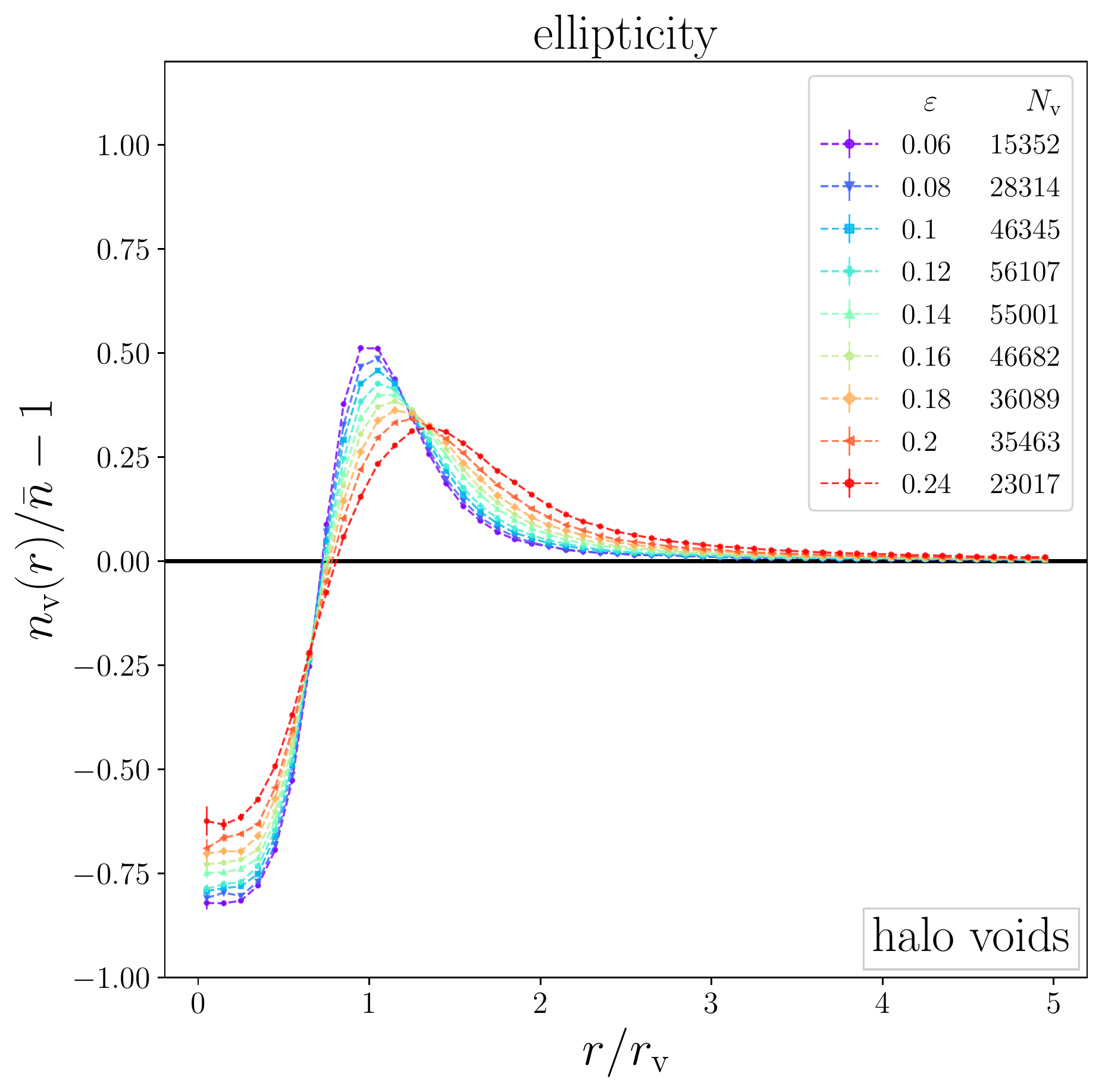}}

               \resizebox{\hsize}{!}{

                               \includegraphics[trim=0 10 0 5, clip]{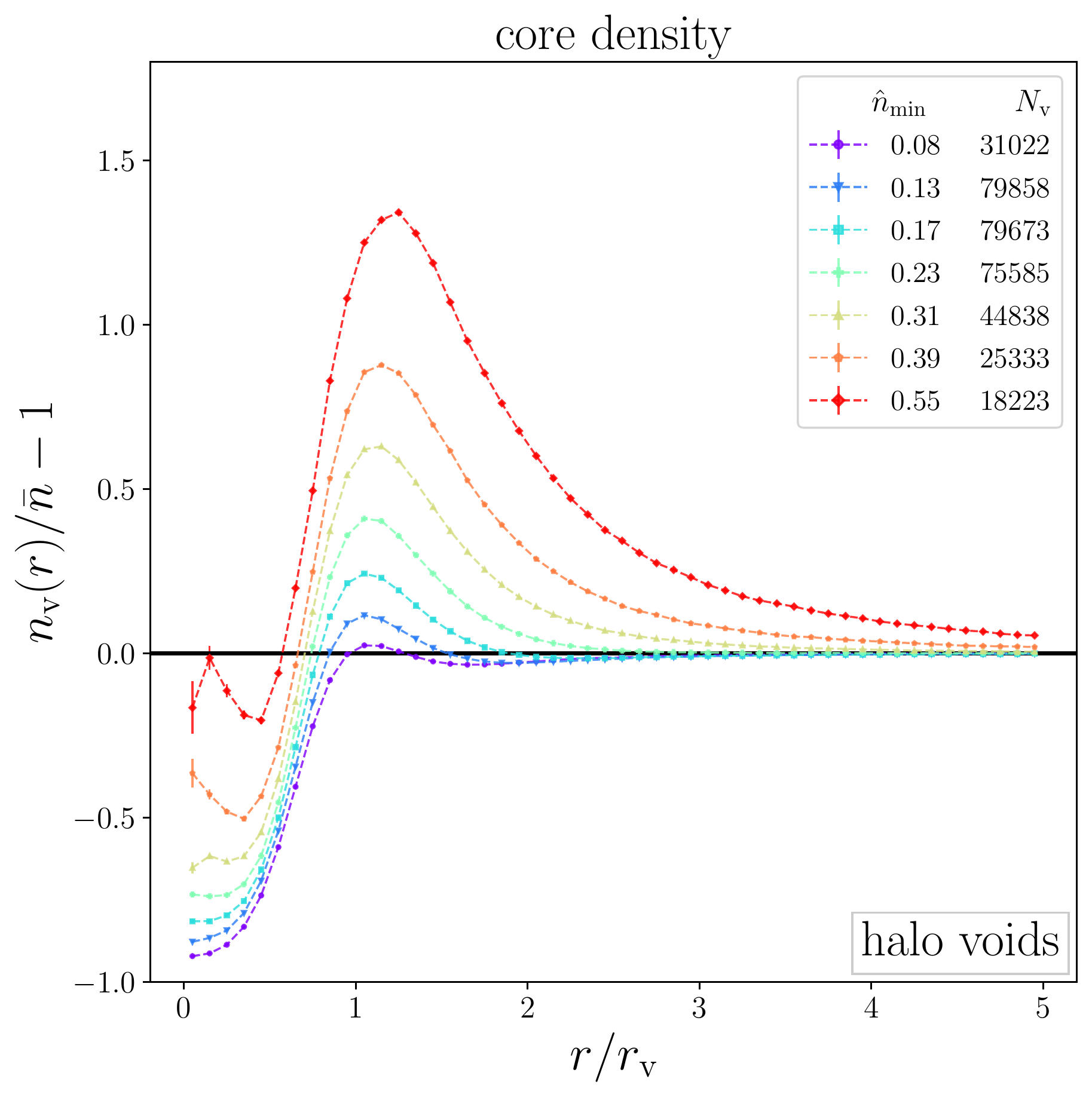}

                               \includegraphics[trim=0 10 0 5, clip]{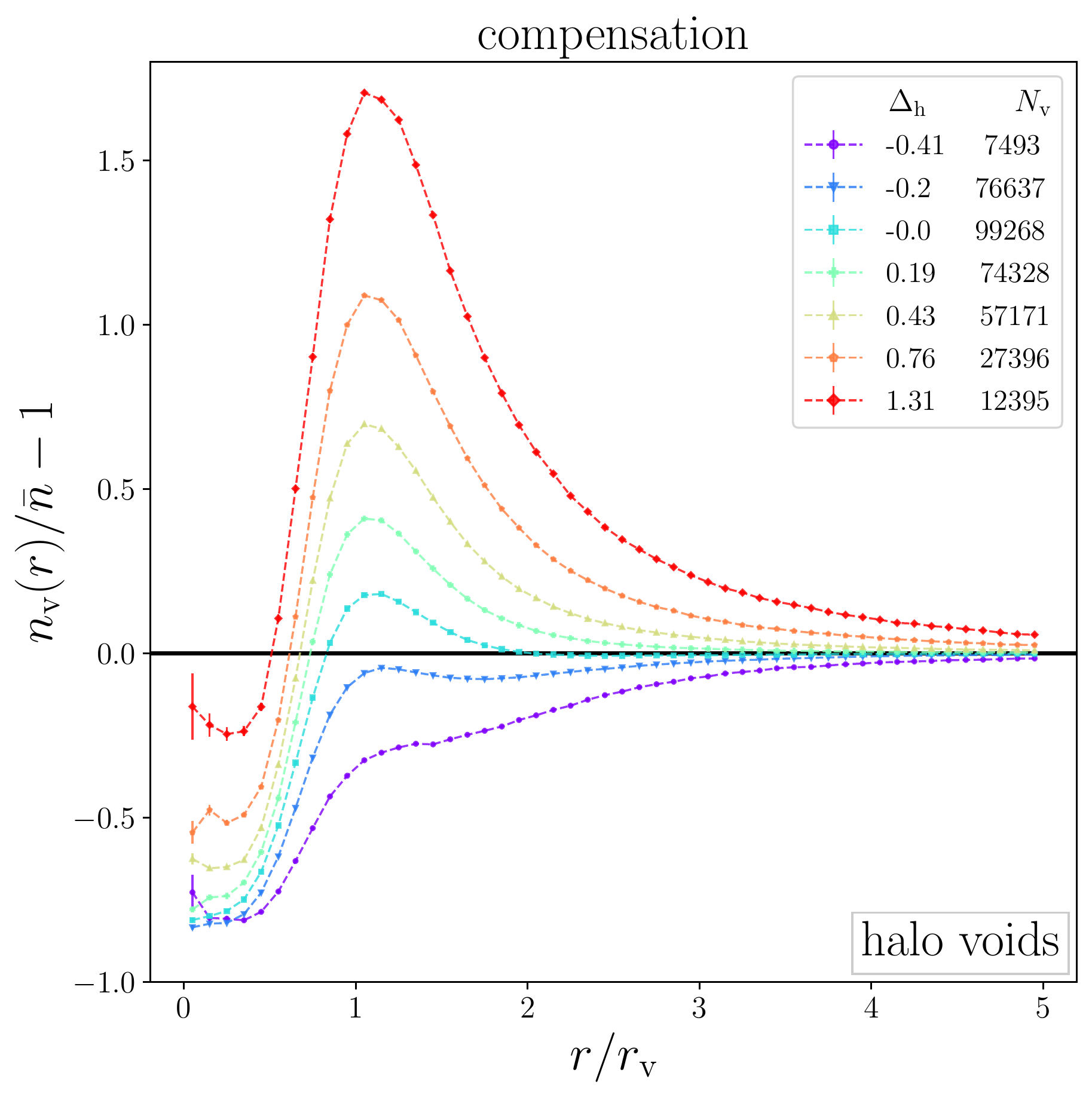}}

               \caption{Stacked number density profiles of \emph{isolated} halo voids from the \MR{} simulation in bins of their radius (top left, as in figure~\ref{fig_density_mr_CDM_halo_merging}), ellipticity (top right), core density (bottom left) and compensation (bottom right). All voids are selected within a radius range from $5\,\hMpc$ to $50\,\hMpc$.}

               \label{fig_number_density_mr_halo_ellipticity_coreDens_compensation_halos}

\end{figure}

Figure~\ref{fig_matter_density_mr_hr_haloCDM} presents the matter density profiles, $\rho_\void(r) / \bar{\rho} - 1$, of \emph{isolated} halo voids in both \MR{} and \HR{}, i.e. instead of using halos as tracers to calculate their profiles (as in figures~\ref{fig_density_mr_CDM_halo_merging} and~\ref{fig_density_hr_CDM_halo_merging}), we now use subsampled CDM particles for this. The differences to the number density profiles of halo voids calculated using halos, $n_\void(r) / \bar{n} -1$, are depicted on the bottom panel of each plot. We notice similar trends as for the CDM void profiles, although less pronounced. The profiles are generally less dense in their centers, while at the same time their compensation walls are attenuated~\cite{Pollina2017}. This is due to the CDM particles being more evenly distributed in space than the halos (i.e., halo bias), making their matter density profiles more similar to the ones of CDM voids.

So far we have exclusively focused on density profiles in bins of void radius. Figure~\ref{fig_number_density_mr_halo_ellipticity_coreDens_compensation_halos} additionally presents the number density profiles of \emph{isolated} halo voids from the \MR{} simulation in bins of ellipticities $\ellipticity$, core densities $\coreDens$, and compensations $\Delta_\halo$. As before, we limit ourselves to voids with radii between $5\,\hMpc$ and $50\,\hMpc$ for the sake of comparability. Density profiles in bins of void ellipticity are shown in the top right panel of figure~\ref{fig_number_density_mr_halo_ellipticity_coreDens_compensation_halos}. As already seen in figure~\ref{fig_radius_ellipticity_coreDens_compensation_dependance}, the ellipticity of voids is only weakly correlated with their radius, especially in the presented range of $\varepsilon$ between $0.0$ and $0.26$. Because the void size function peaks at small radii, small voids dominate in every bin of ellipticity, so their stacked profiles exhibit relatively high compensation walls. The least elliptical voids feature the sharpest walls and the deepest cores in their stacked density profiles, and vice versa. This is a simple consequence of spherical averaging in shells around the void center: if regions of similar density in its vicinity are non-spherical, they overlap with shells at different distances from it, which effectively smooths out the density profile. To account for this effect, an alternative stacking method has been proposed by the authors of reference~\cite{Cautun2016} to follow the geometry of void boundaries when constructing void density profiles. This also leads to sharper compensation walls and deeper cores, in agreement with our findings.

The lower left panel of figure~\ref{fig_number_density_mr_halo_ellipticity_coreDens_compensation_halos} depicts density profile stacks in bins of core density within a range from $0.0$ to $0.8$. Unsurprisingly, voids of lower core density exhibit deeper density profiles and vice-versa. Moreover, the core density is anti-correlated with the height of the compensation wall, which results in a close resemblance with the stacks from radius bins, where a similar pattern is present for \emph{isolated} voids. This is consistent with the upper right plot of figure~\ref{fig_radius_ellipticity_coreDens_compensation_dependance}, where smaller voids feature higher core densities and compensations.

Lastly, compensation bins are shown on the bottom right of figure~\ref{fig_number_density_mr_halo_ellipticity_coreDens_compensation_halos}, covering values from $-1$ (undercompensated) to $2$ (overcompensated). Undercompensated voids tend to have the deepest cores and no clear compensation wall. In fact, the stacked density profile of the most undercompensated voids gradually increases outwards and only approaches the mean background density at a large distance from the void center. Yet, there is a hint for a void boundary at $r = r_\void$, where the slope in density becomes flatter. The higher the compensation of voids, the more pronounced their compensation wall becomes. Here the most overcompensated voids barely reach below the mean density in their cores, resembling the behavior of small voids. These on average exhibit higher compensation values, while large voids have a tendency to be undercompensated, as already evident from figure~\ref{fig_radius_ellipticity_coreDens_compensation_dependance}. The compensation is a measure of the environment a void is located in, since it depends on its size and total number of member particles. Undercompensated voids happen to be within underdense environments, whereas overcompensated voids in regions of higher local density. These two phenomena have been coined ``void-in-void'' and ``void-in-cloud'' scenario, respectively~\cite{Sheth2004}. This defines an interesting transition point for exactly compensated voids with $\Delta_\halo=0$~\cite{Hamaus2014a}. Their density profile converges to the background density at the closest distance from the void center among all stacks in bins of compensation, around $r\simeq2r_\void$.

\subsection{Mass weighting \label{subsec:mass_weight_profiles}}

\begin{figure}[ht]
\centering
\resizebox{\hsize}{!}{
    \includegraphics{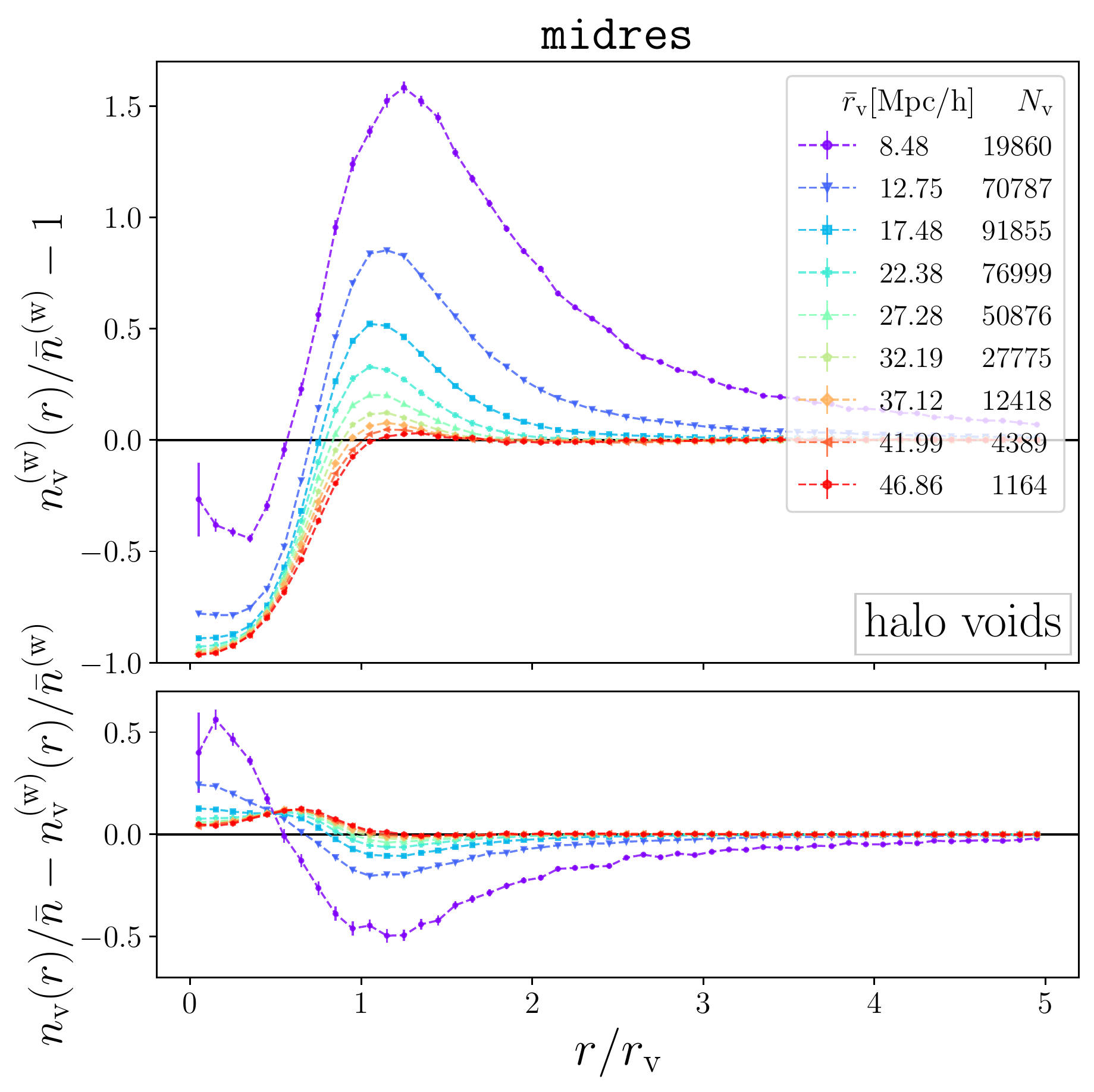}
    \includegraphics{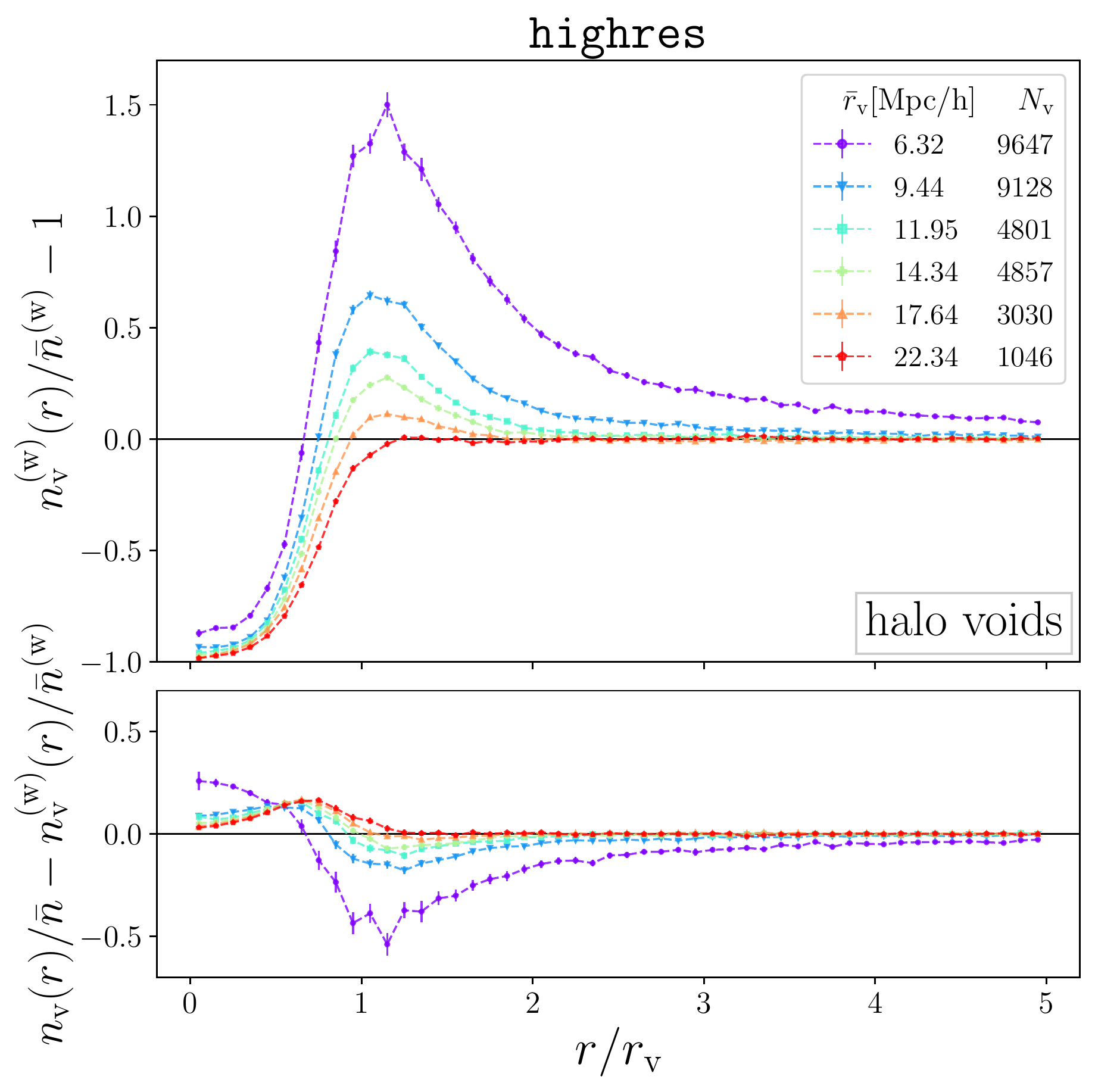}}
\caption{Stacked mass-weighted density profiles of \emph{isolated} halo voids from the \MR{} (left) and \HR{} simulations (right). Lower panels compare this with the unweighted number density profiles of halo voids from figures~\ref{fig_density_mr_CDM_halo_merging} and~\ref{fig_density_hr_CDM_halo_merging} in identical void radius bins.}
\label{fig_massweight_mr_hr_halo}
\end{figure}

Apart from the imposed mass cuts on our halo catalogs, we have so far neglected the masses of individual halos for the calculation of void density profiles. When used as weights $w_j = M_j$ for every halo $j$ in equation~(\ref{eq:general_density_profile}), one can estimate the mass-weighted density profile around voids from all the matter contained in halos. This is different from using CDM particles for the profile estimation, because not all of the CDM is confined inside halos. Yet, it yields a special type of matter density profile around voids, which additionally probes the spatial distribution of halos depending on their mass. While individual halo masses are difficult to obtain in large-scale structure surveys, the magnitudes of their hosted galaxies can provide an observational proxy for them~\cite{Seljak2009,Hamaus2010}.

In figure~\ref{fig_massweight_mr_hr_halo} we present the mass-weighted density profiles of halo voids from \MR{} and \HR{} boxes, with differences to their unweighted number density profiles on the bottom. The profiles are stacked in void radius bins identical to the ones used in figures~\ref{fig_density_mr_CDM_halo_merging} and~\ref{fig_density_hr_CDM_halo_merging}. Mass-weighted profiles are amplified with respect to number density profiles, featuring higher compensation walls and deeper cores, most notably for small voids. This implies that the least massive halos tend to reside closer to the void centers, whereas more massive halos are rather located in the more clustered regions at the void boundary, as expected from previous studies~\cite[e.g.,][]{Zhang2020,Verza2022}. For larger voids, however, this effect diminishes. Alternatively, one can interpret the effect of mass weighting as a boost of the halo bias~\cite{Hamaus2010}.

\subsection{Velocity profiles \label{subsec:velocity_profiles}}
\begin{figure}[t]
\centering
\resizebox{\hsize}{!}{
    \includegraphics{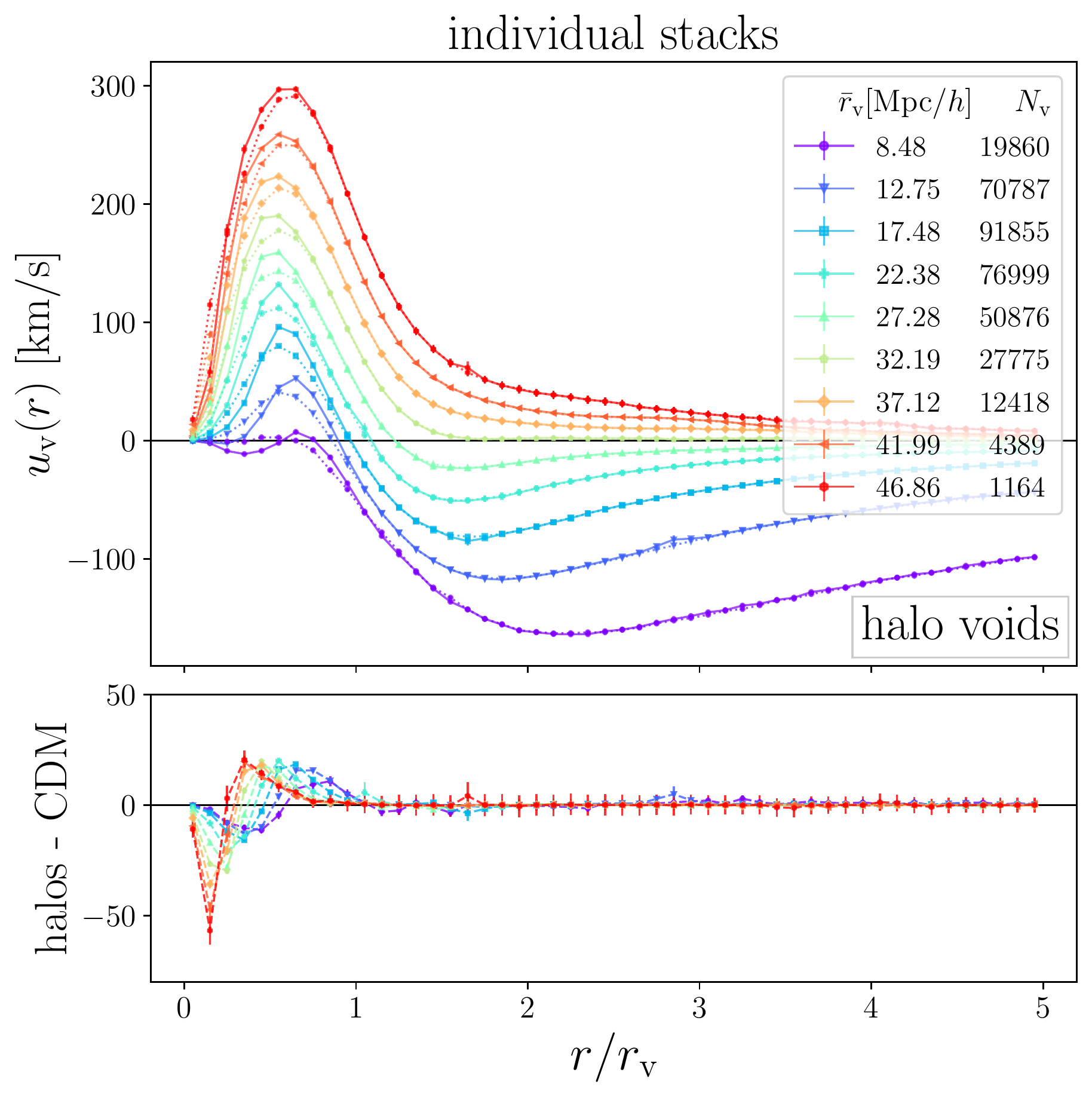}
    \includegraphics{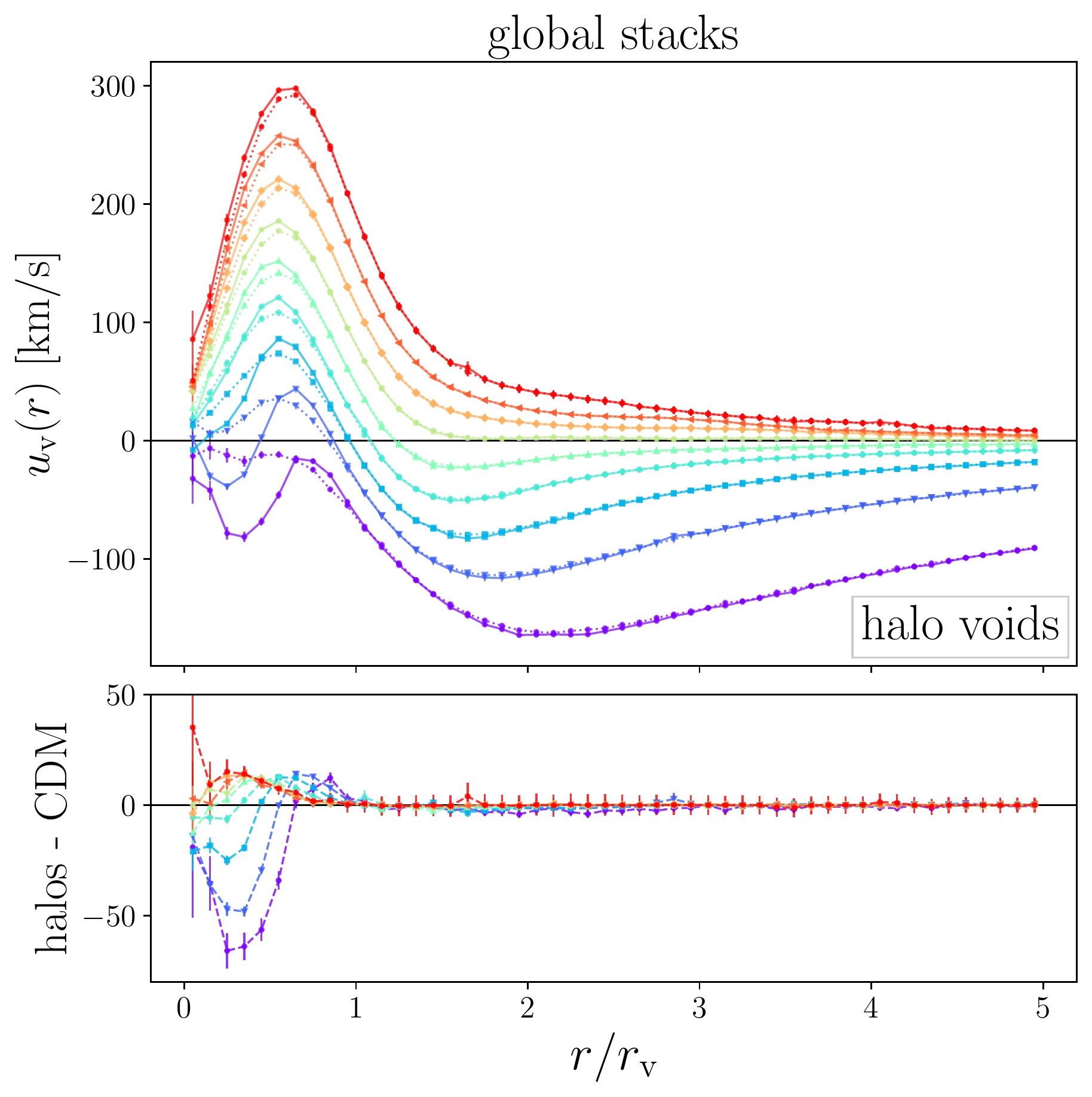}}
\caption{Velocity profiles from individual stacks (left) and global stacks (right) around \emph{isolated} halo voids from the \MR{} simulation with identical void radius bins as in figure~\ref{fig_density_mr_CDM_halo_merging}. Solid lines indicate the velocity of halos, dotted lines the velocity of CDM, their differences are shown in the bottom panels.}
\label{fig_velocity_mr_halo_CDM}
\end{figure}

\begin{figure}[t]
\centering
\resizebox{\hsize}{!}{
    \includegraphics{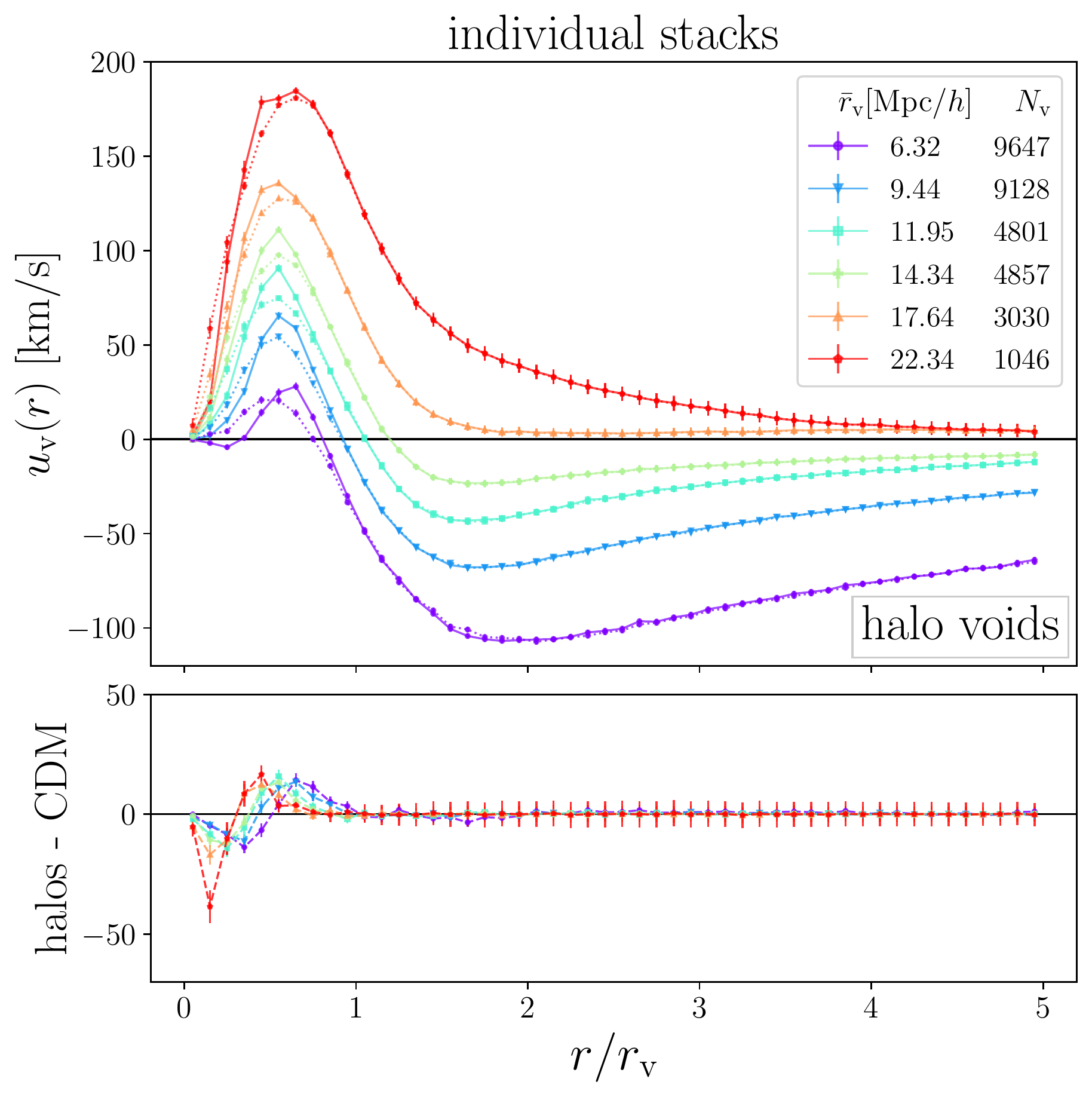}
    \includegraphics{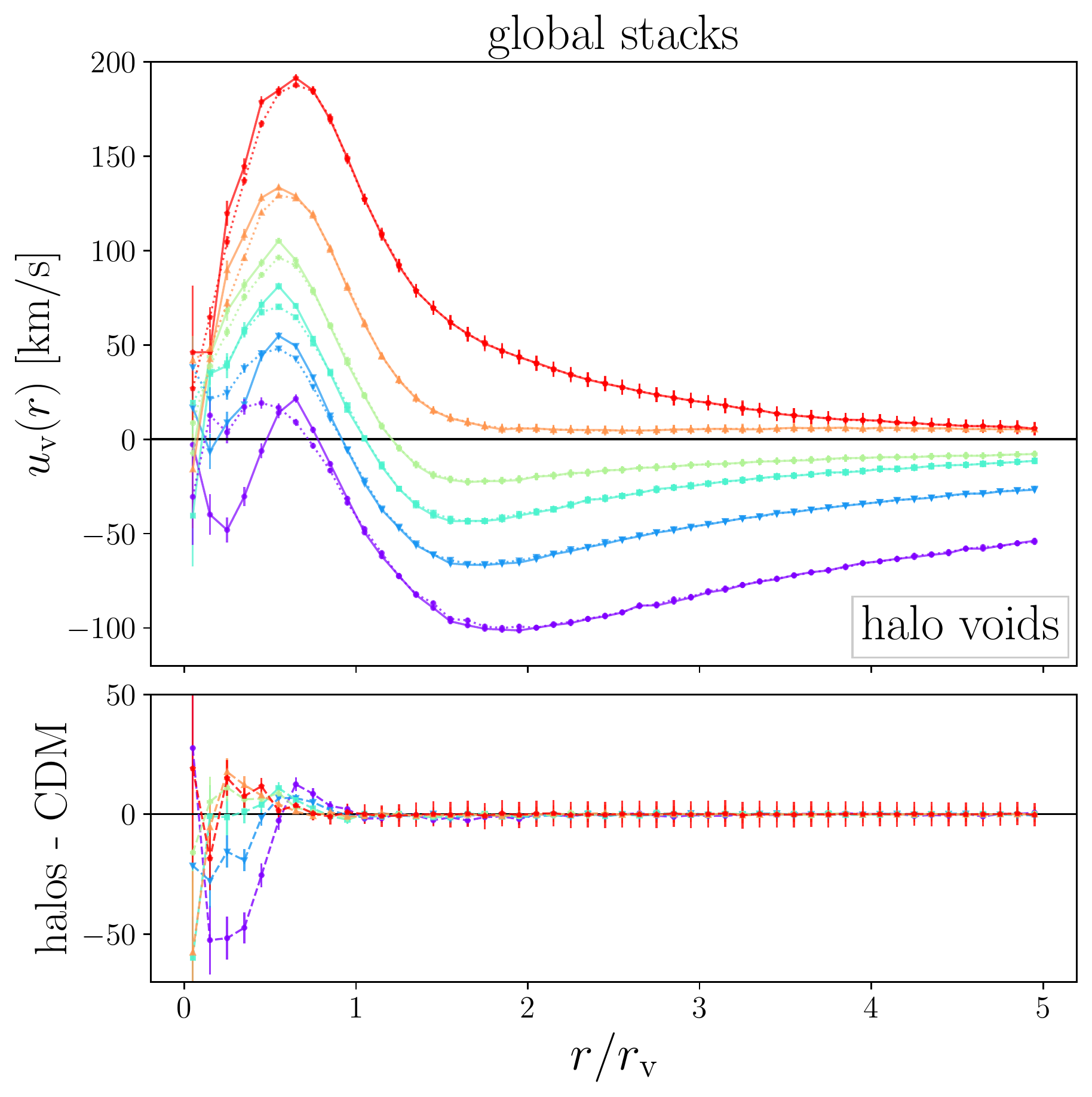}}
\caption{Same as figure~\ref{fig_velocity_mr_halo_CDM}, but in the \HR{} simulation. Identical void radius bins as in figure~\ref{fig_density_hr_CDM_halo_merging}.}
\label{fig_velocity_hr_halo_CDM}
\end{figure}

Having investigated the spatial distribution of CDM and halos inside voids, we now turn to their movements. Void dynamics are characterized by the coherent, radially directed flow of matter around their centers, which can be quantified via stacked velocity profiles. As described in section~\ref{subsec:void_profiles}, we distinguish two different ways of calculating these, using individual stacks via equation~(\ref{eq:individual_stack_velocity_profile}) and using global stacks via equation~(\ref{eq:global_stack_velocity_profile}). Figures~\ref{fig_velocity_mr_halo_CDM} and~\ref{fig_velocity_hr_halo_CDM} present the two methods for \emph{isolated} halo voids from the \MR{} and \HR{} simulations with identical void radius bins as those used in figures~\ref{fig_density_mr_CDM_halo_merging} and~\ref{fig_density_hr_CDM_halo_merging} for number density profiles. The velocities of both halos, as well as the CDM are shown, with differences highlighted in the bottom panels.

First of all, there is an exquisite agreement between the velocities of halos and CDM particles around halo voids for both stacking methods. This is as expected from the equivalence principle, stating that test particles in a common gravitational potential fall with the same speeds, irrespective of their mass and composition. Differences are becoming more visible when approaching the void centers, where the tracer statistics are sparser. Large voids are characterized by outflows, which steadily increase from close to zero at their center until a maximum near the compensation wall, then drop again to approach zero in the large distance limit. This is consistent with the deep and extended underdensity of matter in the vicinity of those voids, as seen in figure~\ref{fig_matter_density_mr_hr_haloCDM}. As the compensation wall becomes more pronounced for voids of smaller size, the direction of motion features a turning point and matter flows radially inward towards the compensation wall~\cite{Hamaus2014b}. This influx is most extreme around the smallest voids, which may eventually overcome their interior expansion and eliminate them~\cite{Sheth2004}. One may argue this to be the case for the smallest voids in figure~\ref{fig_velocity_mr_halo_CDM}, but our results from the \HR{} simulation in figure~\ref{fig_velocity_hr_halo_CDM} show that voids of even smaller size experience internal outflows.

While the most salient features in the velocity profiles are similar in both stacking methods, there are some important differences to notice in proximity of the void center. Individual stacks closely converge towards zero velocity towards the void centers, while global stacks reach finite values there. For global stacks the differences between halo and CDM velocities decrease with increasing void size, whereas the opposite trend is manifest in individual stacks, which yield the best agreement for the smallest voids. It is mainly the halo velocities that are affected by the choice of stacking method, CDM velocities are more consistent with each other. Since the same halos are used for the identification of voids, this suggests that the residuals between halo and CDM velocities are caused by the same sparse sampling effects already found in the density profiles of small voids in figures~\ref{fig_density_mr_CDM_halo_merging} and~\ref{fig_density_hr_CDM_halo_merging}, as discussed in section~\ref{subsec:density_profiles}.

In fact, the velocity profiles of CDM voids, which can be found in figures~\ref{fig_mass_conservation_stacked_density} and~\ref{fig_stacks_density_velocity_mr_CDM} of section~\ref{sec:mass_conservation}, experience similar sampling artifacts, because the same tracers are used for the identification of voids and the calculation of their profiles. This becomes most severe for the voids defined by the fewest tracer particles. We should of course bear in mind that a direct comparison between CDM voids and halo voids has a limited scope, as their size functions are very different. One could also argue for other potential causes for a discrepancy between halo and CDM velocities, such as a velocity bias from halos, or some kind of nonlinear dynamics~\cite{Achitouv2017b,Paillas2021}. However, when increasing the simulation resolution in figure~\ref{fig_velocity_hr_halo_CDM} the discrepancy between the velocities of halos and CDM diminishes, so we cannot find any evidence in favor of such effects. Moreover, we have checked that void merging does not impact any of these conclusions either.

\begin{figure}[t]

               \centering

               \resizebox{\hsize}{!}{

                               \includegraphics[trim=0 5 0 5, clip]{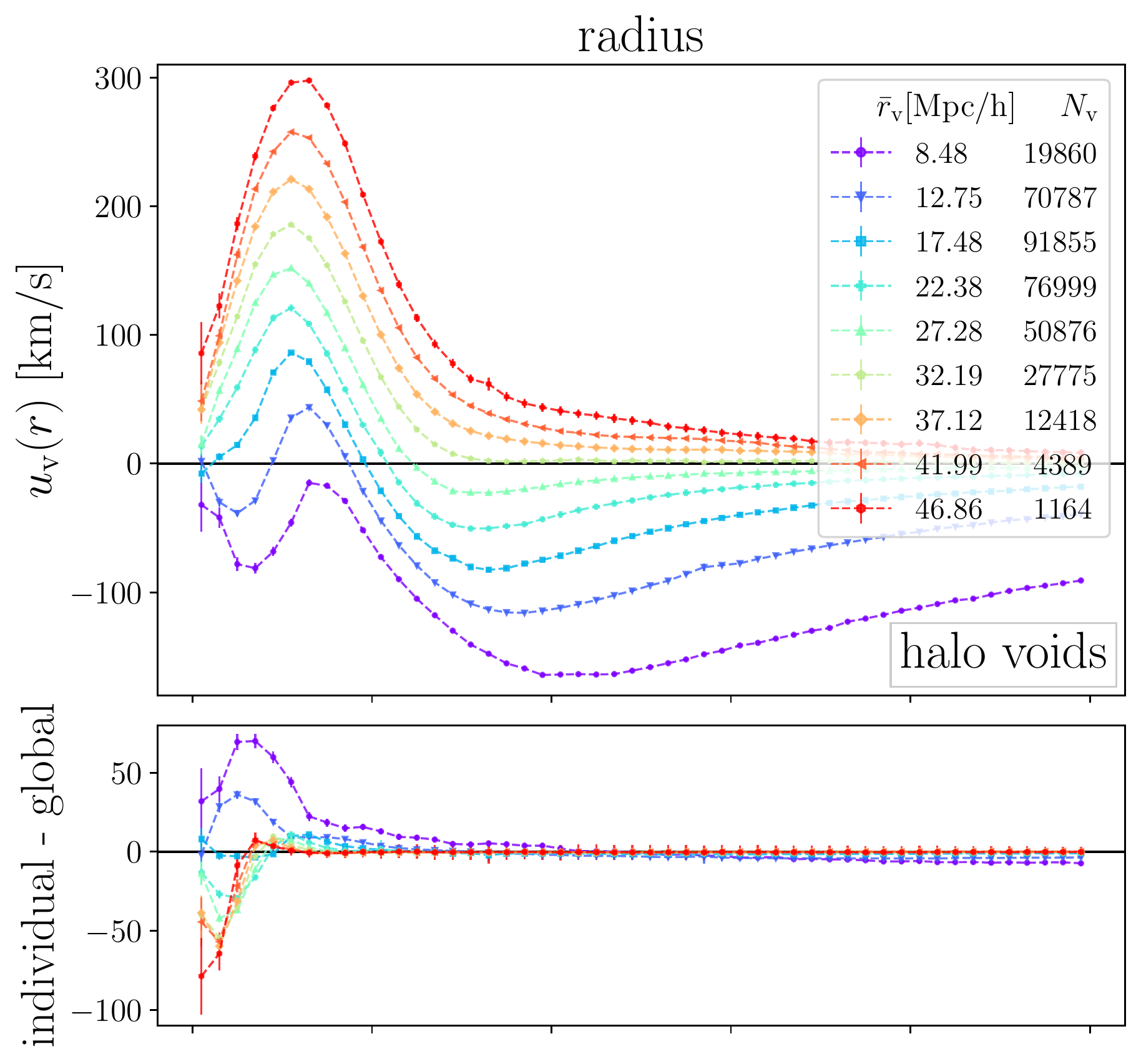}

                               \includegraphics[trim=0 5 0 5, clip]{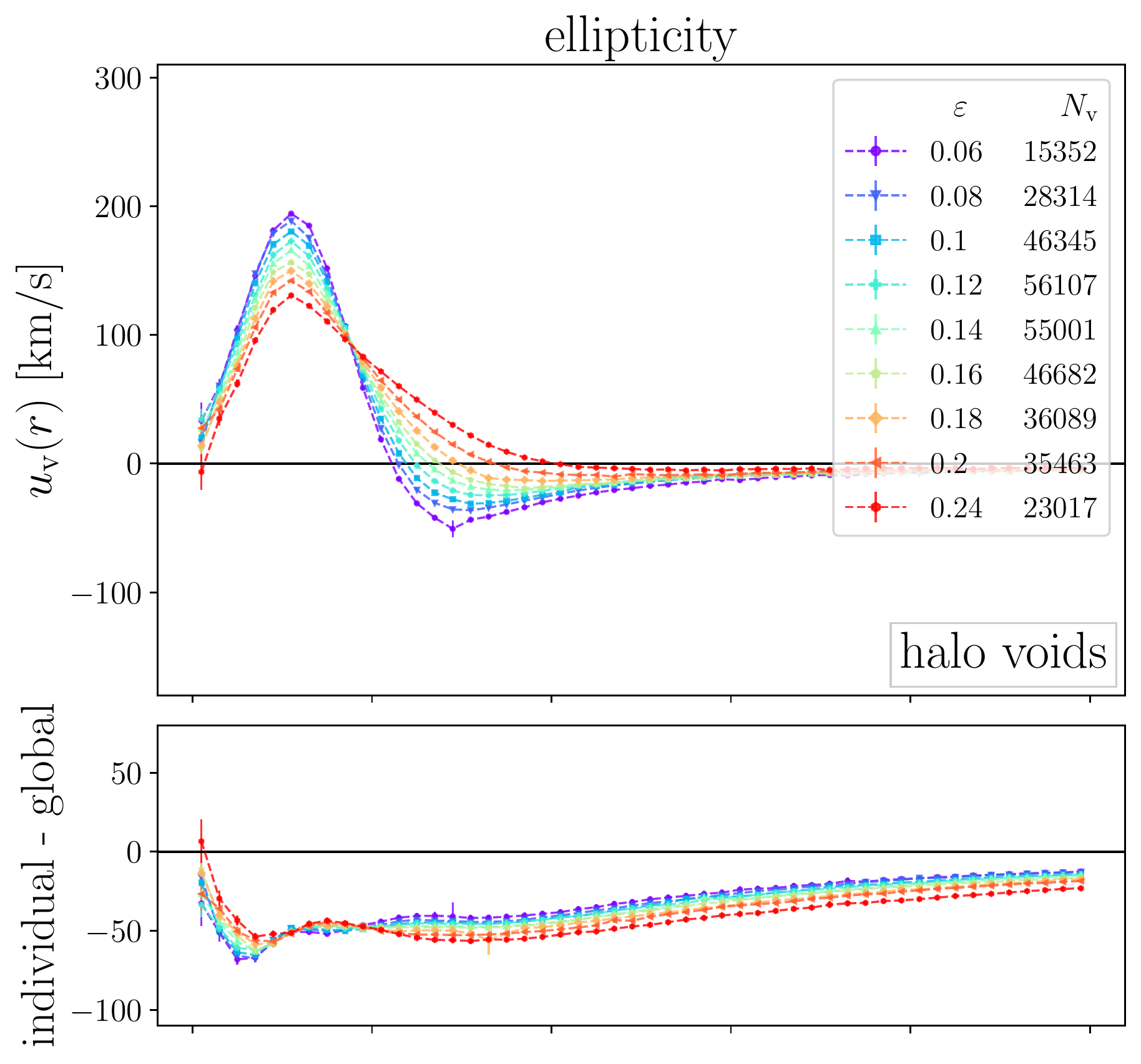}}

               \resizebox{\hsize}{!}{

                               \includegraphics[trim=0 10 0 5, clip]{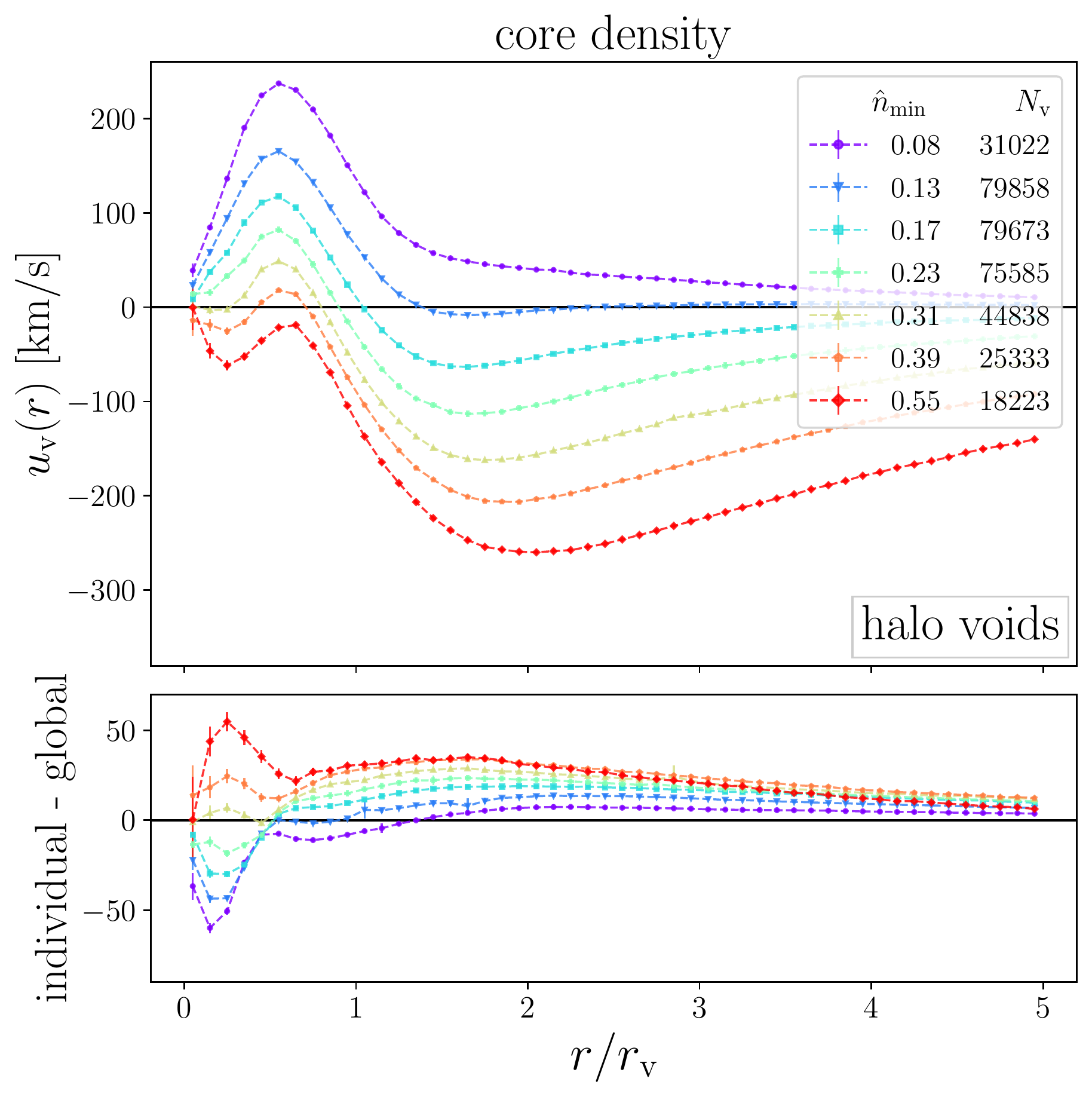}

                               \includegraphics[trim=0 10 0 5, clip]{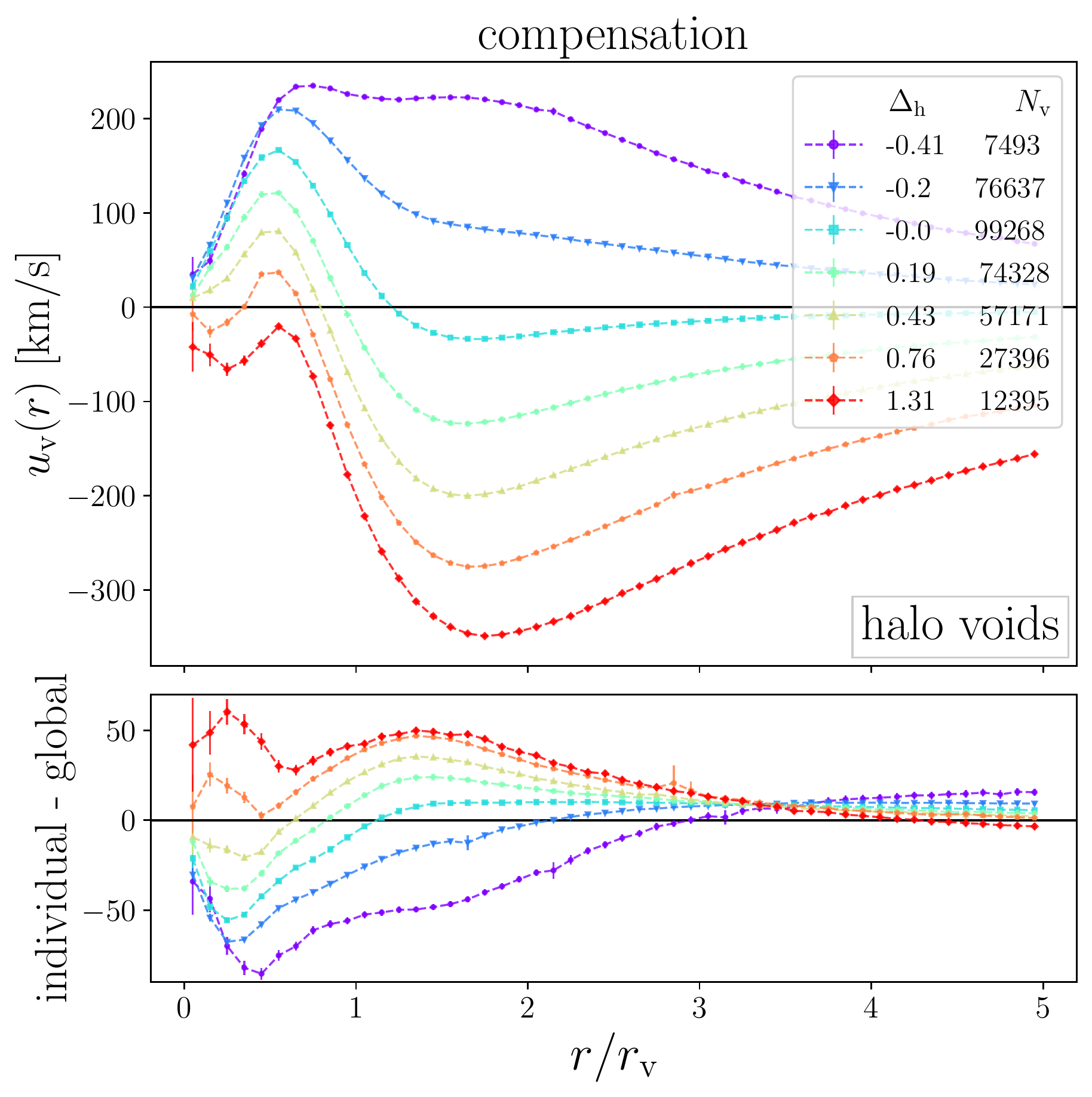}}

               \caption{Velocity profiles from global stacks, using halos around \emph{isolated} halo voids in the \MR{} simulation for bins in void radius (top left), ellipticity (top right), core density (bottom left) and compensation (bottom right). Lower panels show the differences between individual and global stacks. The bins and the void selection are identical to figure~\ref{fig_number_density_mr_halo_ellipticity_coreDens_compensation_halos}.}

               \label{fig_velocity_mr_halo_ellipticity_coreDens_compensation_halos}

\end{figure}

Analogously to the density profiles in figure~\ref{fig_number_density_mr_halo_ellipticity_coreDens_compensation_halos}, we can choose void properties other than their radius to stack velocity profiles. Figure~\ref{fig_velocity_mr_halo_ellipticity_coreDens_compensation_halos} presents this for global stacks of \emph{isolated} halo voids from the \MR{} simulation in bins of void ellipticity $\ellipticity$, core density $\coreDens$, and compensation $\Delta_\halo$, in addition to the bins in void radius $r_\void$. As before, we only select voids with radii between $5\,\hMpc$ and $50\,\hMpc$ for the sake of comparability, but now also compare both methods for stacking velocity profiles in the lower panels.

When considering ellipticity bins in the upper right plot of figure~\ref{fig_velocity_mr_halo_ellipticity_coreDens_compensation_halos}, we notice that all bins experience the characteristic outflow of halos from the void center, with the maximum velocity steadily decreasing with increasing ellipticity, in correspondence with the shape of density profiles from figure~\ref{fig_number_density_mr_halo_ellipticity_coreDens_compensation_halos}. This is true for both stacking methods, although individual stacks result in generally lower velocities than global stacks, even at large distances from the void center. In global stacks the most elliptical voids are dominated by outflows of halos, only the more spherical voids exhibit inflows towards their compensation wall. In contrast, individual stacks feature inflows in all bins of ellipticity. This difference can be understood considering the definition of the two stacking methods in equations~(\ref{eq:individual_stack_velocity_profile}) and~(\ref{eq:global_stack_velocity_profile}) together with  figure~\ref{fig_radius_ellipticity_coreDens_compensation_dependance}, which shows that void ellipticities and radii are largely independent from each other, such that each ellipticity bin covers a wide range of void sizes. Therefore, an average over individual velocity profiles from equation~(\ref{eq:individual_velocity_profile}) is biased towards small voids, which are more numerous than large ones. On the other hand, equation~(\ref{eq:global_stack_velocity_profile}) sums up the volume-weighted tracer velocities around the full void sample before normalizing by the total Voronoi volume of all tracers per shell, which is biased towards shells containing more tracers and hence large voids. The dependence of velocity profiles on void radius, as shown in figure~\ref{fig_velocity_mr_halo_CDM}, then translates into the differences of the two stacking methods appearing in the upper right panel of figure~\ref{fig_velocity_mr_halo_ellipticity_coreDens_compensation_halos}.

The velocity stacks for bins in core density are depicted in the lower left panel of figure~\ref{fig_velocity_mr_halo_ellipticity_coreDens_compensation_halos}. The correspondence with the associated number density profiles from figure~\ref{fig_number_density_mr_halo_ellipticity_coreDens_compensation_halos} is evident: the most underdense voids experience the strongest outflows and vice versa. Differences between the two stacking methods are present, but do not affect the general trends already manifest in figures~\ref{fig_number_density_mr_halo_ellipticity_coreDens_compensation_halos} and~\ref{fig_velocity_mr_halo_CDM}. This is because the core densities of voids are more strongly correlated with their radii than ellipticities are, so the averaging effects discussed above are of lower importance here.

For the velocity profiles in compensation bins, as shown in the lower right panel of figure~\ref{fig_velocity_mr_halo_ellipticity_coreDens_compensation_halos}, the general trends are similar to the previous case, except that compensation has a stronger impact on the void environment than its core density, as expected from figure~\ref{fig_number_density_mr_halo_ellipticity_coreDens_compensation_halos}. Overcompensated voids are dominated by influx of matter, while undercompensated voids expand out to large distances from their center. At the boundary between these two regimes, the peculiar motions around compensated voids with $\Delta_\halo=0$ vanish at the smallest distance from the void center. Test particles near a region of average background density experience no net force and simply move with the Hubble flow.

In summary, an important conclusion from figure~\ref{fig_velocity_mr_halo_ellipticity_coreDens_compensation_halos} is the fact that different estimators for stacked velocity profiles around voids can be biased in different ways, depending on the diversity of the considered void sample and the selection property used to bin the stacks. Individual stacks are biased towards the more numerous small voids, whereas global stacks are biased towards larger voids that are sampled with more tracer particles. The issue can be partially mitigated by limiting the range of void sizes per stack, but a comparison of both stacking methods is helpful in revealing residual biases the estimators may encounter.

\subsection{Sampling effects  \label{subsec:sampled_profiles}}
A comparison of the different profile estimators in the previous two subsections revealed some of their limitations in certain regimes. We expect these limitations to arise from the sparse statistics of tracers or voids, which are unavoidable when approaching scales near the average tracer separation at any resolution. However, we can artificially amplify the impact of sparse statistics by removing the lowest-mass halos, or by randomly subsampling the tracers used in the profile calculations. Nevertheless, we do not repeat the void identification step on modified tracer samples, because this would render a direct comparison between different void samples impossible. For example, this can be seen when comparing the number density profiles of \emph{isolated} halo voids in the \MR{} and \HR{} simulations from figures~\ref{fig_density_mr_CDM_halo_merging} and~\ref{fig_density_hr_CDM_halo_merging}. The distribution of void sizes is different in these two cases, and voids of the same size do not necessarily share the same properties.

\begin{figure}[t]
\centering
\resizebox{\hsize}{!}{
    \includegraphics{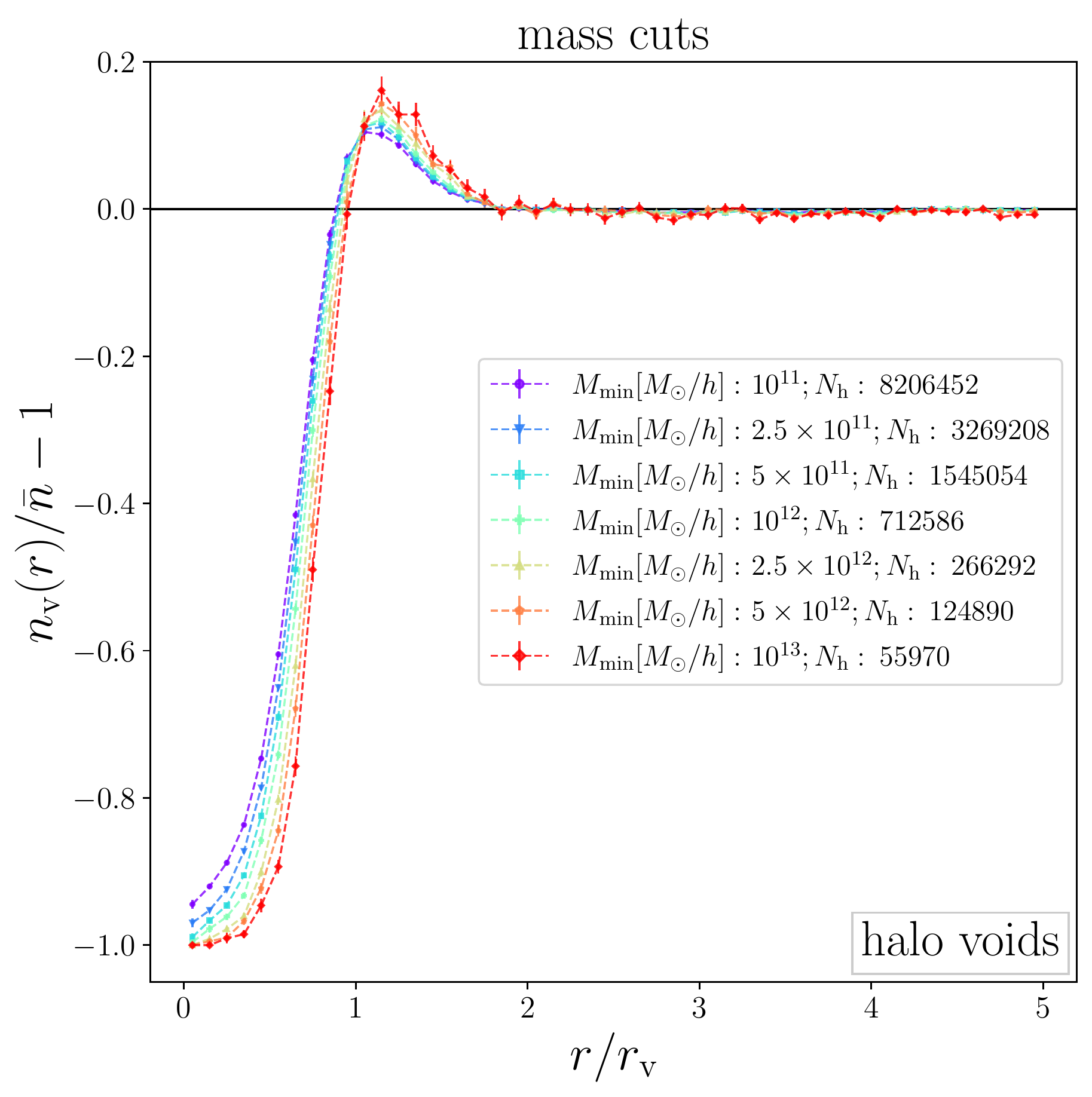}
    \includegraphics{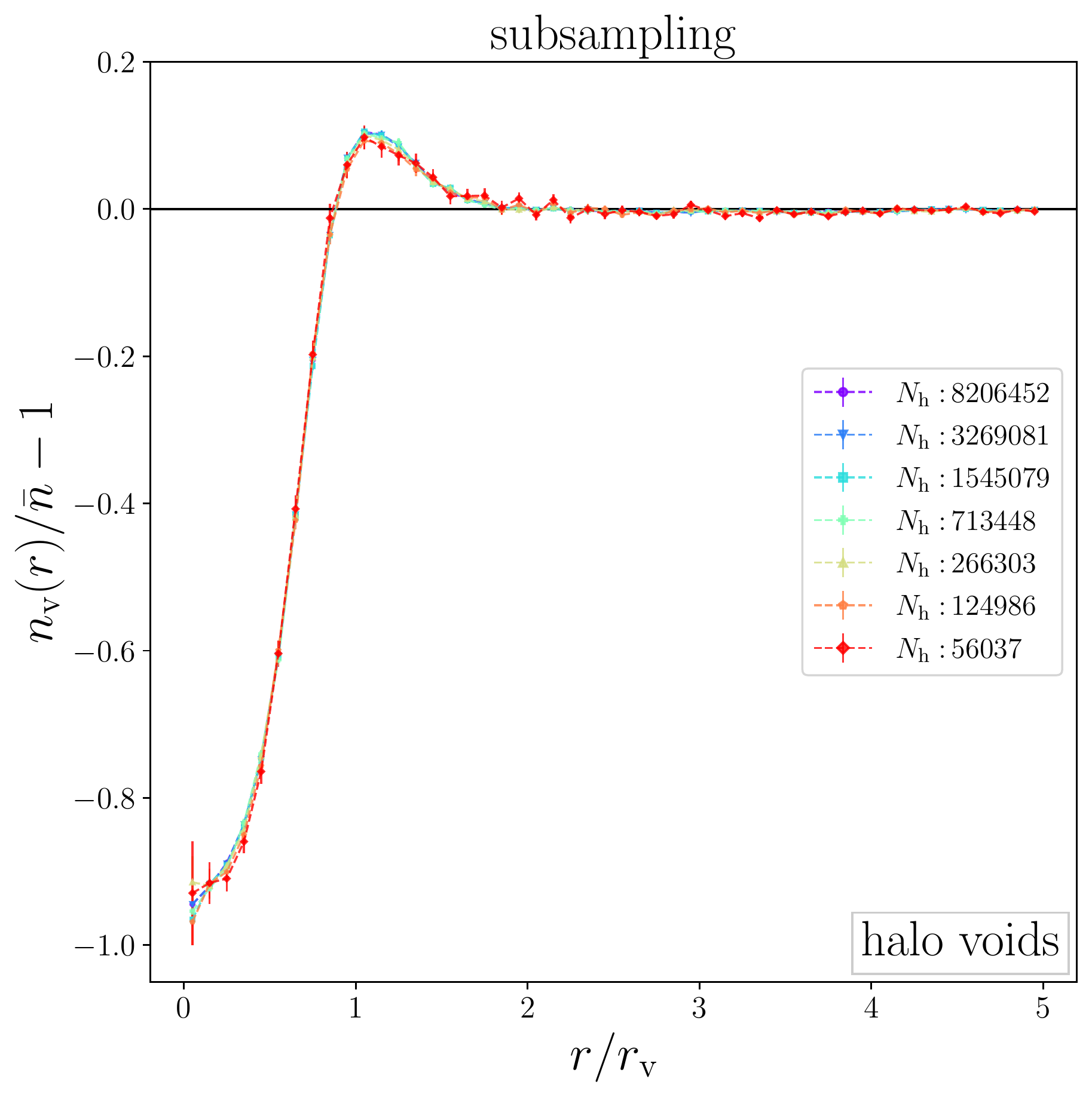}}
\caption{Stacked number density profiles of \emph{isolated} halo voids in the \HR{} simulation after applying mass cuts (left) and subsamplings (right) to the halo sample. Voids are always identified in the full halo distribution and selected in a radius range from $16\,\hMpc$ to $20\,\hMpc$. Legends indicate the minimum halo mass $M_\mathrm{min}$ (left) and the number of halos $N_\halo$ after subsampling (right).}
\label{fig_num_density_massbins_subsampling_hr}
\end{figure}

Without loss of generality we restrict ourselves to \emph{isolated} halo voids from the \HR{} simulation with radii ranging from $16\,\hMpc$ to $20\,\hMpc$, and subsequently apply mass cuts and subsamplings to the halo sample for estimating the void profiles. Figure~\ref{fig_num_density_massbins_subsampling_hr} presents the stacked number density profiles after selecting halos above different mass thresholds $M_\mathrm{min}$ and after random subsamplings that match the previously selected number of halos. Mass cuts affect the density profiles in a similar way as mass weights (c.f. figure~\ref{fig_massweight_mr_hr_halo}), revealing the tendency of more massive halos to be distributed within the voids' compensation wall, but being scarcer in the voids' core. On the contrary, random subsamplings have no significant effect on the number density profiles of voids, except for an increase of the error bars, as expected. This certifies that our density estimator from equation~(\ref{eq:general_density_profile}) is not biased for sparse tracer samples.

\begin{figure}[t]

               \centering

               \resizebox{\hsize}{!}{

                               \includegraphics[trim=0 50 0 5, clip]{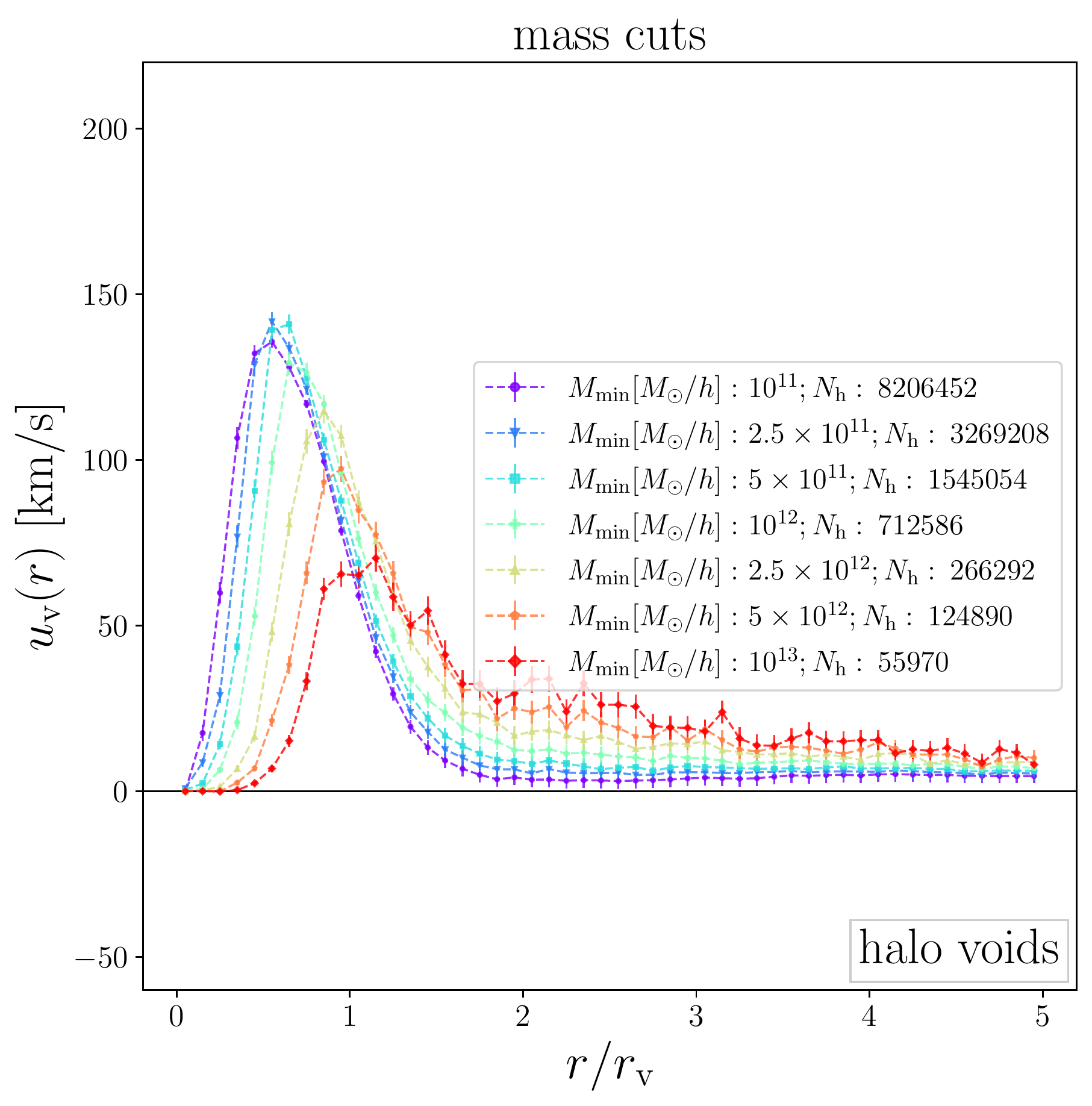}

                               \includegraphics[trim=0 50 20 5, clip]{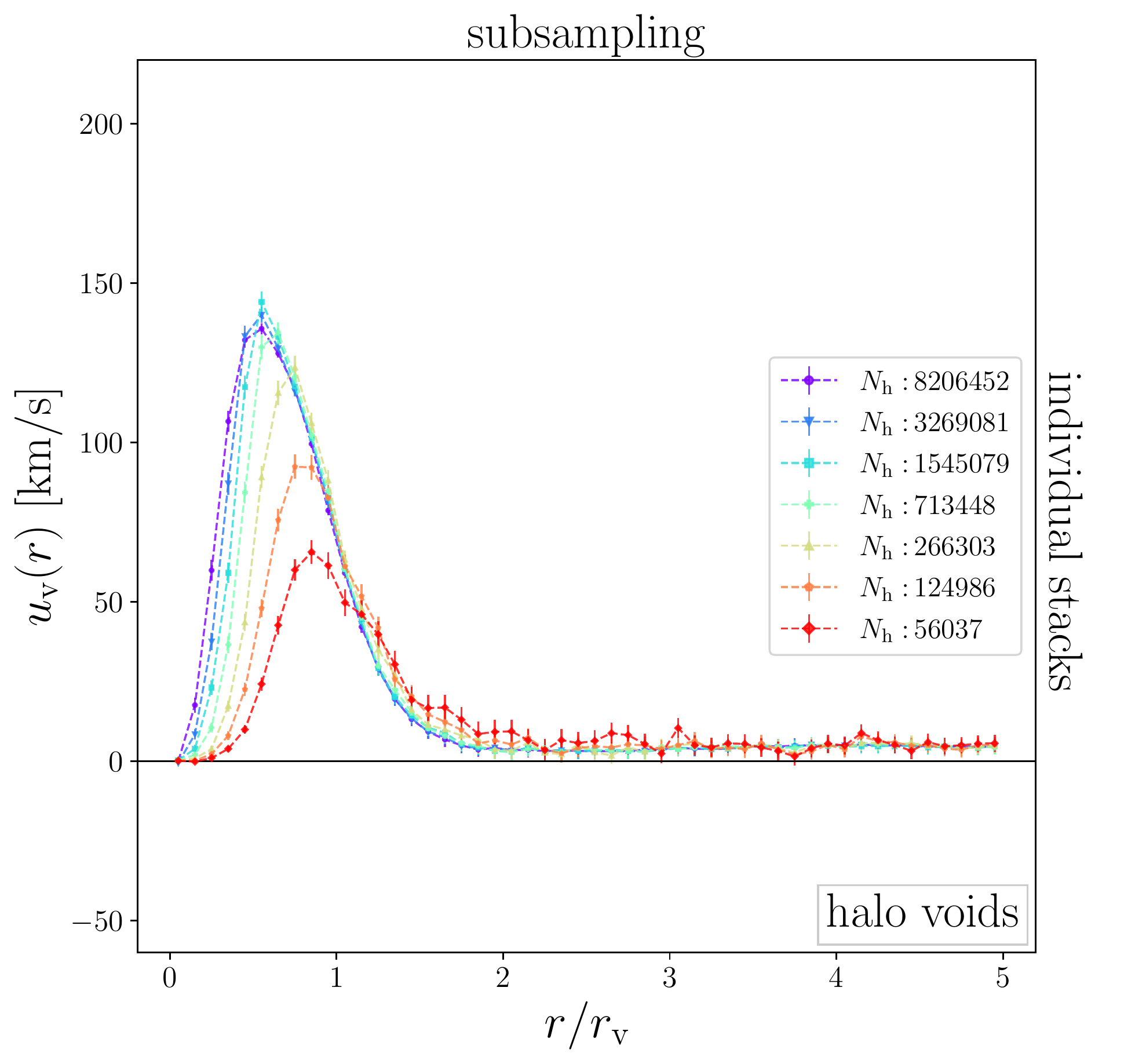}}

               \resizebox{\hsize}{!}{

                               \includegraphics[trim=0 10 0 29, clip]{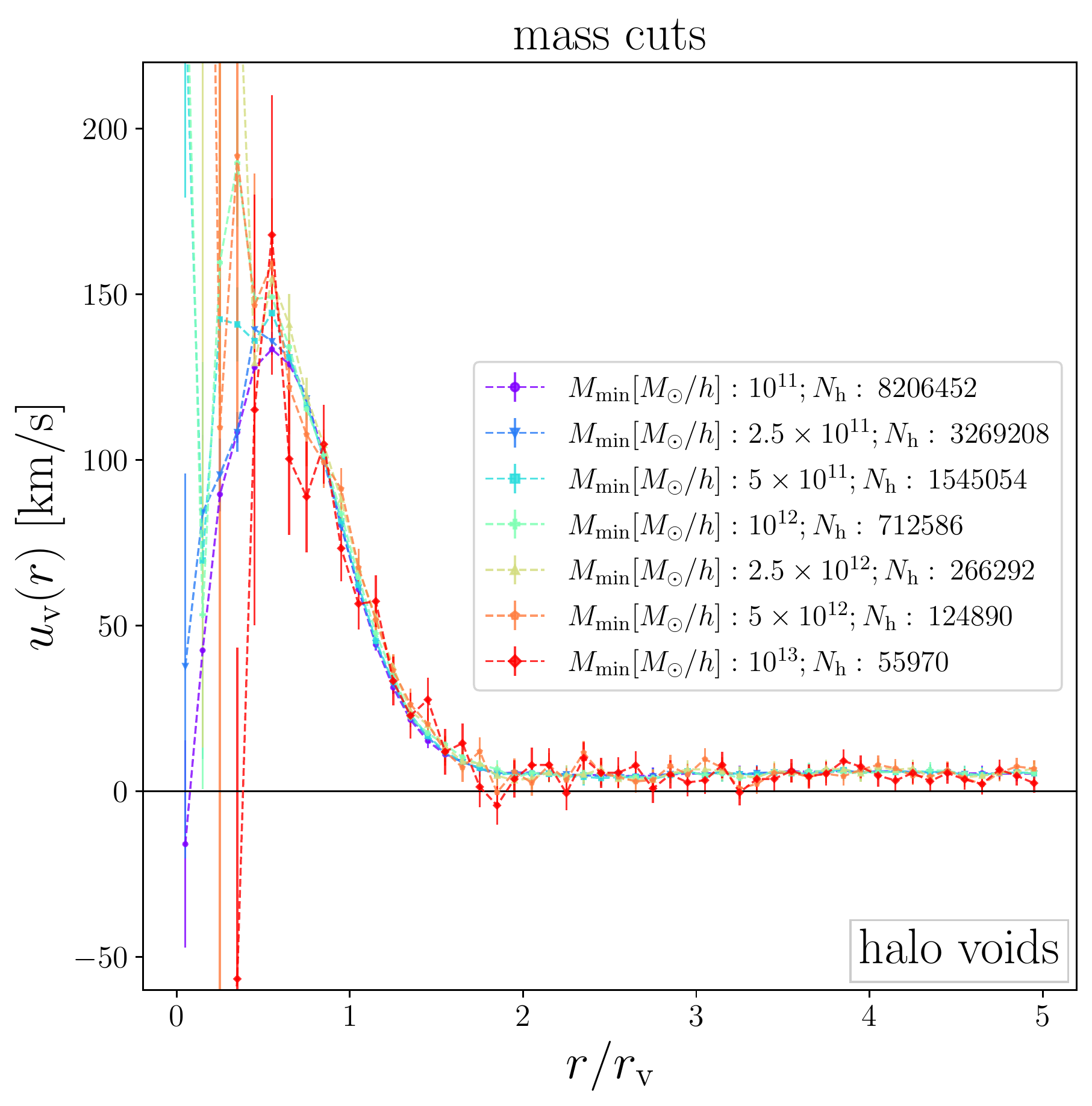}

                               \includegraphics[trim=0 10 20 29, clip]{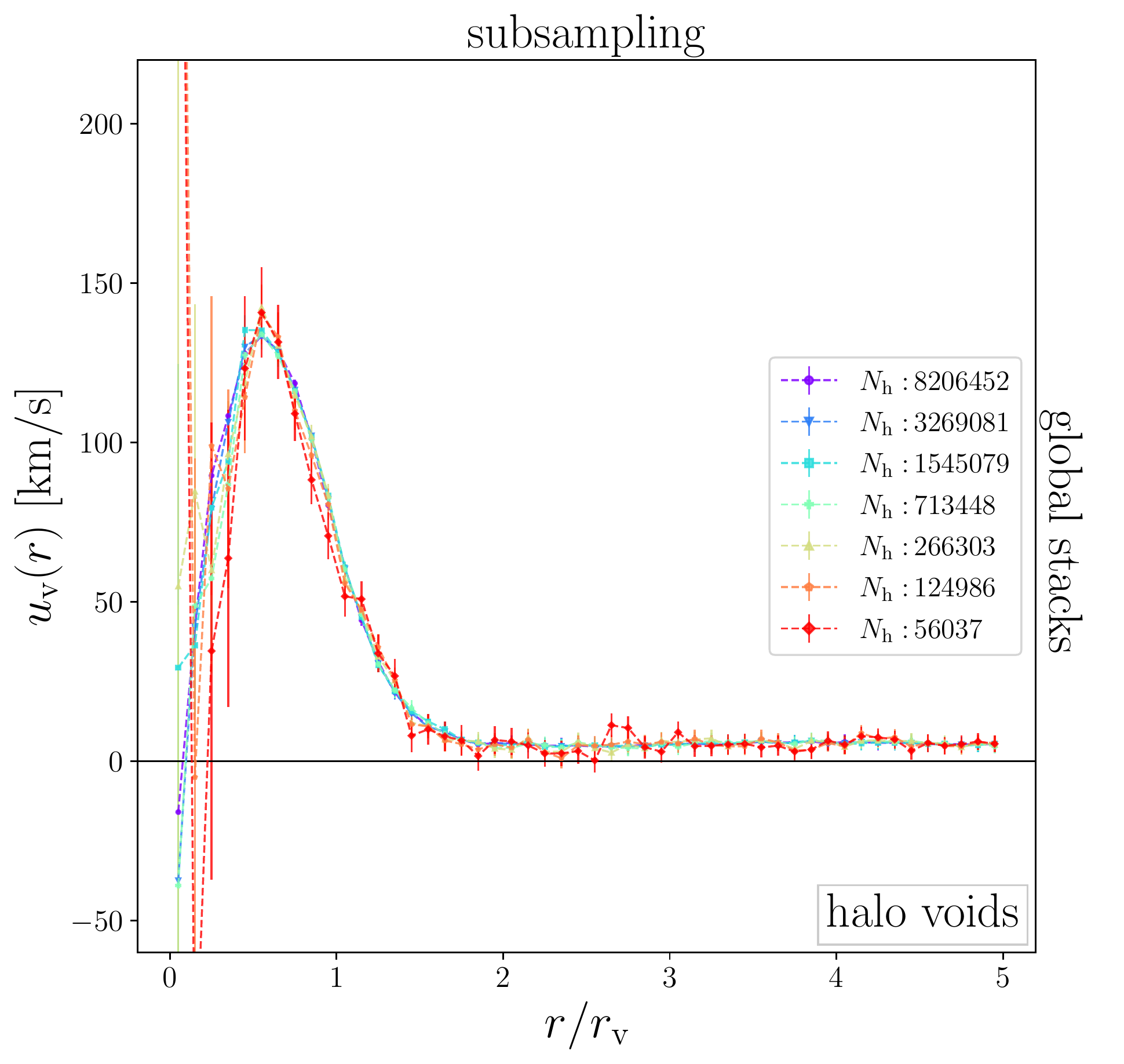}}

               \caption{Same as figure~\ref{fig_num_density_massbins_subsampling_hr}, but for velocity profiles using halo velocities in individual stacks (top) based on equation~(\ref{eq:individual_stack_velocity_profile}) and in global stacks (bottom) based on equation~(\ref{eq:global_stack_velocity_profile}).}

               \label{fig_velocity_massbins_subsamplings_hr}

\end{figure}

The corresponding velocity profiles are presented in figure~\ref{fig_velocity_massbins_subsamplings_hr}. Evidently, individual stacks and global stacks are affected very differently by mass cuts and subsampling. While the velocity profiles from individual stacks continuously diminish when reducing the number of tracers, the ones from global stacks remain stable within their error bars. In this regard the global stacking method to estimate velocity profiles is preferred, because it does neither generate a bias from sparser tracer samples, nor depend on the tracer mass. A dependence on tracer mass would violate the equivalence principle adopted in the gravity solver of the simulation, so it must be spurious. The issue arises whenever tracers become too scarce to faithfully sample the velocity field, yielding a too low velocity estimate. It is particularly severe near the void center in individual stacks, where massive halos and other tracer particles are scarcest.

\section{Linear mass conservation \label{sec:mass_conservation}}
After examining the stacked density and velocity profiles of voids separately in the previous section, we now want to investigate their interrelation via the linear continuity equation~(\ref{eq:velocity_relation}). For a given density profile, we refer to the velocity profiles calculated via that equation as \emph{linear theory profiles}. We have to determine one free parameter in equation~(\ref{eq:velocity_relation}) when using halos as tracers: their bias $b_\tracer$. We achieve this by fitting the linear theory profiles to the velocity profiles estimated via equation~(\ref{eq:individual_velocity_profile}) for individual voids, and via equations~(\ref{eq:individual_stack_velocity_profile}) and~(\ref{eq:global_stack_velocity_profile}) for stacked voids, treating the bias as a free (inverse) amplitude.

\subsection{Individual voids \label{subsec:lin_theory_individual_voids}}

\begin{figure}[t]
\centering
\resizebox{\hsize}{!}{
    \includegraphics{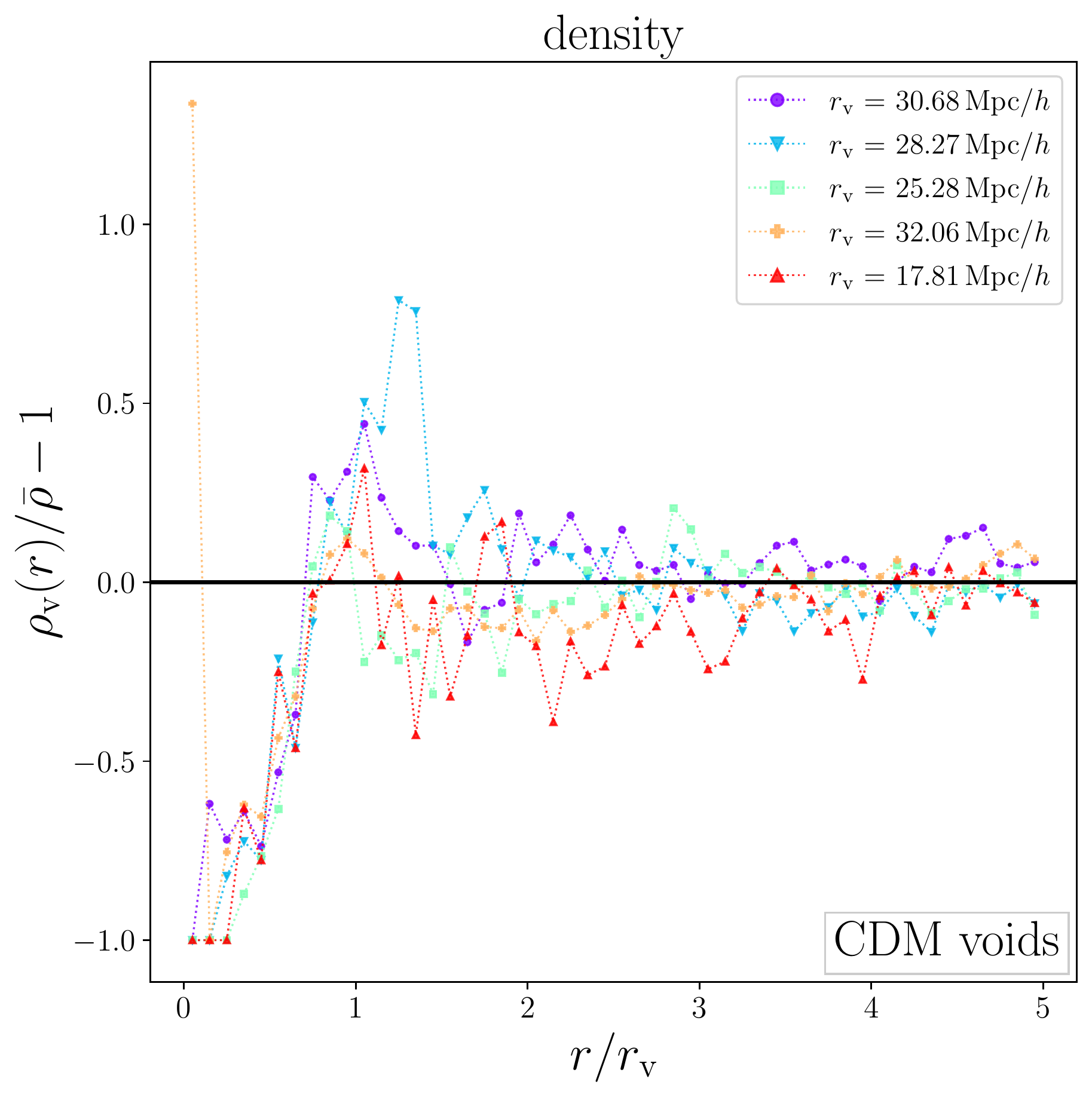}
    \includegraphics{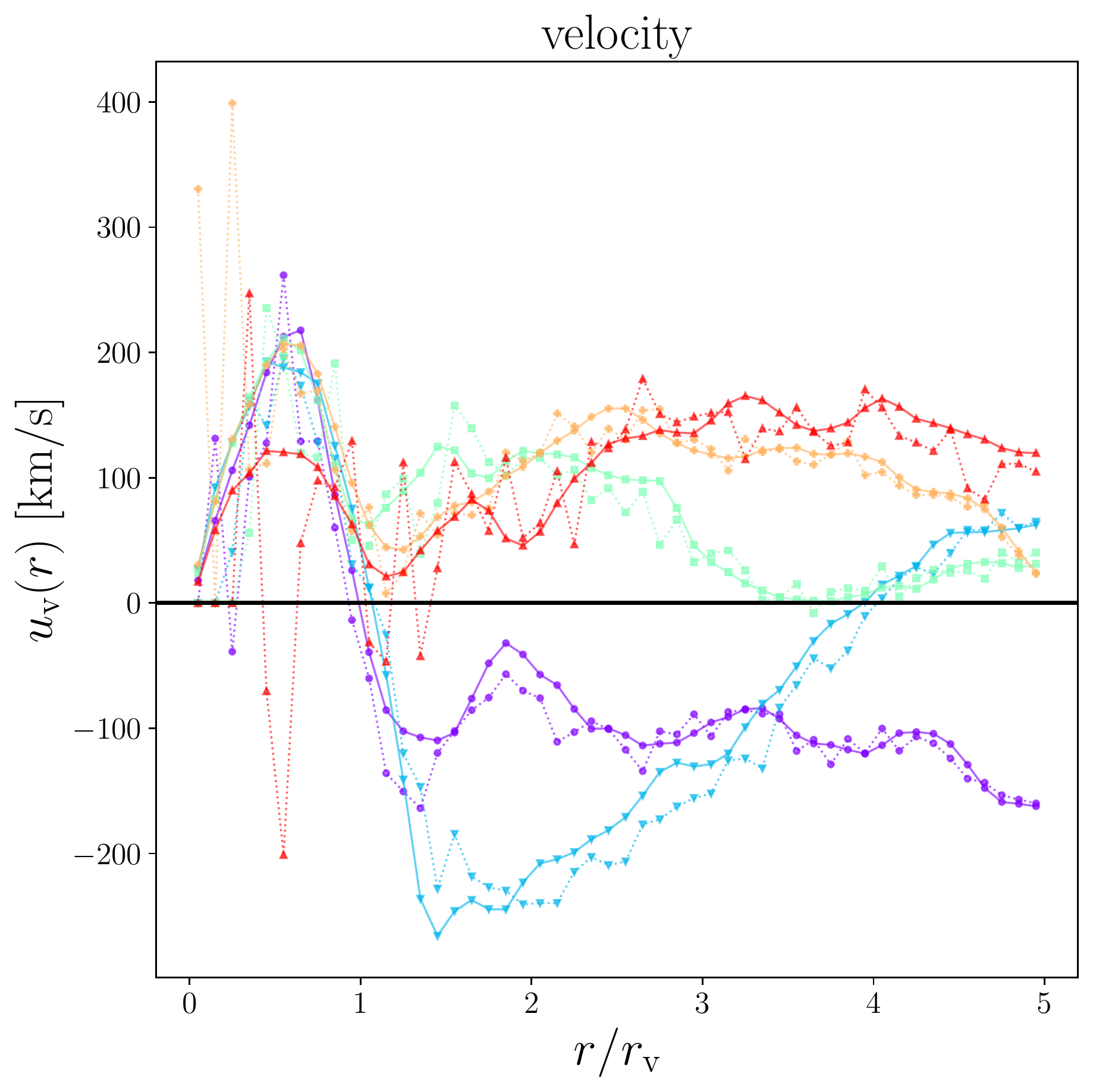}}
\caption{Matter density (left) and velocity (right) profiles of individual \emph{isolated} CDM voids from the \MR{} simulation. Dotted lines show measured density and velocity profiles based on equations~(\ref{eq:general_density_profile}) and~(\ref{eq:individual_velocity_profile}), respectively. Solid lines on the right panel show the linear theory predictions based on equation~(\ref{eq:velocity_relation}), with $b_\tracer = 1 $ for CDM. The legend provides the individual void radii.}
\label{fig_individual_vel_relation_mr_CDM}
\end{figure}

\begin{figure}[ht]

               \centering

               \resizebox{\hsize}{!}{

                               \includegraphics[trim=0 50 0 5, clip]{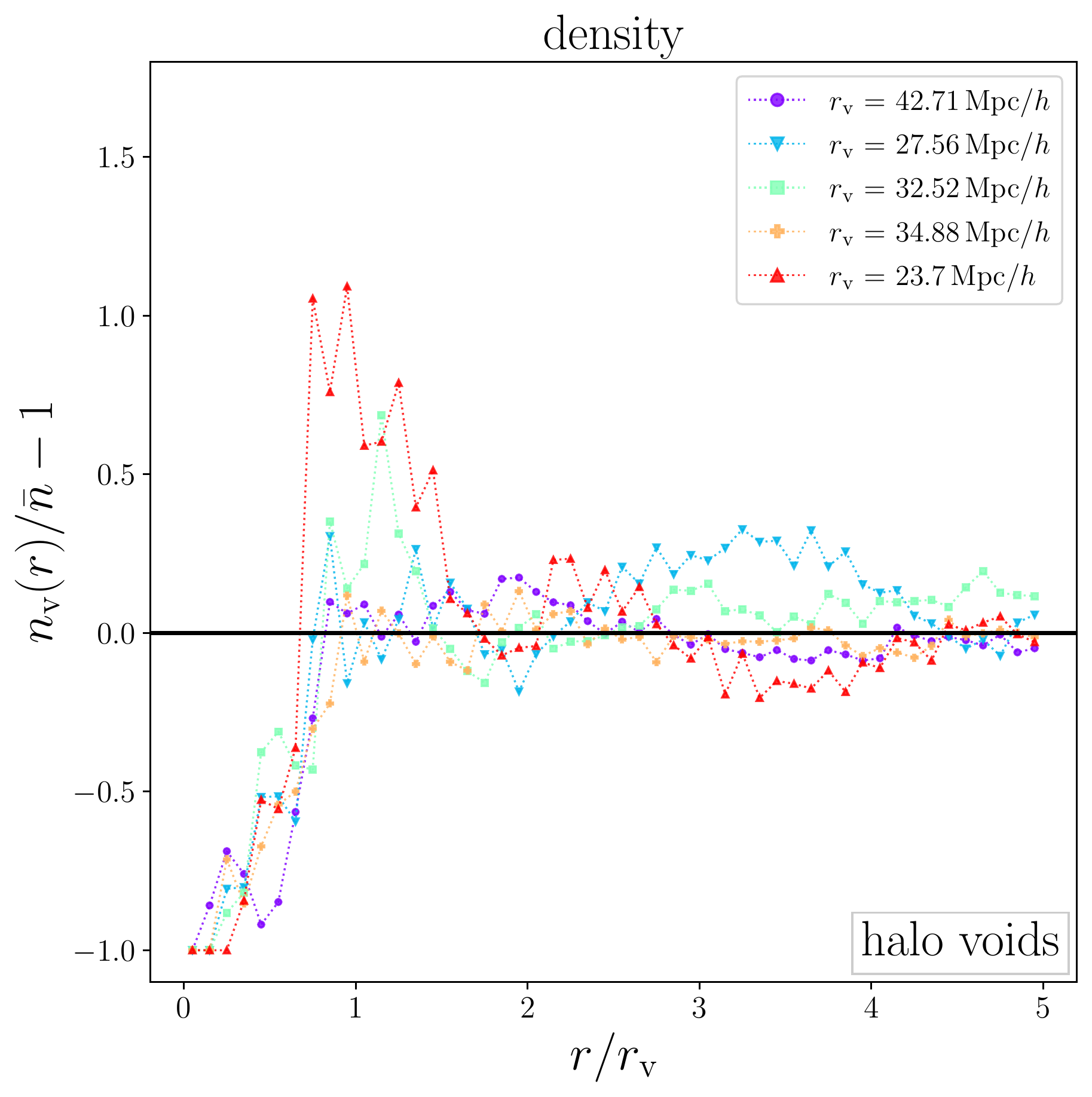}

                               \includegraphics[trim=0 50 0 5, clip]{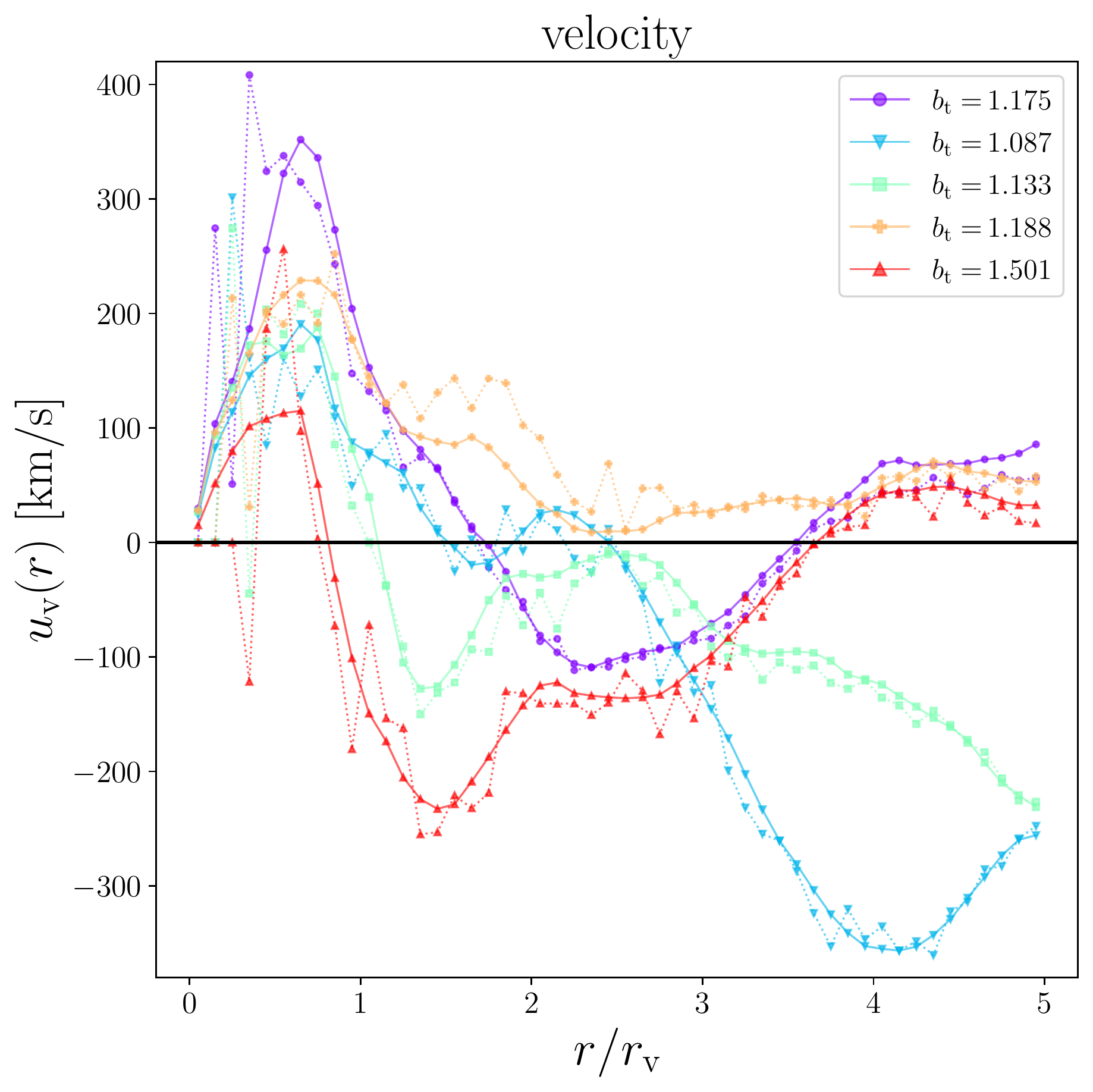}}

               \resizebox{\hsize}{!}{

                               \includegraphics[trim=7 10 0 29, clip]{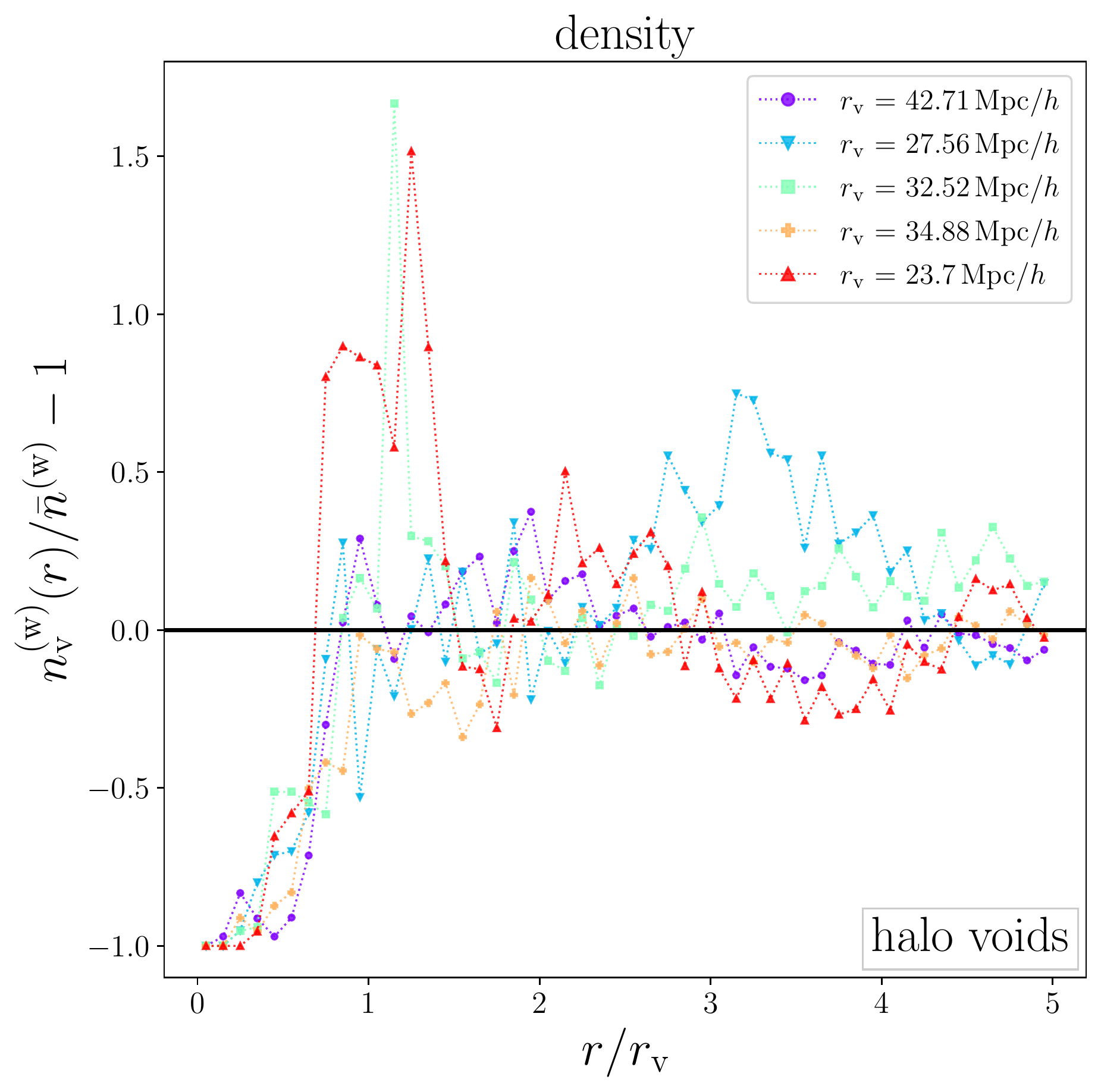}

                               \includegraphics[trim=0 10 0 29, clip]{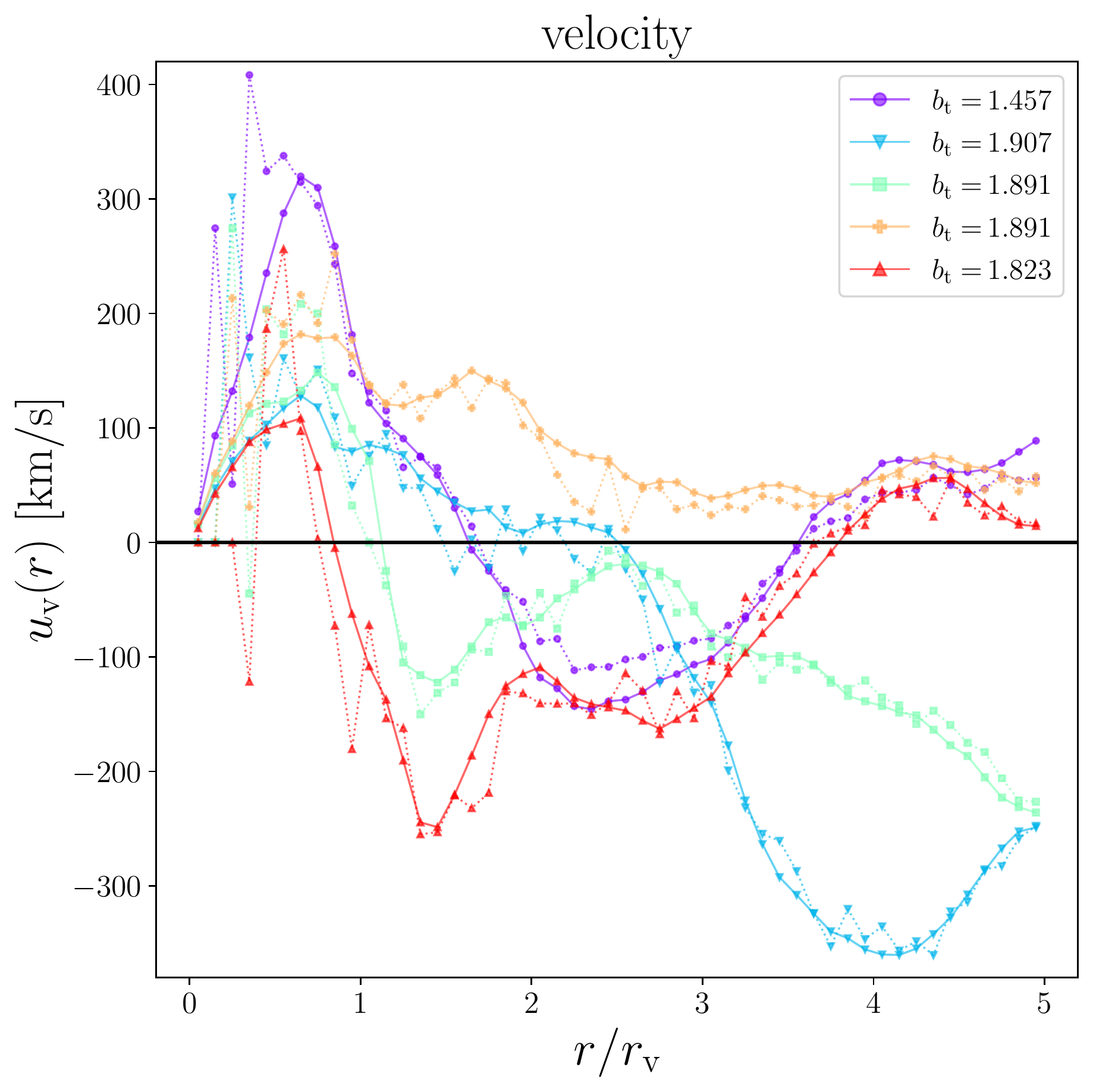}}

               \caption{Same as figure~\ref{fig_individual_vel_relation_mr_CDM}, but for \emph{isolated} halo voids with both unweighted (top) and mass-weighted (bottom) number density profiles of halos. The bias $b_\tracer$ is now a free parameter and fitted to the measured velocity profiles from halo velocities, with best-fit values shown in the legend.}
               
               \label{fig_individual_density_velocity_mr_halo}

\end{figure}

\begin{figure}[t]

               \centering

               \resizebox{\hsize}{!}{

                               \includegraphics[trim=0 50 0 5, clip]{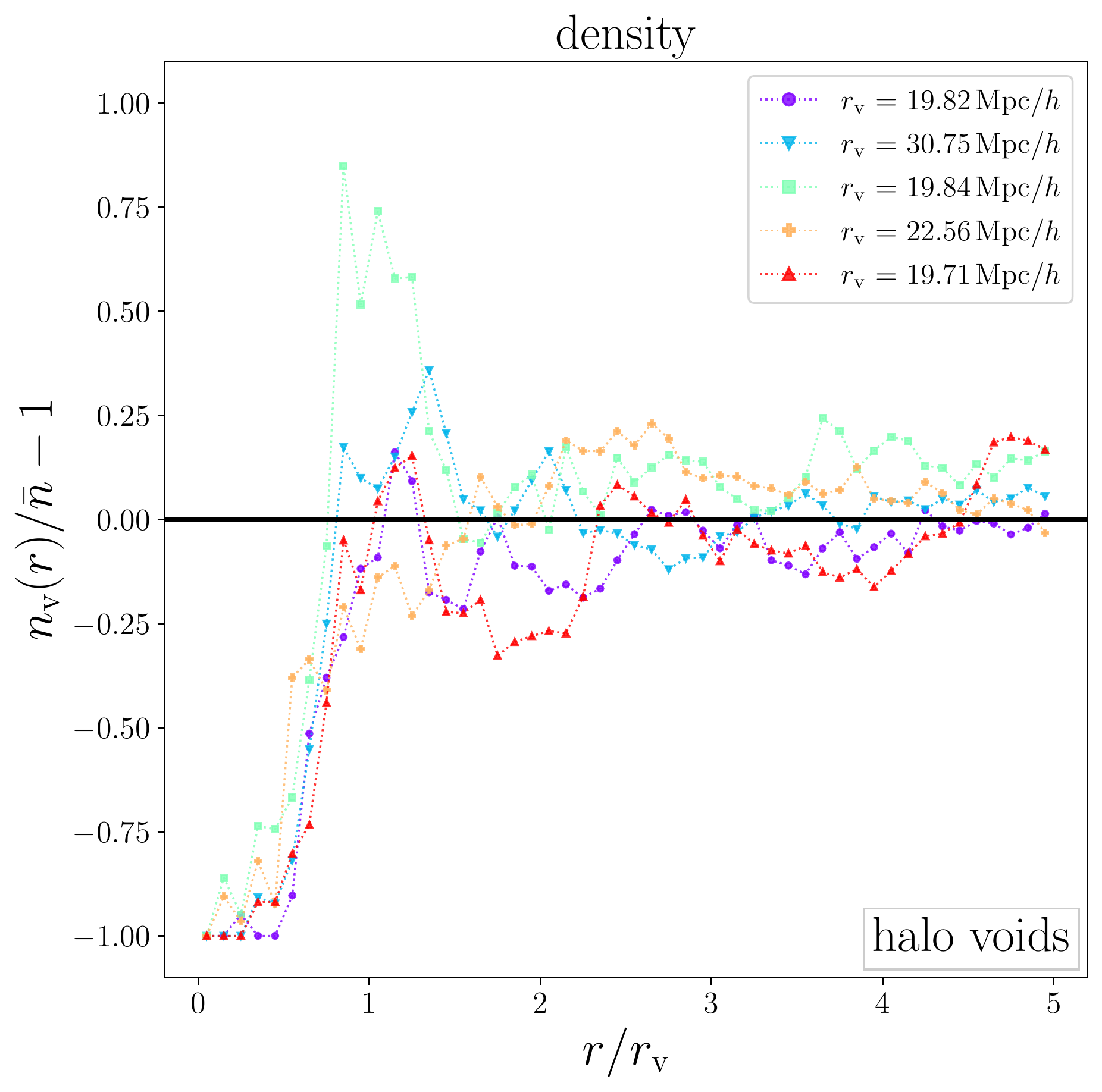}

                               \includegraphics[trim=0 50 0 5, clip]{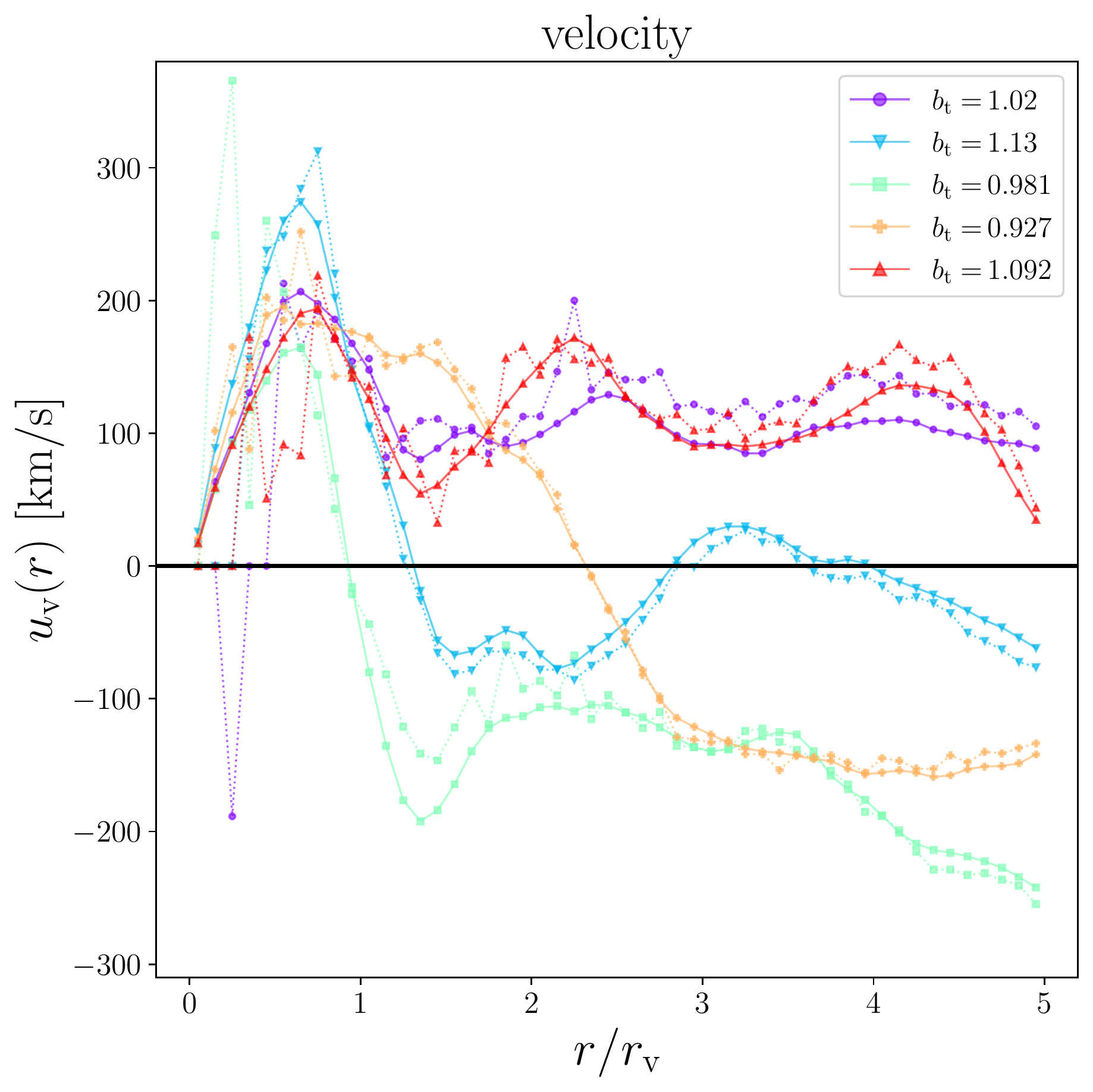}}

               \resizebox{\hsize}{!}{

                               \includegraphics[trim=7 10 0 29, clip]{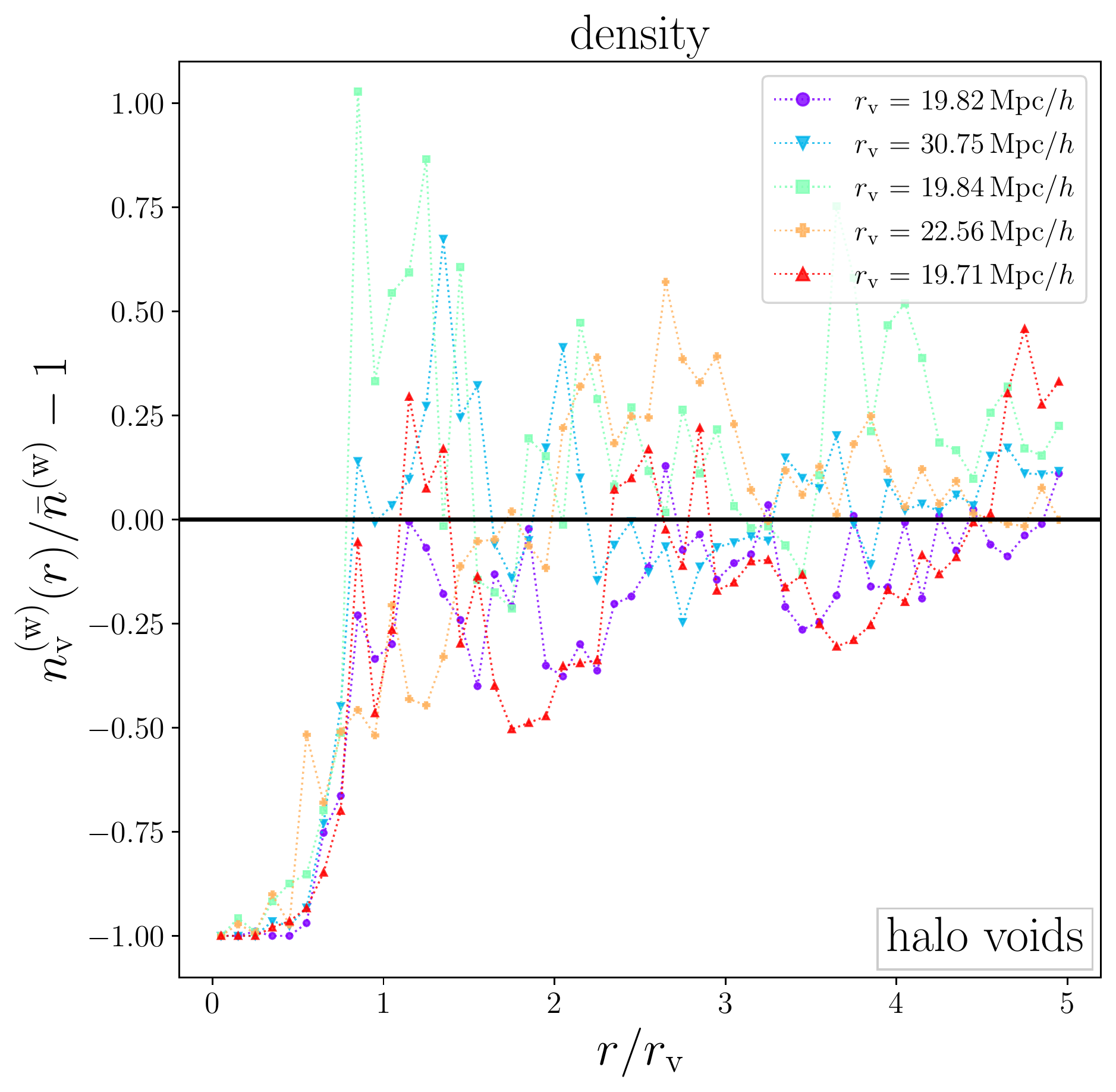}

                               \includegraphics[trim=0 10 0 29, clip]{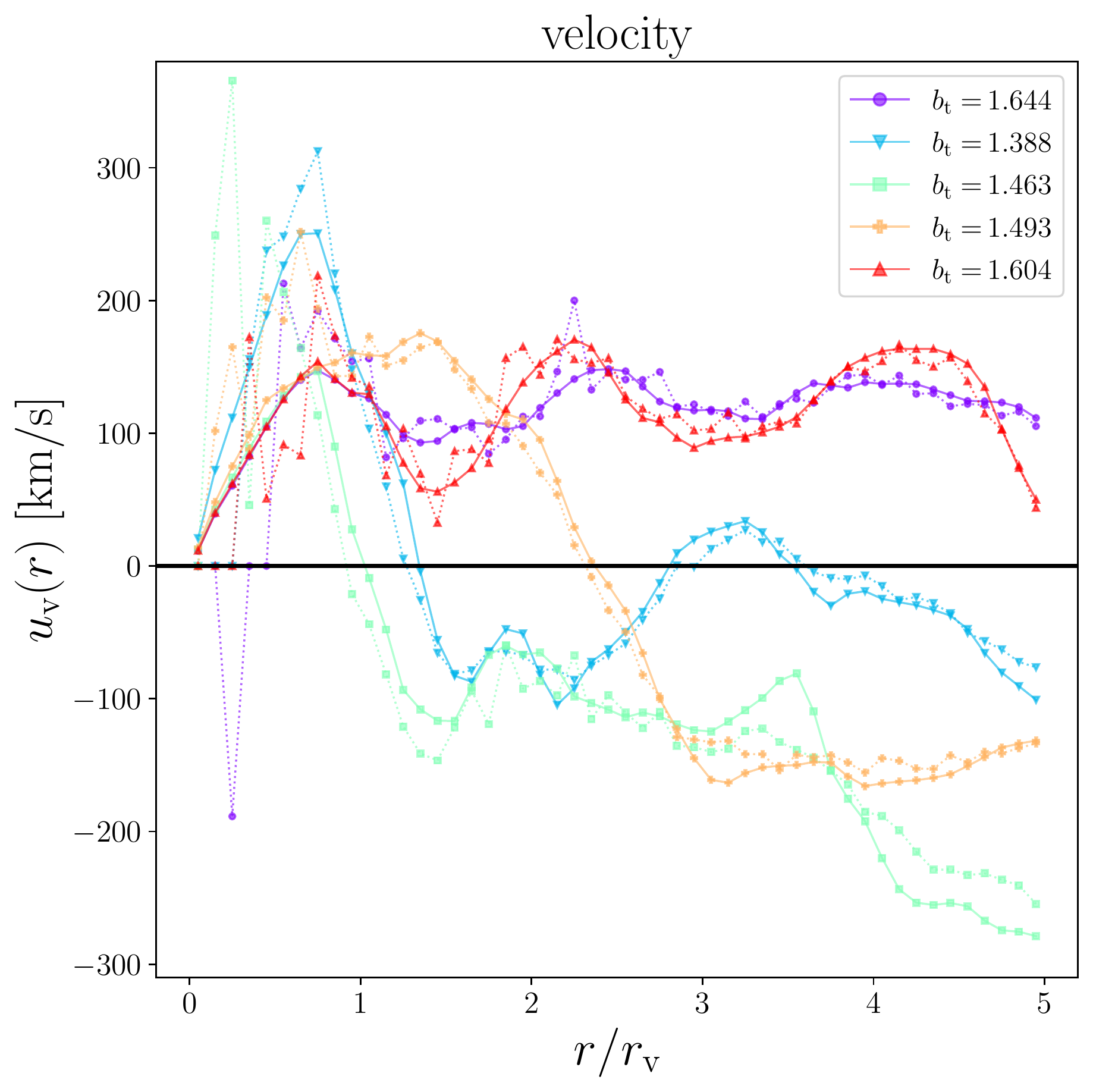}}

               \caption{Same as figure~\ref{fig_individual_density_velocity_mr_halo}, but in the \HR{} simulation.}

               \label{fig_individual_density_velocity_hr_halo}

\end{figure}

We begin with testing the validity of equation~(\ref{eq:velocity_relation}) on density and velocity profiles from individual voids, estimated via equations~(\ref{eq:general_density_profile}) and~(\ref{eq:individual_velocity_profile}), respectively. The left panel of figure~\ref{fig_individual_vel_relation_mr_CDM} depicts the individual matter density profiles of five \emph{isolated} CDM voids from the \MR{} simulation, while their velocity and linear theory profiles (with $b_\tracer = 1$) are shown on the right. These five voids are a random draw from our catalog, with the only condition to sample a range of different void radii. Although the individual density profiles are subject to high sample variance, one can perceive the characteristic underdensity near the voids' center, a compensation wall around $r_\void$, and the trend towards the mean background density at large scales.

Their velocity profiles mirror this fluctuating structure, with some voids dominated by outward motion on all scales and others experiencing infall towards their compensation wall from large distances. One particular void with $r_\void=32.06\,\hMpc$ (in yellow) exhibits a density peak in its very center, which serves as a good example for the sparsity effects that can occur in the estimator from equation~(\ref{eq:general_density_profile}) at the smallest inner shells. Nevertheless, the linear theory profiles match the measured velocity profiles with a remarkable accuracy in every single case. Due to the integral over the density profiles appearing in equation~(\ref{eq:velocity_relation}), the linear theory profiles become smoother than the measured velocity profiles. Near the void centers the differences are more significant due to the tracer sparsity inside voids, but still consistent with the scatter of the measurement. We emphasize that no free parameters have been adjusted in this procedure, the tracer bias of CDM is simply fixed to $b_\tracer = 1$ here.

When using halos as tracers we can no longer assume $b_\tracer = 1$, but have to determine their bias separately. The individual number density profiles of five randomly selected \emph{isolated} halo voids of the \MR{} simulation are presented in figure~\ref{fig_individual_density_velocity_mr_halo}. We additionally revisit mass-weighted density profiles on the bottom panel for the same five voids, as introduced in section~\ref{subsec:mass_weight_profiles}. The two types of number density profiles follow similar shapes, but the mass-weighted ones have a tendency for higher peaks and deeper troughs, because more massive halos tend to reside in regions of higher density, as discussed before. One void with $r_\void=23.7\,\hMpc$ (in red) is embedded within a larger underdensity located between $r = 3r_\void $ and $r = 4r_\void$, a nice example for the void-in-void scenario.

The individual velocity profiles reflect the density structure of all five voids. We now leave the bias $b_\tracer$ in equation~(\ref{eq:velocity_relation}) as a free parameter to fit the linear theory profiles to the measured velocity profiles. Again, the match between the two is striking, even for the mass-weighted case. The only difference from the latter is an enhanced value of the best-fit bias parameter, which corroborates our conclusion from section~\ref{subsec:mass_weight_profiles}.

For completeness we present additional profiles from five \emph{isolated} halo voids of the \HR{} simulation in figure~\ref{fig_individual_density_velocity_hr_halo}. This time we intentionally selected four voids of similar size to point out that their profiles can still be very diverse. The void with $r_\void = 19.71\,\hMpc$ (in red) could be ascribed to a ``void-in-void-in-void'' scenario, exemplifying how deeper hierarchies with multiple levels of sub-voids can occur. This is substantiated by its velocity profile, which exhibits three distinct peaks of enhanced outflow near the boundaries of each sub-void in the density profile. No matter how complex the void structure is, linear theory successfully reproduces the dynamics within each individual void. In the \HR{} simulation the only difference to \MR{} is that lower bias values are obtained, due to the lower mass cut of $10^{11} \Msun$.

To summarize, mass conservation based on the linear continuity equation~(\ref{eq:velocity_relation}) provides an extremely accurate description for the dynamics operating around individual voids that are well resolved, independently of whether CDM or halos are used as tracer particles for void finding. To the best of our knowledge, this unique behavior has not been found in any environment of large-scale structure before.

\subsection{Stacked voids \label{subsec:lin_theory_stacked_voids}}

\begin{figure}[t]
\centering
\resizebox{\hsize}{!}{
    \includegraphics{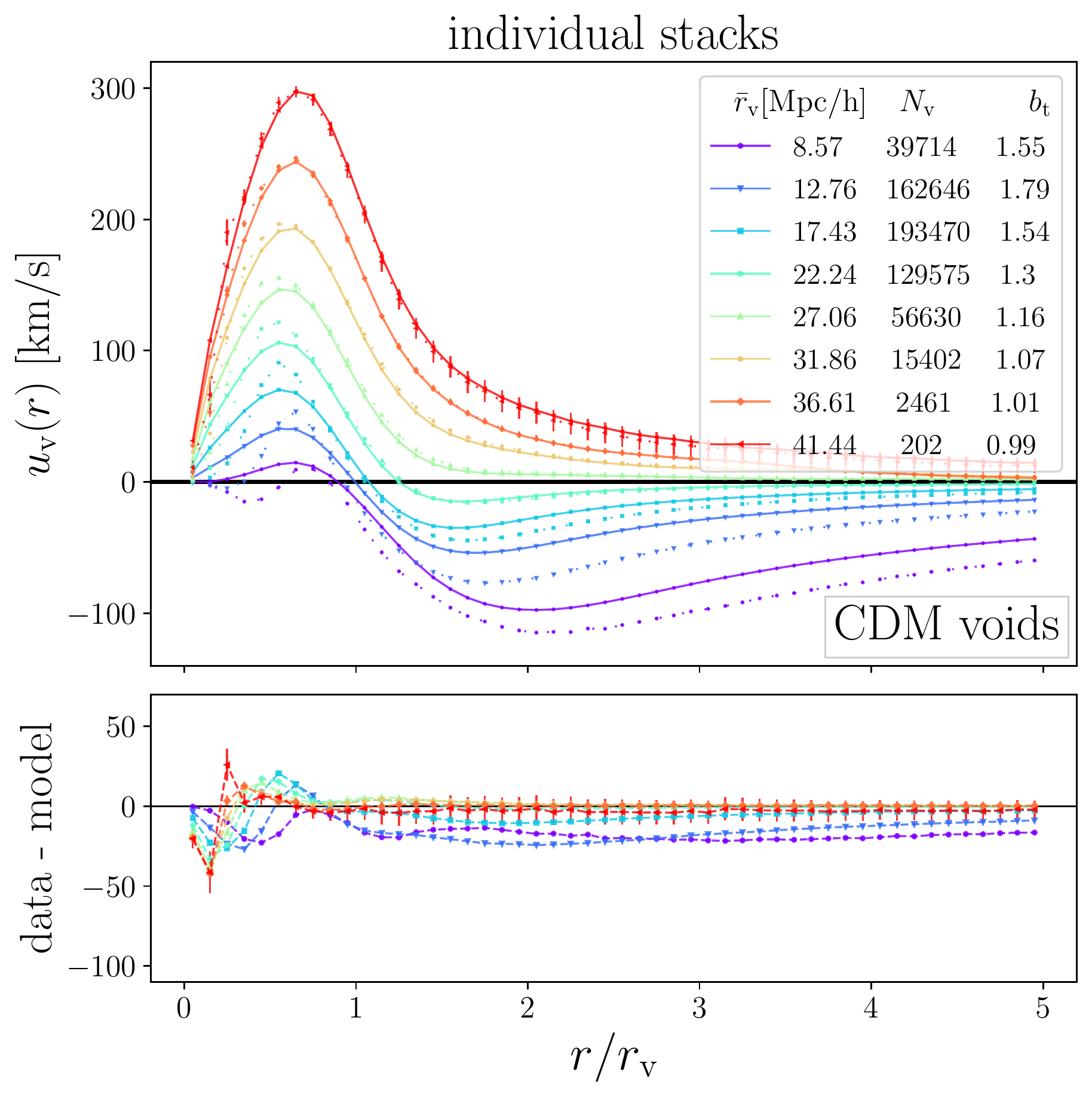}
    \includegraphics{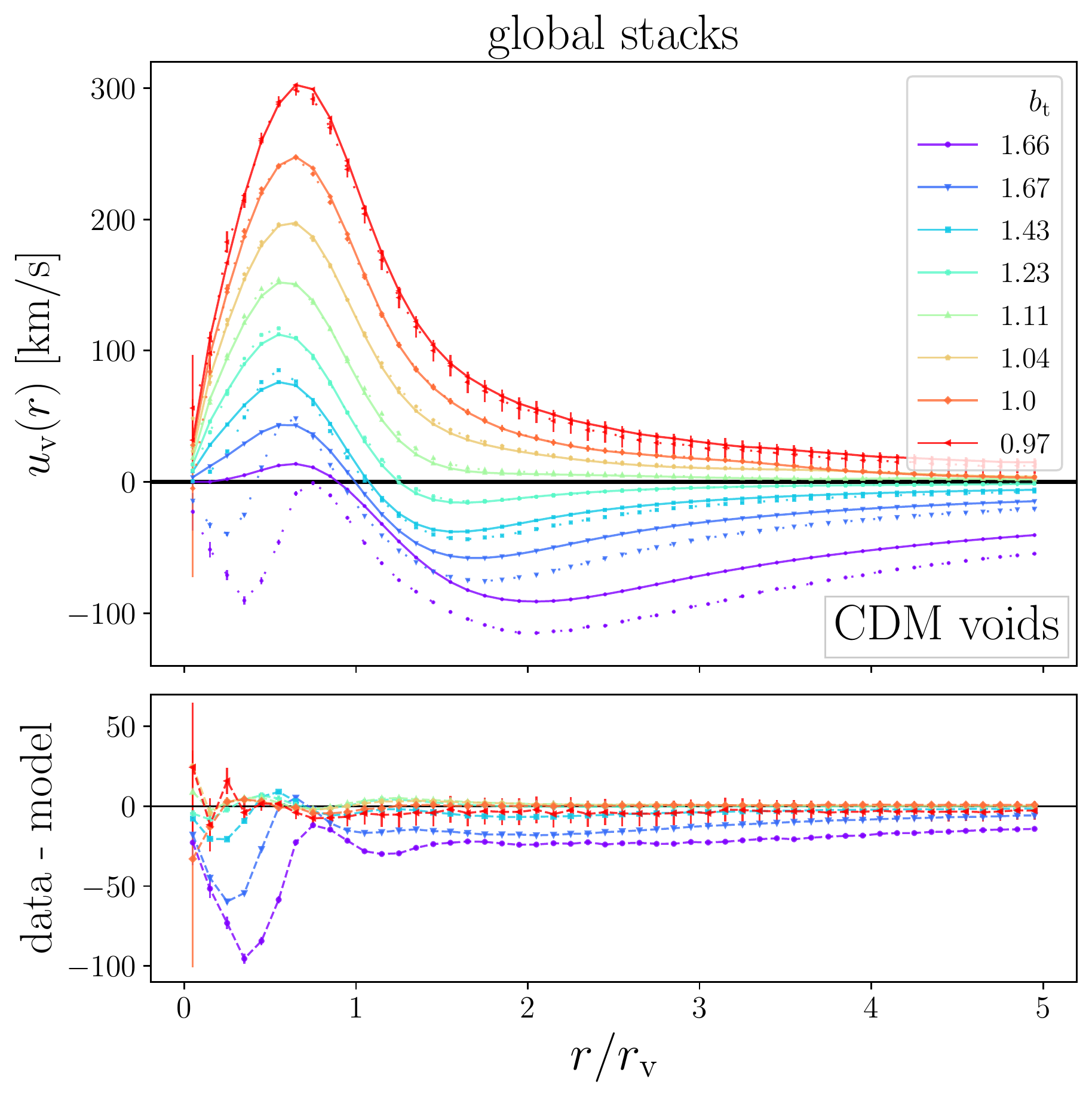}}
\caption{Application of equation~(\ref{eq:velocity_relation}) on stacked matter density profiles of \emph{isolated} CDM voids in the \MR{} simulation (from the top left of figure~\ref{fig_density_mr_CDM_halo_merging}). Solid lines show the linear theory predictions and dotted lines the measured velocity profiles for individual (left) and global stacks (right). The bottom panels highlight differences between measured velocity (data) and linear theory profiles (model).}
\label{fig_mass_conservation_stacked_density}
\end{figure}

\begin{figure}[t]

               \centering

               \resizebox{\hsize}{!}{

                               \includegraphics[trim=0 50 42 5, clip]{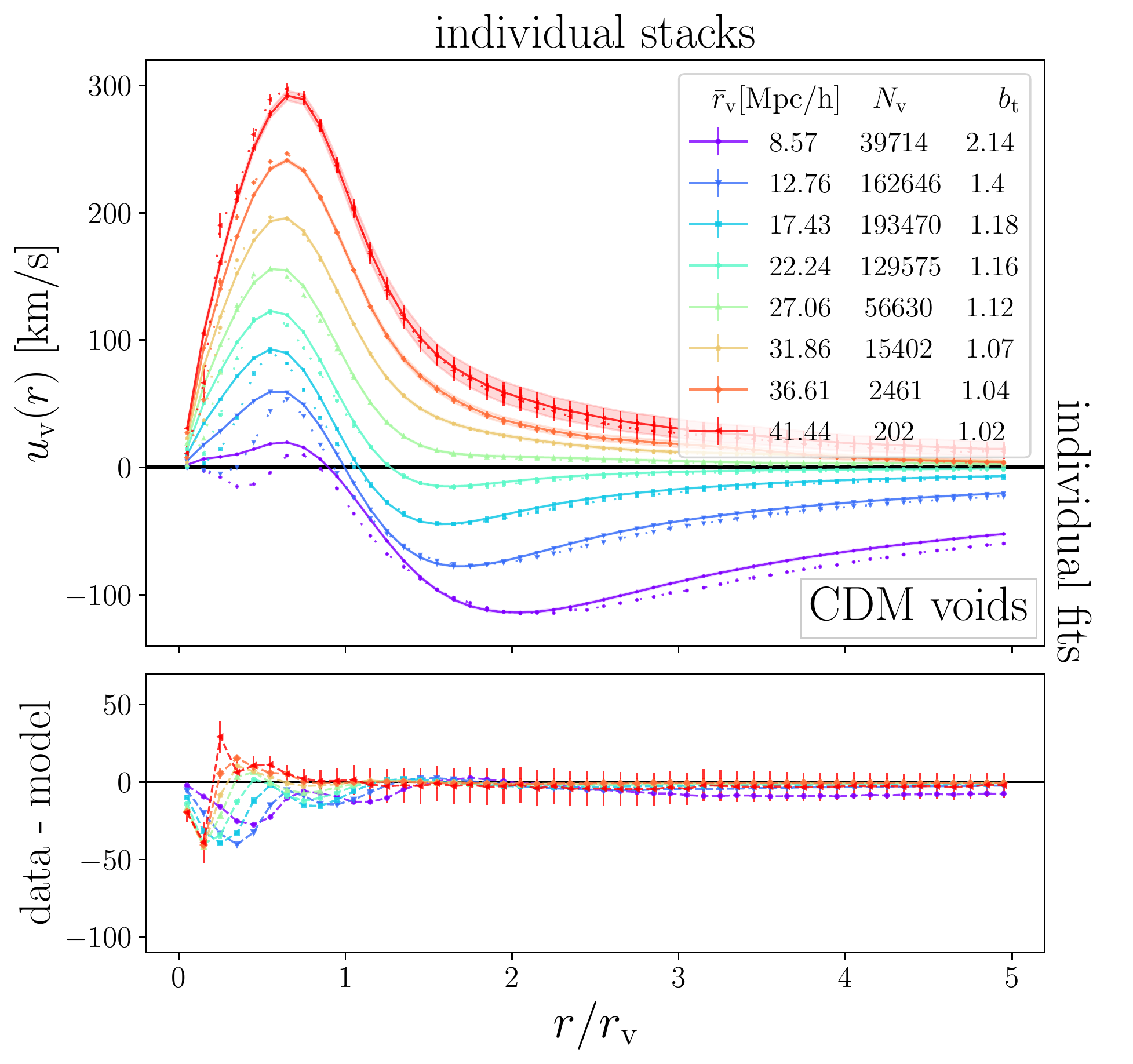}

                               \includegraphics[trim=0 50 20 5, clip]{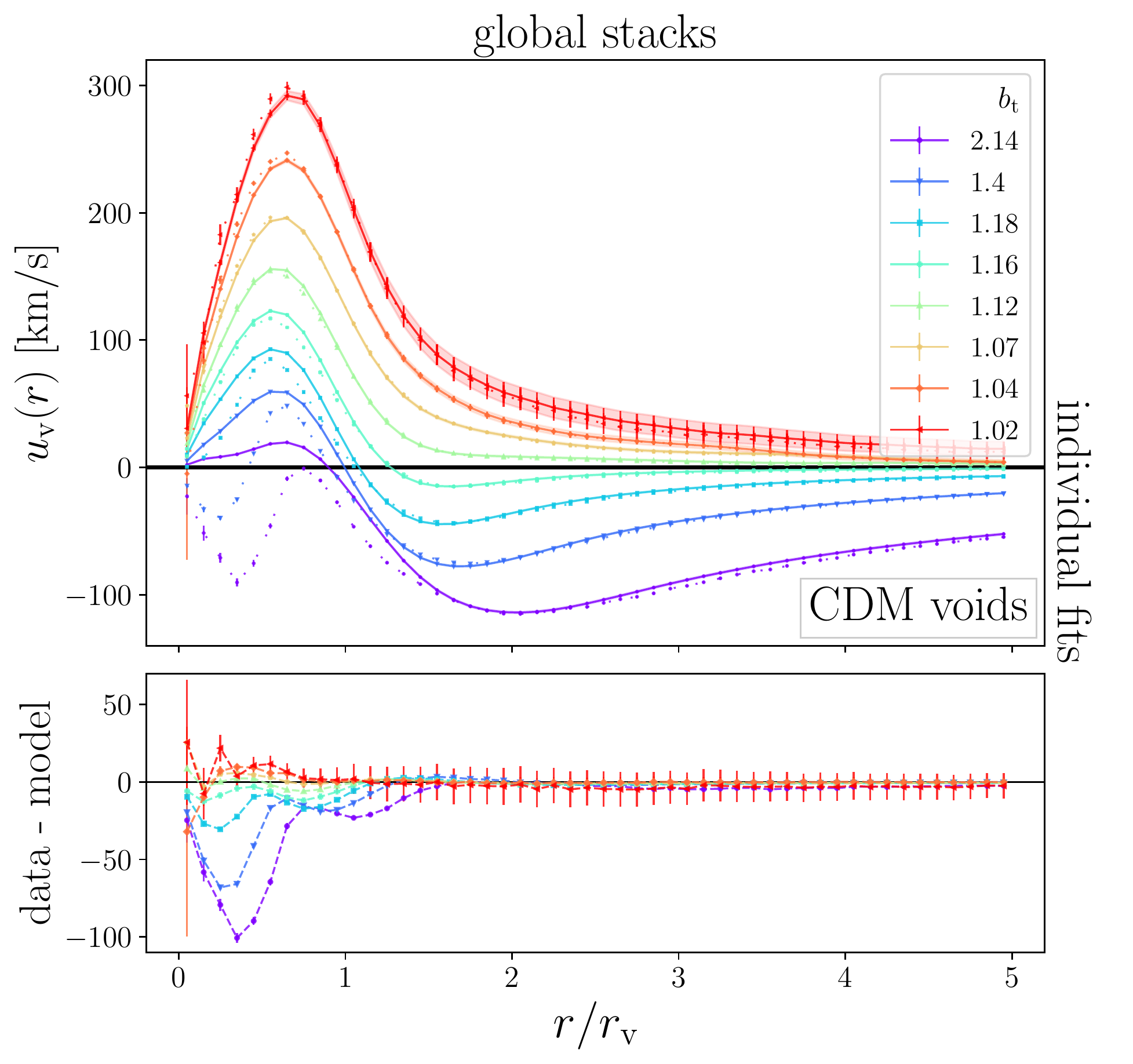}}

                 \resizebox{\hsize}{!}{

                               \includegraphics[trim=0 10 42 29, clip]{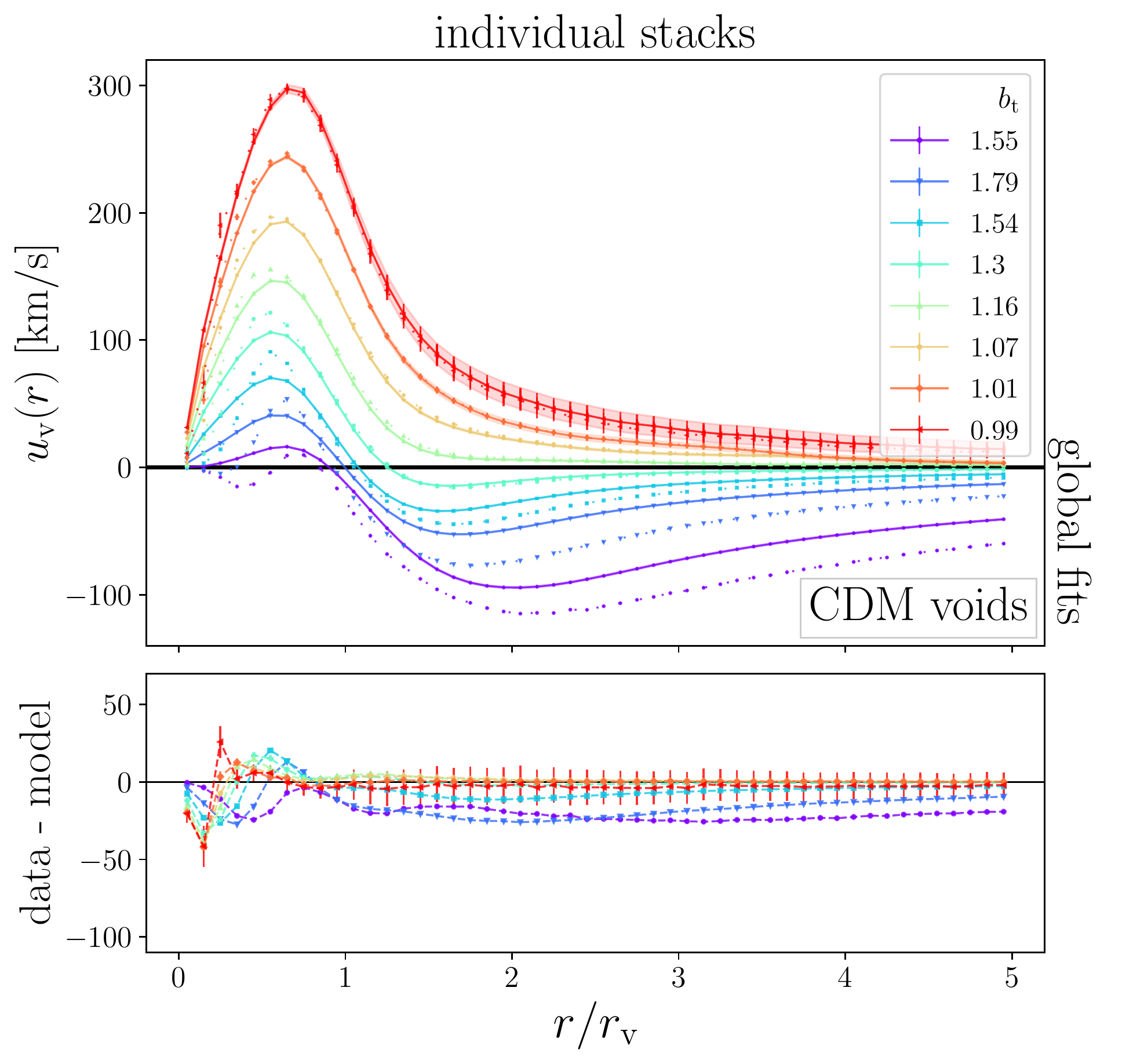}

                               \includegraphics[trim=0 10 20 29, clip]{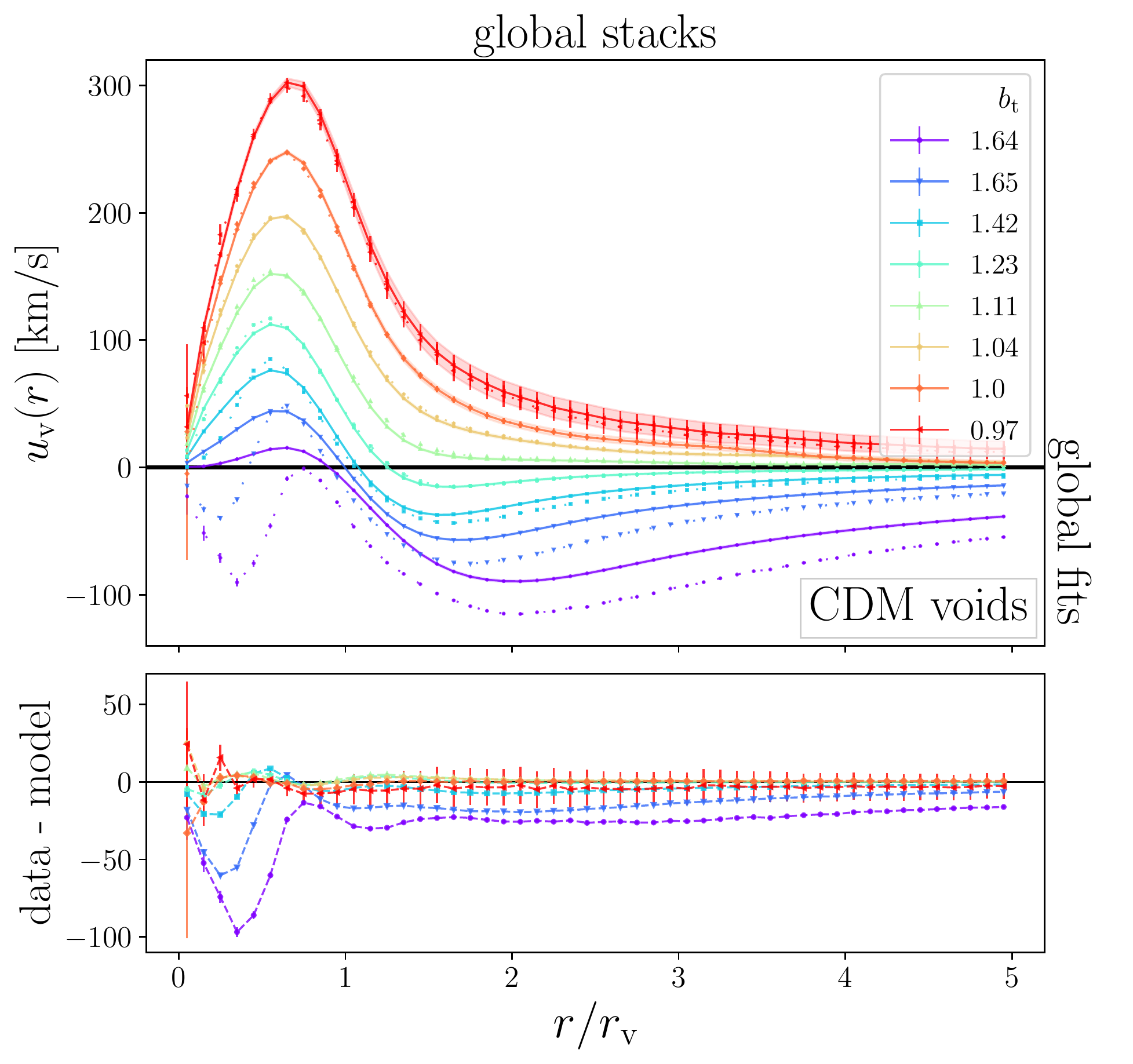}}

               \caption{Same as figure~\ref{fig_mass_conservation_stacked_density}, but for individual fits (top) and global fits (bottom) of the tracer bias. For individual fits the tracer bias is determined via the linearized continuity equation~(\ref{eq:velocity_relation}) for each individual void before averaging. For global fits a single tracer bias parameter is determined after averaging the individual linear theory profiles with $b_\tracer=1$ and applying equation~(\ref{eq:velocity_relation}). Shaded regions indicate standard deviations around the mean linear theory profiles.}

               \label{fig_stacks_density_velocity_mr_CDM}

\end{figure}

\begin{figure}[t]

               \centering

               \resizebox{\hsize}{!}{

                               \includegraphics[trim=0 50 42 5, clip]{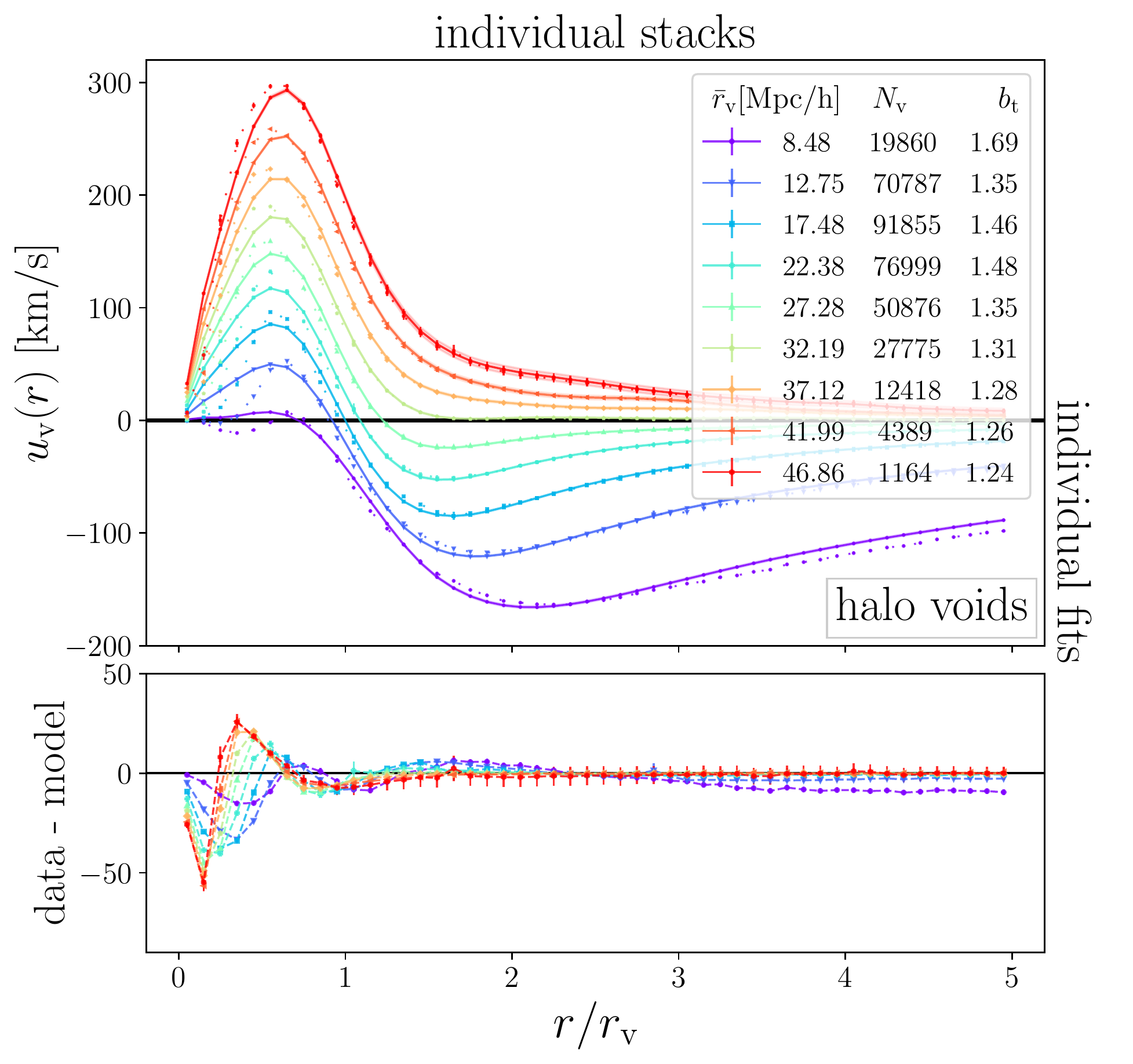}

                               \includegraphics[trim=0 50 20 5, clip]{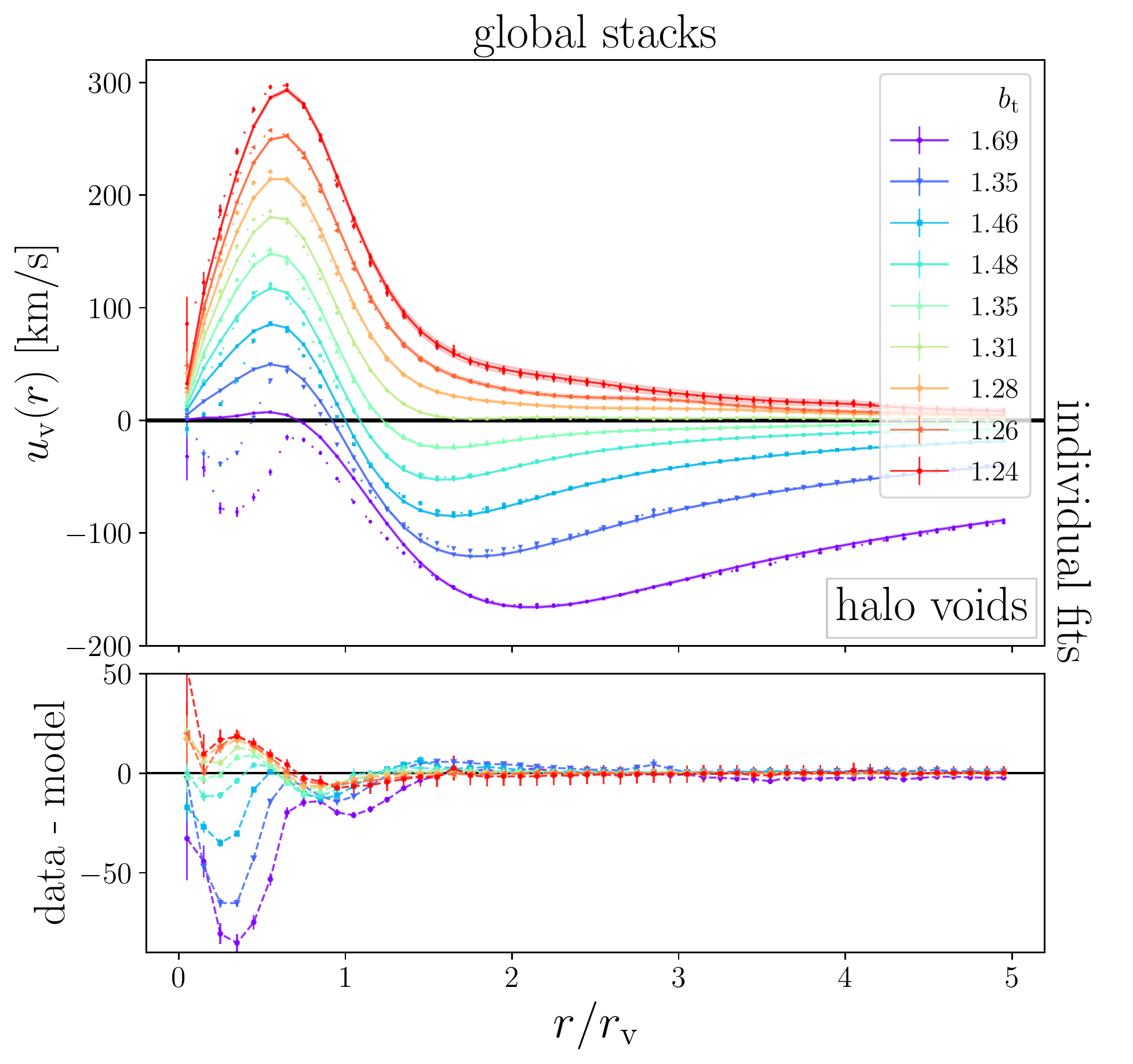}}

                 \resizebox{\hsize}{!}{

                               \includegraphics[trim=0 10 42 29, clip]{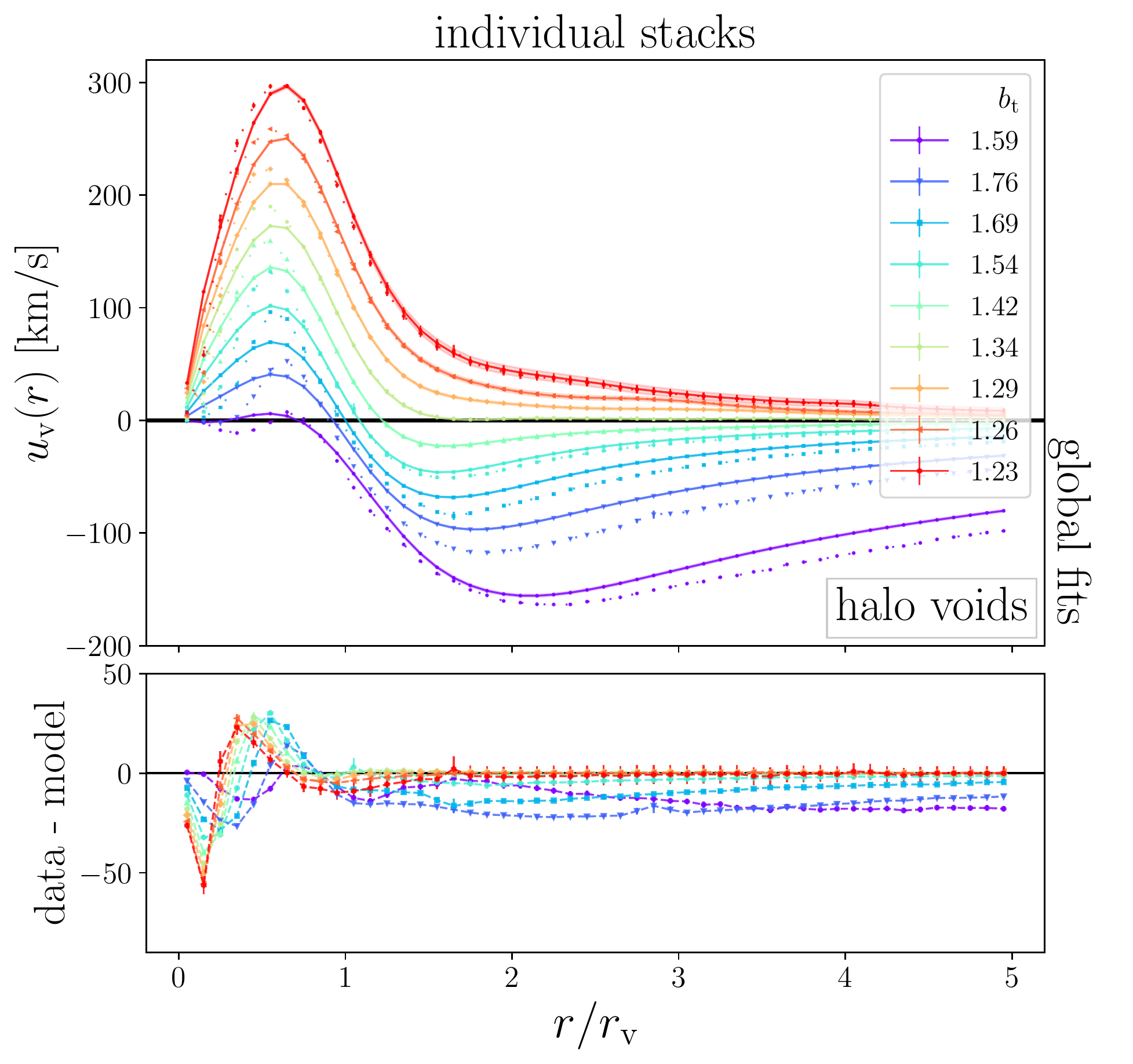}

                               \includegraphics[trim=0 10 20 29, clip]{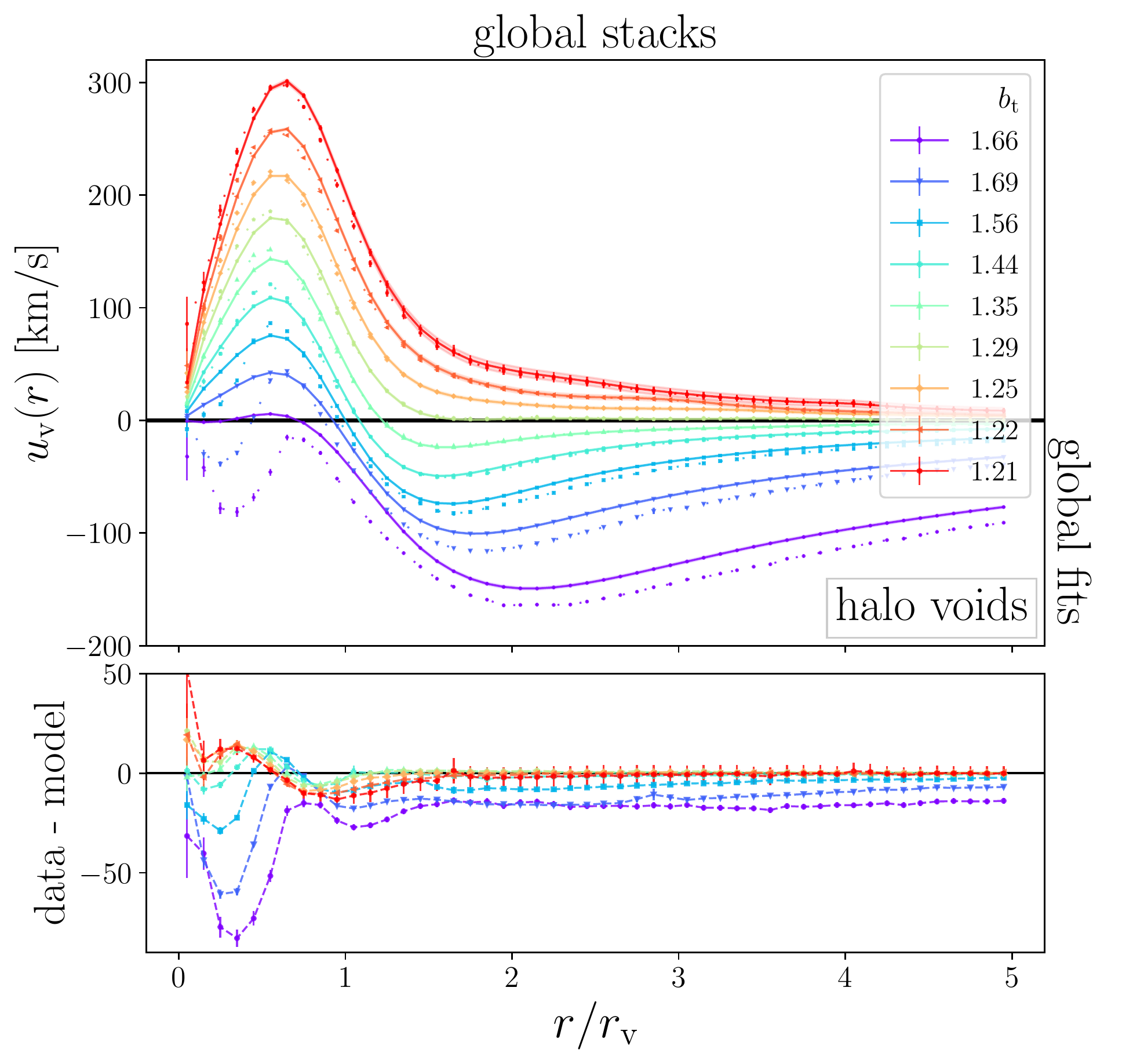}}

               \caption{Same as figure~\ref{fig_stacks_density_velocity_mr_CDM}, but for \emph{isolated} halo voids in the \MR{} simulation.}

               \label{fig_stacks_density_velocity_mr_halos}

\end{figure}

\begin{figure}[t]

               \centering

               \resizebox{\hsize}{!}{

                               \includegraphics[trim=0 50 42 5, clip]{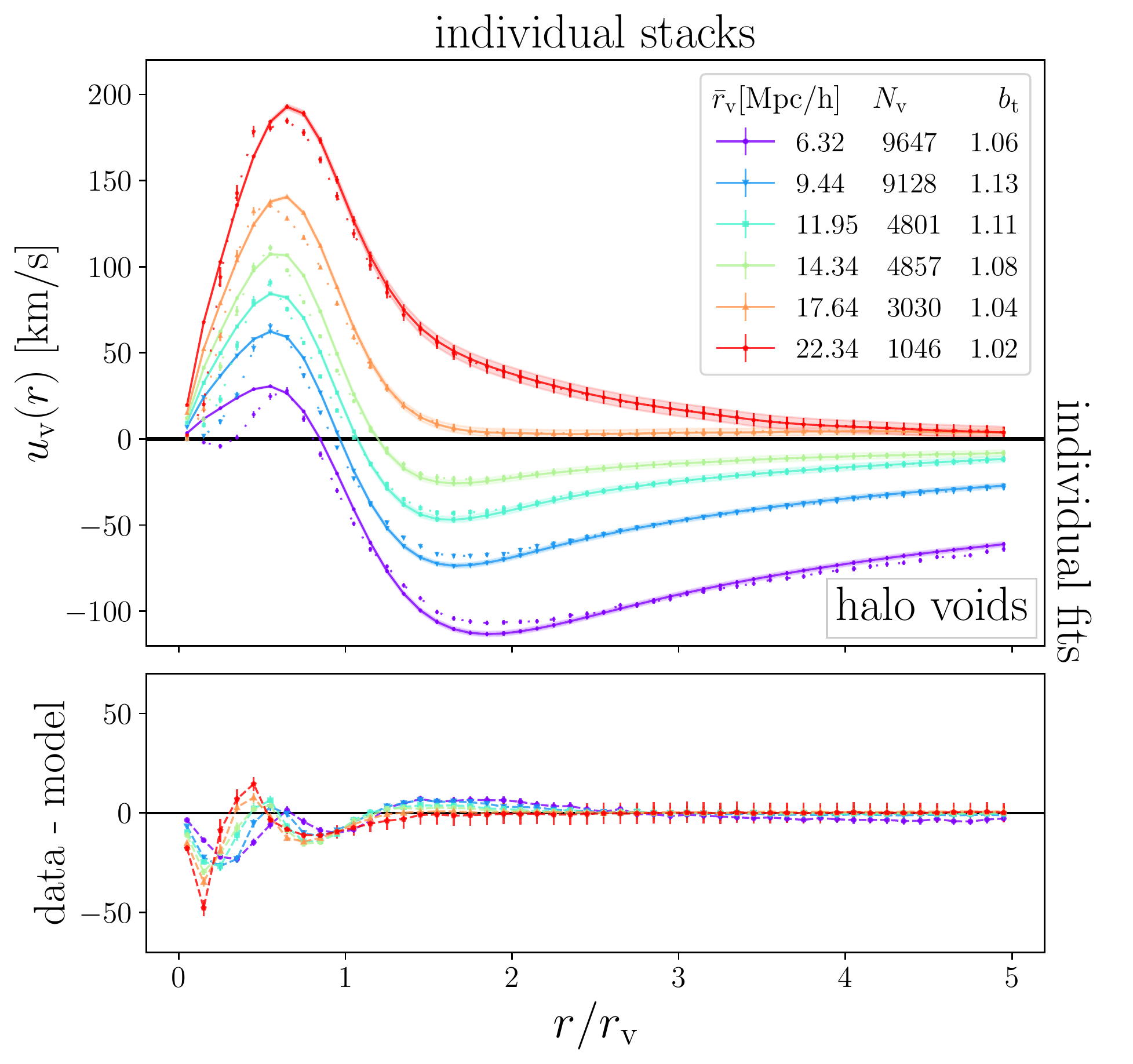}

                               \includegraphics[trim=0 50 20 5, clip]{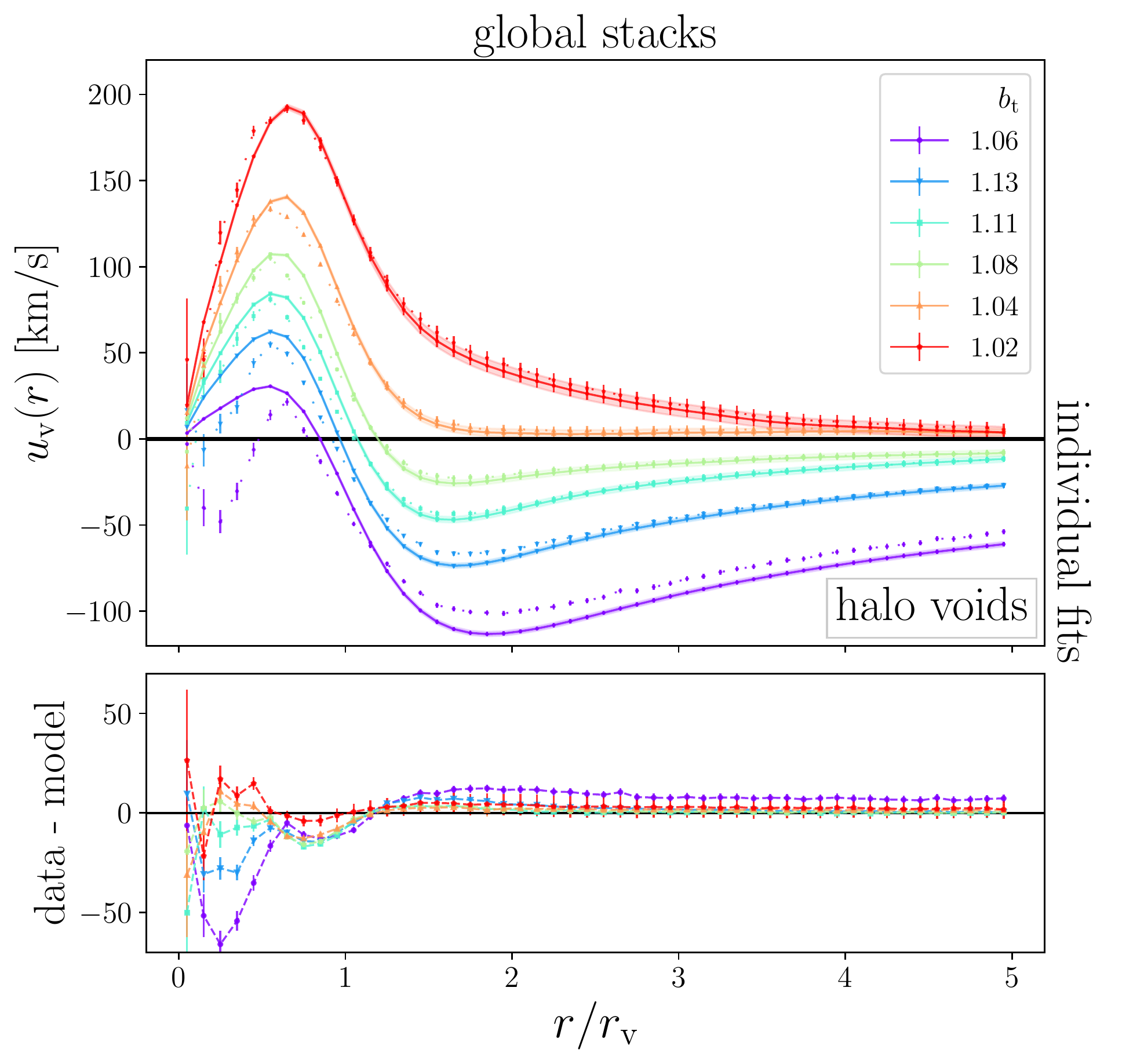}}

                 \resizebox{\hsize}{!}{

                               \includegraphics[trim=0 10 42 29, clip]{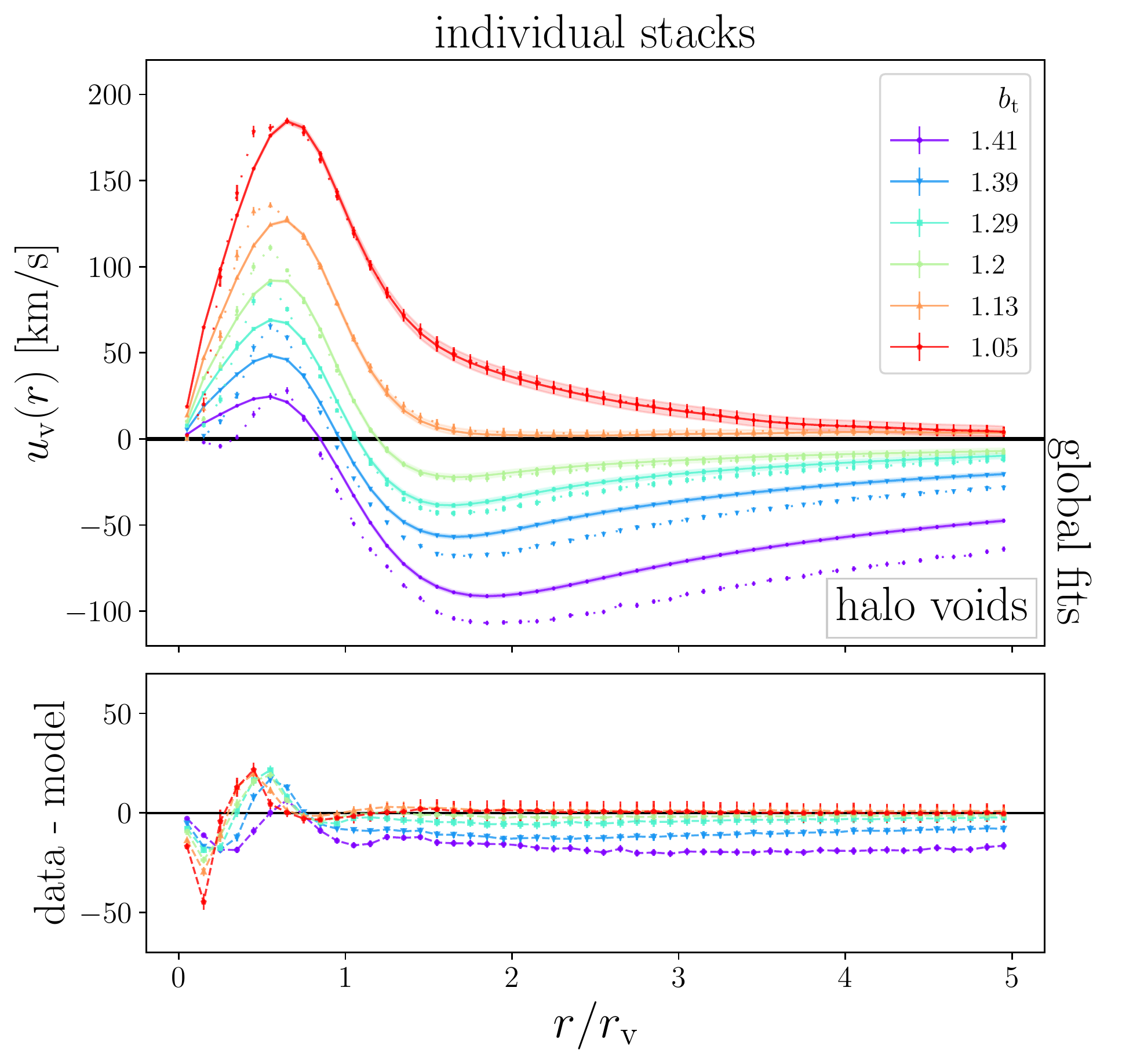}

                               \includegraphics[trim=0 10 20 29, clip]{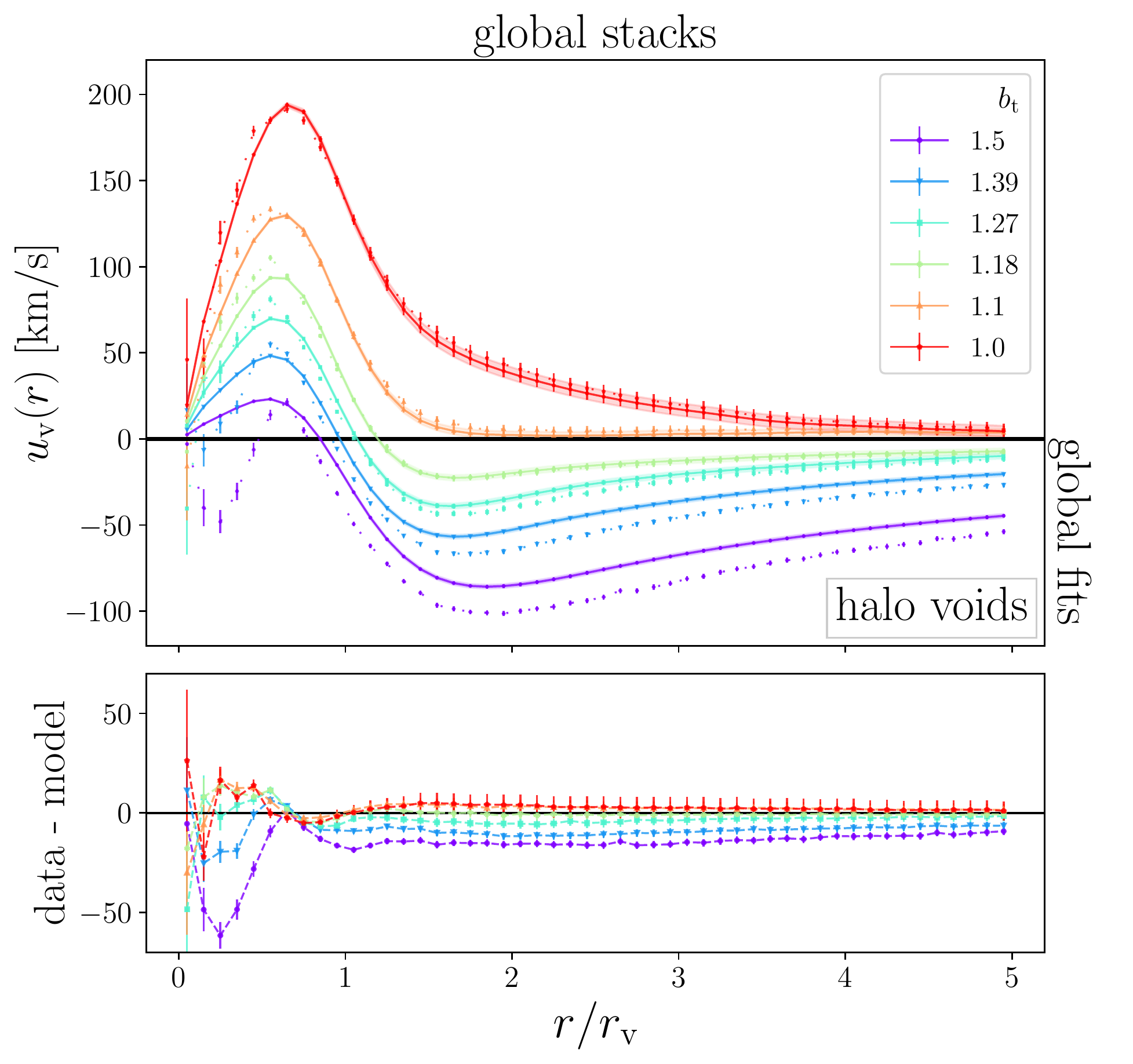}}

               \caption{Same as figure~\ref{fig_stacks_density_velocity_mr_halos}, but for the \HR{} simulation.}

               \label{fig_stacks_density_velocity_hr_halos}

\end{figure}

\begin{figure}[t]

               \centering

               \resizebox{\hsize}{!}{

                               \includegraphics[trim=0 50 42 5, clip]{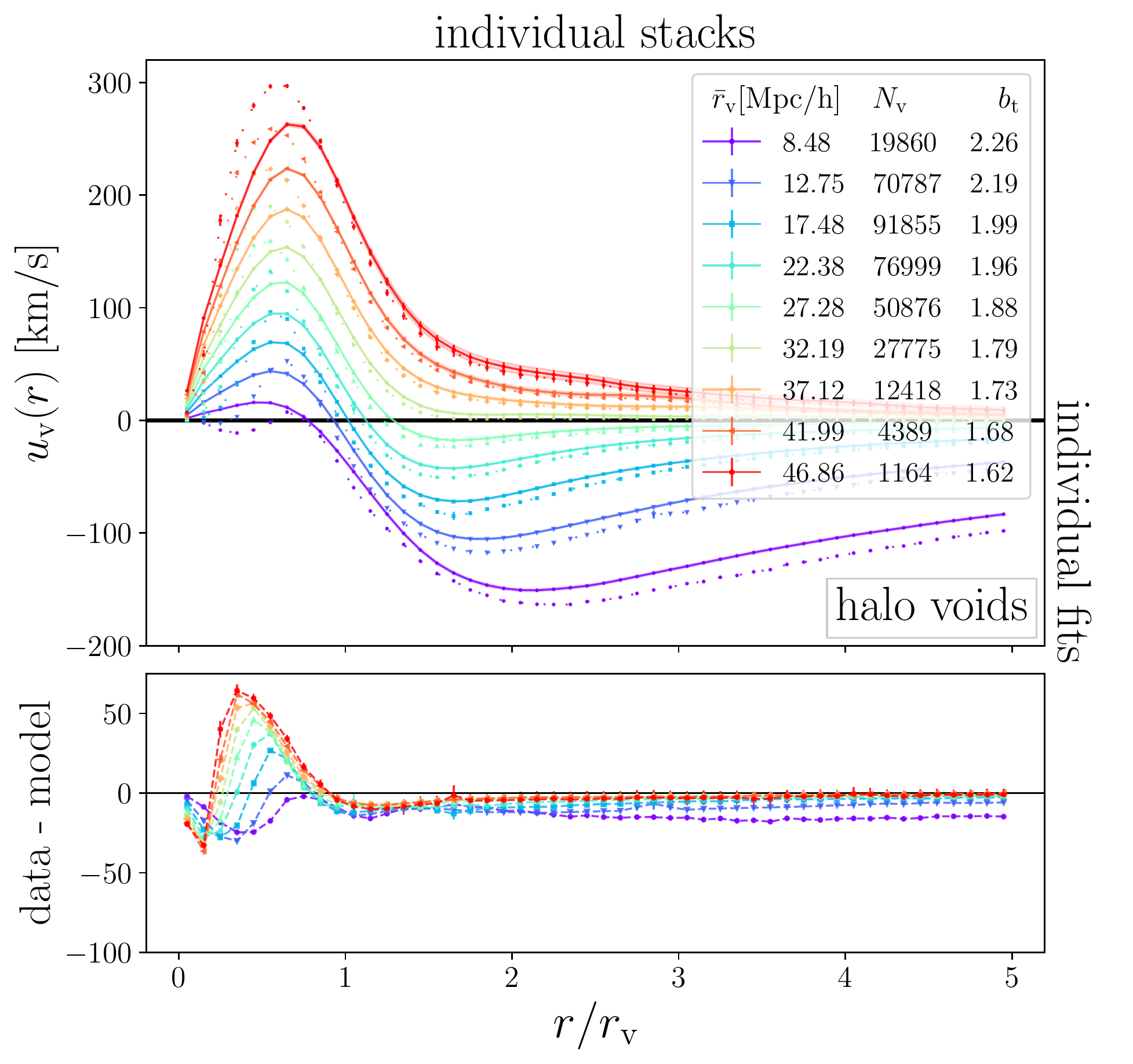}

                               \includegraphics[trim=0 50 20 5, clip]{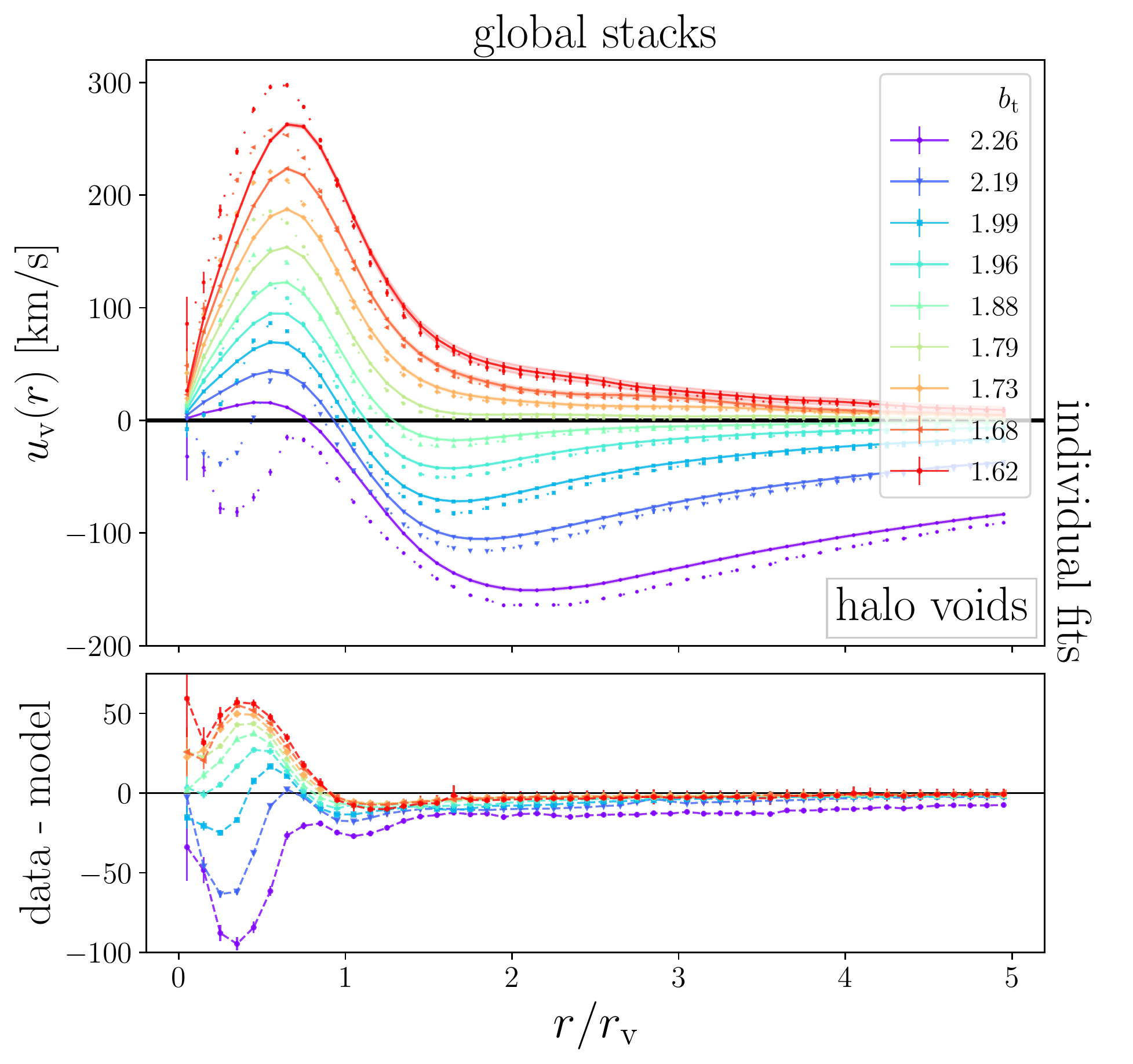}}

                 \resizebox{\hsize}{!}{

                               \includegraphics[trim=0 10 42 29, clip]{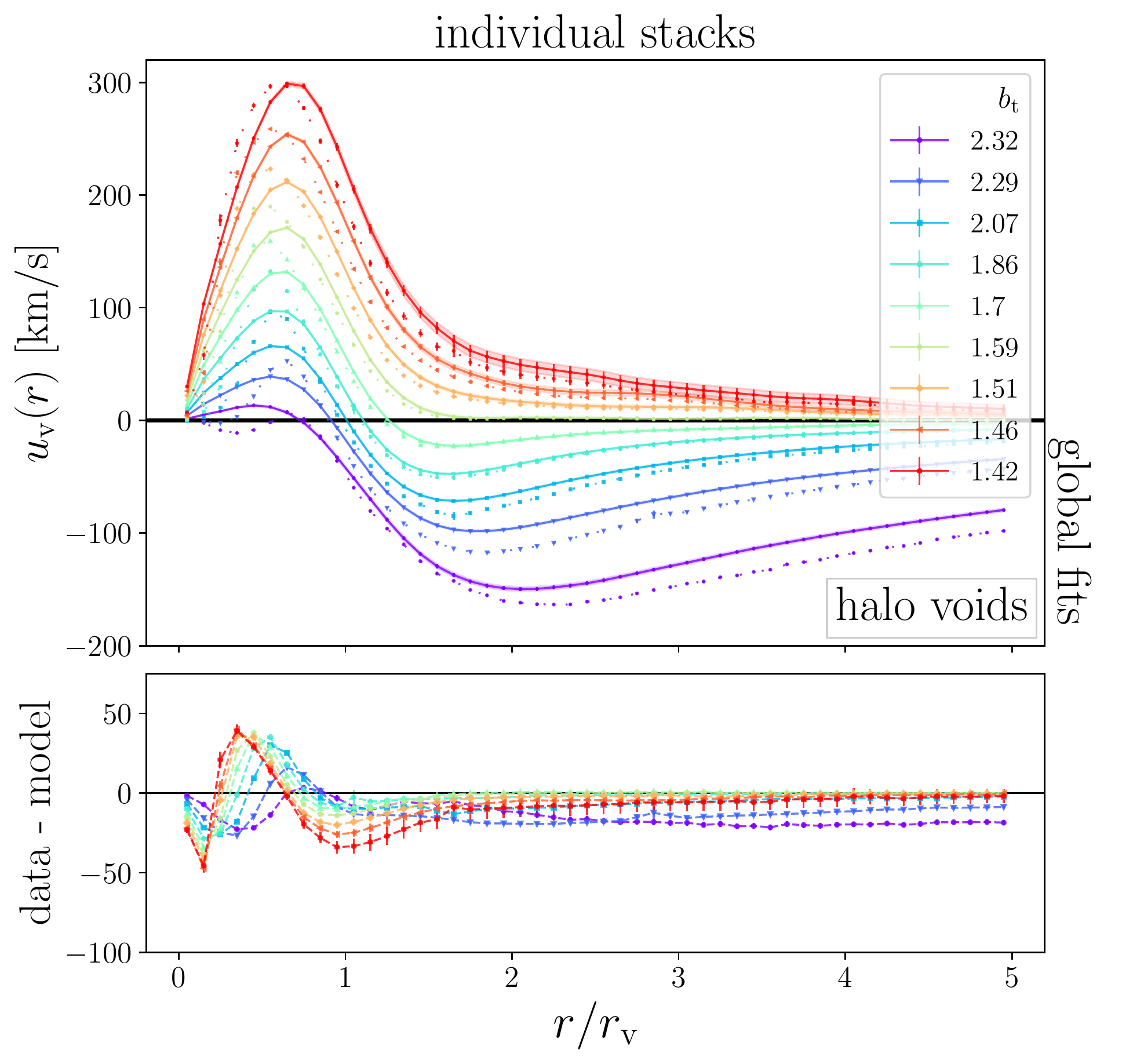}

                               \includegraphics[trim=0 10 20 29, clip]{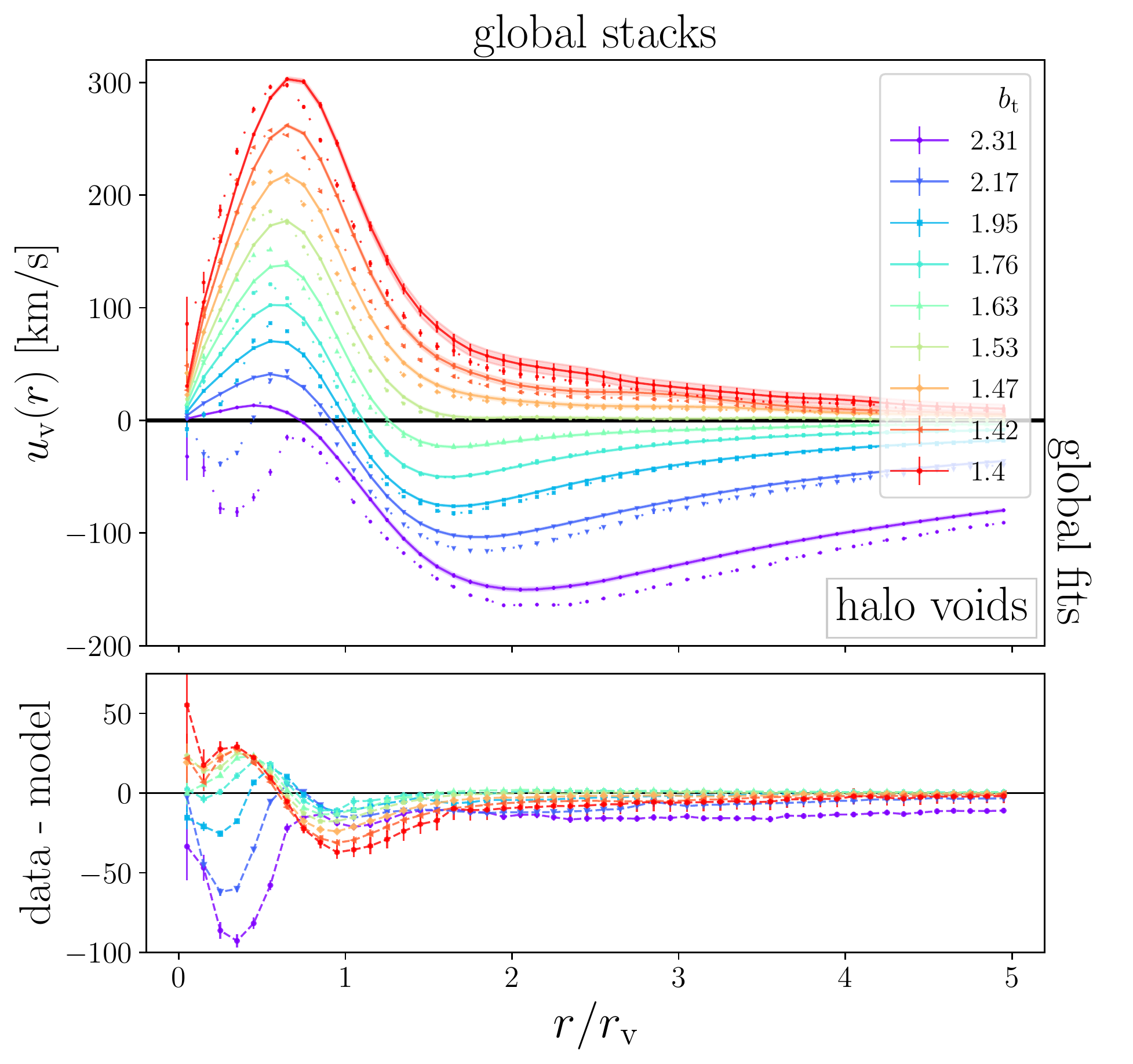}}

               \caption{Same as figure~\ref{fig_stacks_density_velocity_mr_halos}, but using mass-weighted density profiles in equation~(\ref{eq:velocity_relation}).}

               \label{fig_stacks_density_velocity_mr_halos_massweight}

\end{figure}

In the same fashion as performed for individual voids, we may apply the linear continuity equation~(\ref{eq:velocity_relation}) for stacked voids. To this end we can make use of the previously examined matter density profiles of CDM voids, number density profiles of halo voids (with or without mass weighting), and their corresponding velocity profiles (individual and global stacks). With all these variants there are various options to test linear dynamics with stacked void profiles. Reference~\cite{Hamaus2014b} already presented such a test based on individually stacked profiles of \emph{merged} CDM voids, which we repeat here with \emph{isolated} CDM voids for both stacking methods in figure~\ref{fig_mass_conservation_stacked_density}. We use identical void radius bins as used in the upper left panel of figure~\ref{fig_density_mr_CDM_halo_merging}. Furthermore, we treat $b_\tracer$ as a single free parameter even for the case of CDM. In that case, deviations from $b_\tracer=1$ can indicate additional biases of estimators.

We confirm a very good agreement between the linear prediction and the velocity profiles for both individual and global stacks, especially for larger voids. The profiles from global stacks exhibit a more linear rise near their center, which is slightly better in line with linear theory than in the case of individual stacks. Deviations become stronger for small voids, especially close to their center due to tracer sparsity, as discussed in section~\ref{subsec:velocity_profiles}. However, also at larger distances from the void center a constant offset remains for small voids. The best-fit values of the tracer bias parameter converge towards $1$ for large voids, as expected for CDM tracers. Smaller voids exhibit values exceedingly higher than unity, in accordance with the results of reference~\cite{Pollina2017}. These conclusions are not affected by the choice of merging threshold and remain valid for halo voids and their number density and velocity profiles, which yield higher values for the tracer bias, as expected from halos.

With the methodology developed for individual void profiles in section~\ref{subsec:lin_theory_individual_voids}, we can now introduce two additional tests for linear dynamics around stacked voids. The first option is to apply equation~(\ref{eq:velocity_relation}) to each individual density profile of a given void sample, fit $b_\tracer$ for every void profile separately, and average the linear theory profiles and best-fit bias parameters in the end. We will refer to this method as \textit{individual fits}. In the second option we also apply equation~(\ref{eq:velocity_relation}) to the individual void density profiles, but fix $b_\tracer$ to $1$. Then we average the resulting linear theory profiles and fit for a single `global' tracer bias parameter in the end. This method will be referred to as \textit{global fits}. Both methods have the advantage that they provide an entire posterior distribution of predictions for the linear theory profile, which can be used to quantify uncertainties (error bars).

Figure~\ref{fig_stacks_density_velocity_mr_CDM} presents the results from these two methods (individual and global fits) for \emph{isolated} CDM voids. In all cases the measured velocity profiles and linear theory predictions match closely, with stronger differences again occurring around small voids. Moreover, differences between data and model gradually diminish towards larger voids in global stacks, whereas in individual stacks those are of the same magnitude for all void radius bins. In general, the results from global fits are very similar to the previous case from figure~\ref{fig_mass_conservation_stacked_density}. In contrast, individual fits feature a slightly better agreement between data and model, particularly at larger distance from the void center. On the other hand, the best-fit values of the tracer bias are somewhat closer to unity in the case of global fits. Note that for individual fits the obtained bias values are identical in both stacking methods, since the fitting is performed before stacking. For global fits the bias values slightly vary between the two stacking methods.

Subsequent figures~\ref{fig_stacks_density_velocity_mr_halos} and~\ref{fig_stacks_density_velocity_hr_halos} present individual and global fits to the profiles of \emph{isolated} halo voids in both the \MR{} and \HR{} simulation. Essentially all our conclusions from CDM voids remain valid for halo voids. Only the tracer bias parameters are higher, as expected, but also decrease towards larger void sizes. A comparison across different resolutions allows us to explore a wider range in void size and potentially identify a limit for the validity of linear theory for voids below some characteristic size. For the \HR{} simulation, we use the following void radius bin edges: $4\,\hMpc$, $8\,\hMpc$, $11\,\hMpc$, $13\,\hMpc$, $16\,\hMpc$, $20\,\hMpc$, and $30\,\hMpc$. Instead, in the \MR{} simulation void radius bins are arranged from $5\,\hMpc$ to $50\,\hMpc$ in steps of $5\,\hMpc$ width. We can now compare voids within a given range in size at different resolution. For example, voids in \HR{} with radii around $12\,\hMpc$ exhibit smaller residuals between data and model than in \MR{}. In order to reach the same level of agreement in \MR{}, one has to look at much larger voids of around $22\,\hMpc$ in radius. Voids of that size are among the largest in \HR{} and exclusively feature outflows, similar to the largest voids above $40\,\hMpc$ in \MR{}. This is a strong indication for the fact that residual mismatches are due the sparsity of tracers in the \MR{} simulation, rather than a limit of linear theory below some fundamental scale.

As a final test we exchange the number density profiles of halo voids with their mass-weighted versions. This is presented in figure~\ref{fig_stacks_density_velocity_mr_halos_massweight} for the \MR{} simulation. The theoretical model still produces the correct shape of stacked velocity profiles, but residuals increase somewhat in comparison to using number density profiles. In individual stacks, the differences now increase with void size, although the mass-weighted density profiles of large voids are the most similar compared to their number density profiles (see figure~\ref{fig_massweight_mr_hr_halo}). In either stacking method we observe that individual fits experience the largest discrepancies inside voids, where the maximal velocity of linear theory profiles considerably falls short of the measured velocities. On larger scales the profiles align again, except for the smallest voids. Global fits instead reproduce the maxima of the velocity profiles more accurately, but disagree more near the compensation walls. In all cases, the retrieved tracer bias values increase, as expected from mass weighting, but the agreement with linear theory is generally worse than for the unweighted case. This can be explained by the fact that this weighting scheme amplifies the impact of high-mass halos, which are much scarcer than the bulk of the halo population. In turn, this leads to a stronger sensitivity to the impact of tracer sparsity in the profile estimators, most notably close to the void centers.

We have repeated the same analysis on the matter density profiles of halo voids and \emph{merged} voids. Furthermore, we tested higher mass cuts when selecting halos for our void finding and using those halos for calculating the profiles. We also performed tests at different redshifts ($z = 0$ to $z \simeq 4$) and for a variety of cosmological parameter values in \MR{} simulations of smaller box size, available within the \texttt{Magneticum} suite. We find identical conclusions, so we refrain from presenting this in additional figures here. One also has the option to stack voids based on properties other than their radius, like in figures~\ref{fig_number_density_mr_halo_ellipticity_coreDens_compensation_halos} and~\ref{fig_velocity_mr_halo_ellipticity_coreDens_compensation_halos}. In that case the linear theory profiles often yield a less accurate prediction due to the large ranges in void size per stack. However, this can be remedied using the methods of individual fits and individual stacks, because then voids are modeled independently from each other.

\subsection{Resolution study \label{subsec:resolution}}

\begin{figure}[t]

               \centering

               \resizebox{\hsize}{!}{

                               \includegraphics[trim=0 50 0 5, clip]{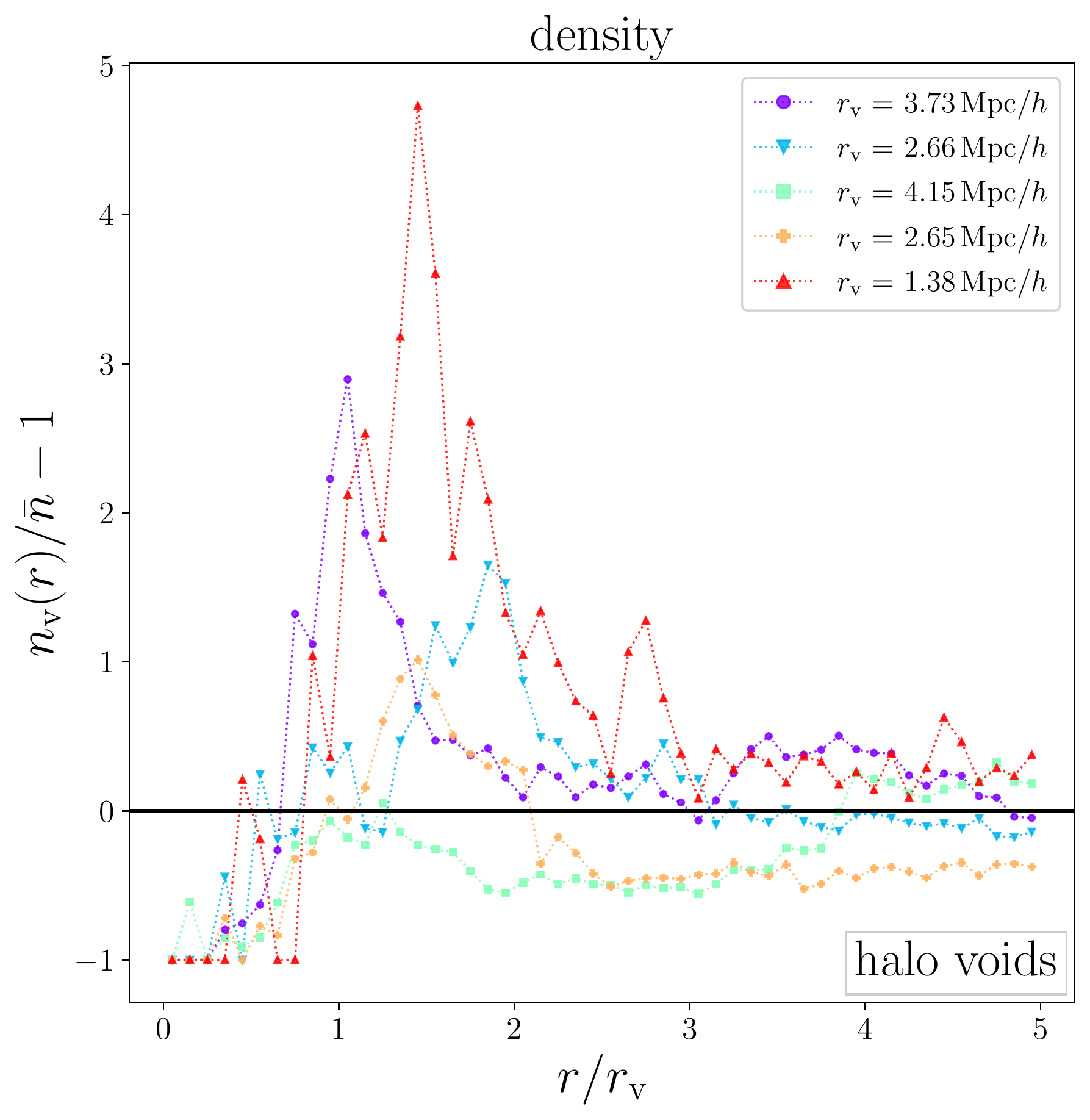}

                               \includegraphics[trim=0 50 0 5, clip]{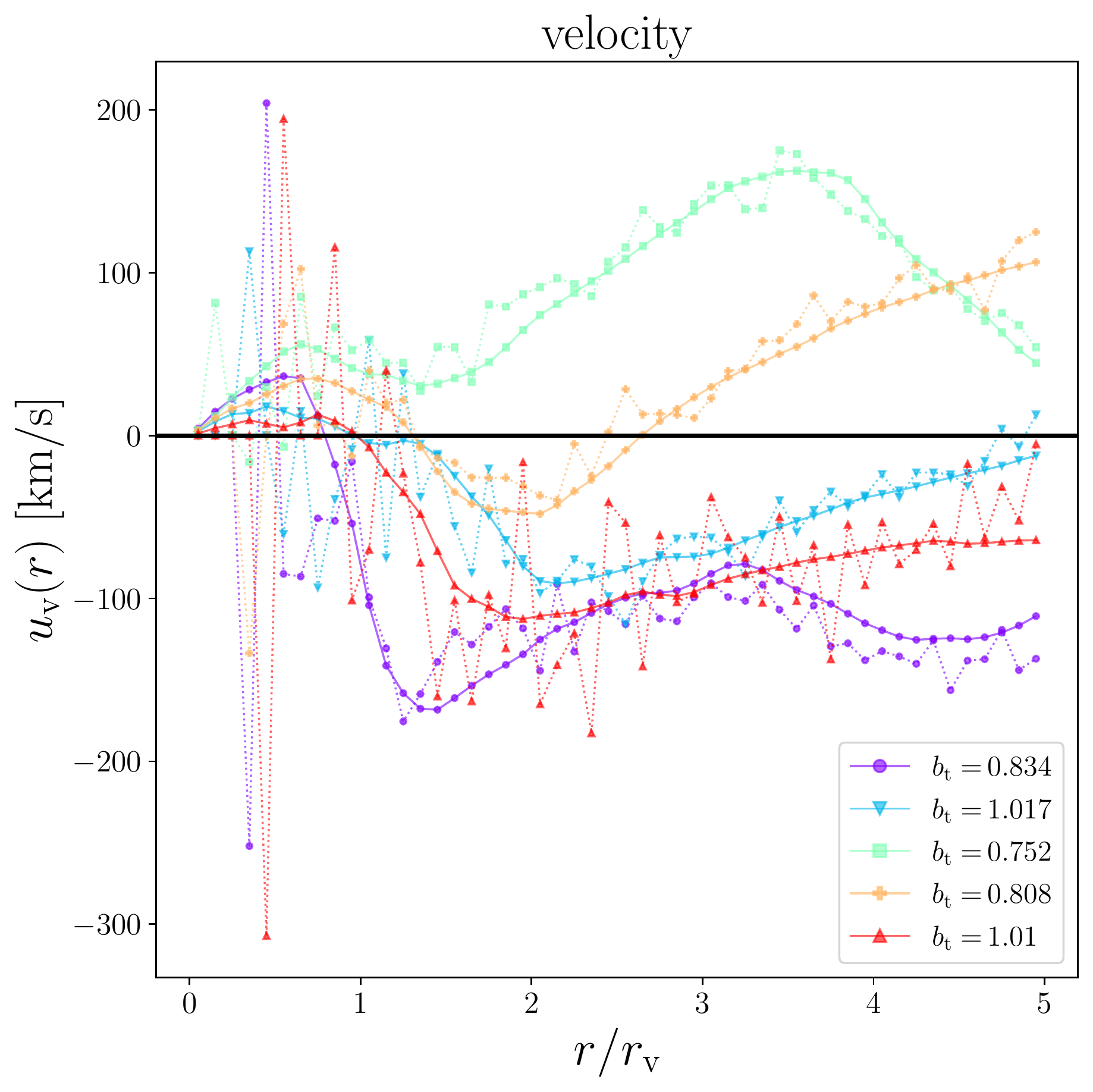}}

                 \resizebox{\hsize}{!}{

                               \includegraphics[trim=11 10 0 29, clip]{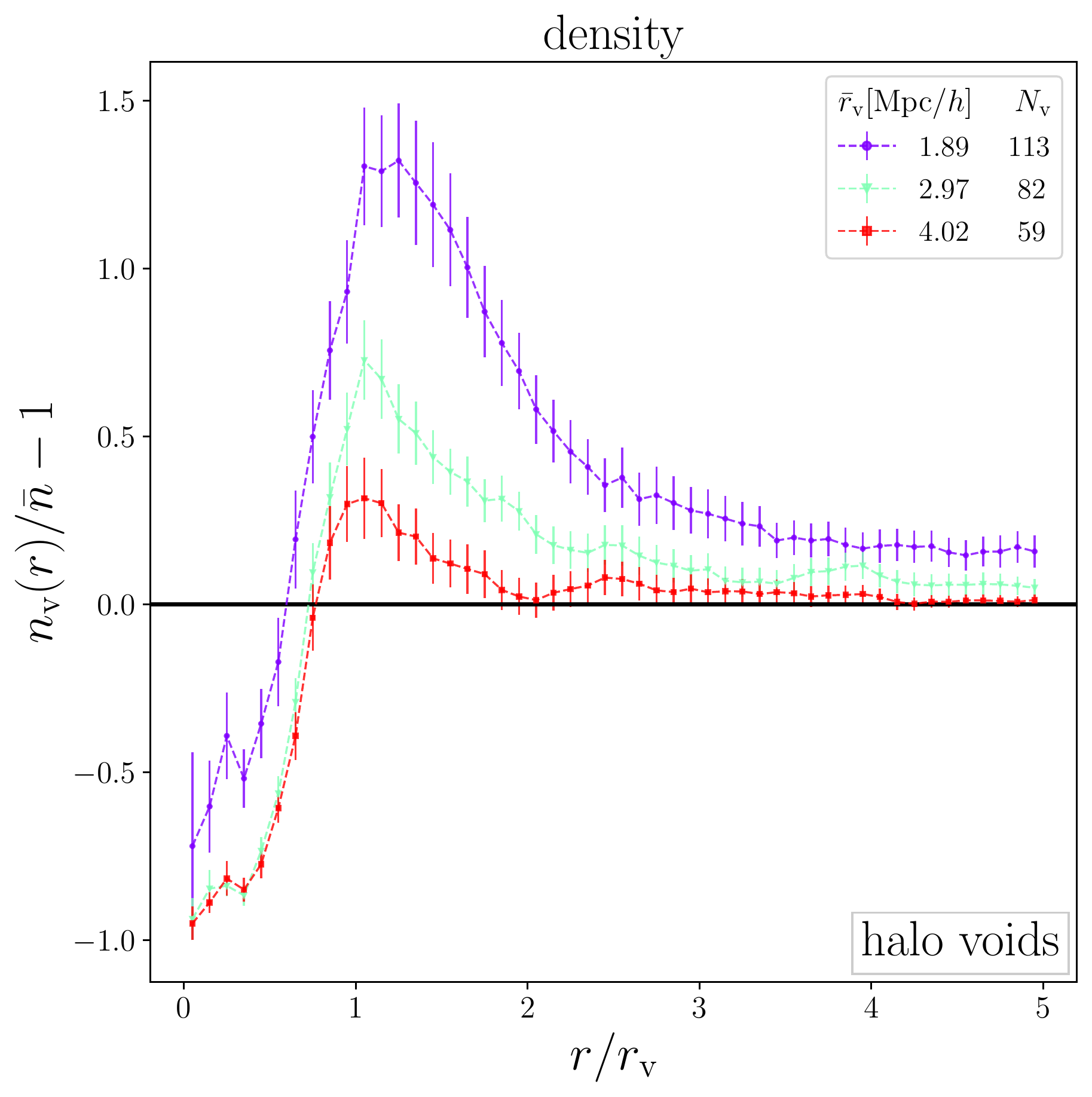}

                               \includegraphics[trim=-11 10 0 29, clip]{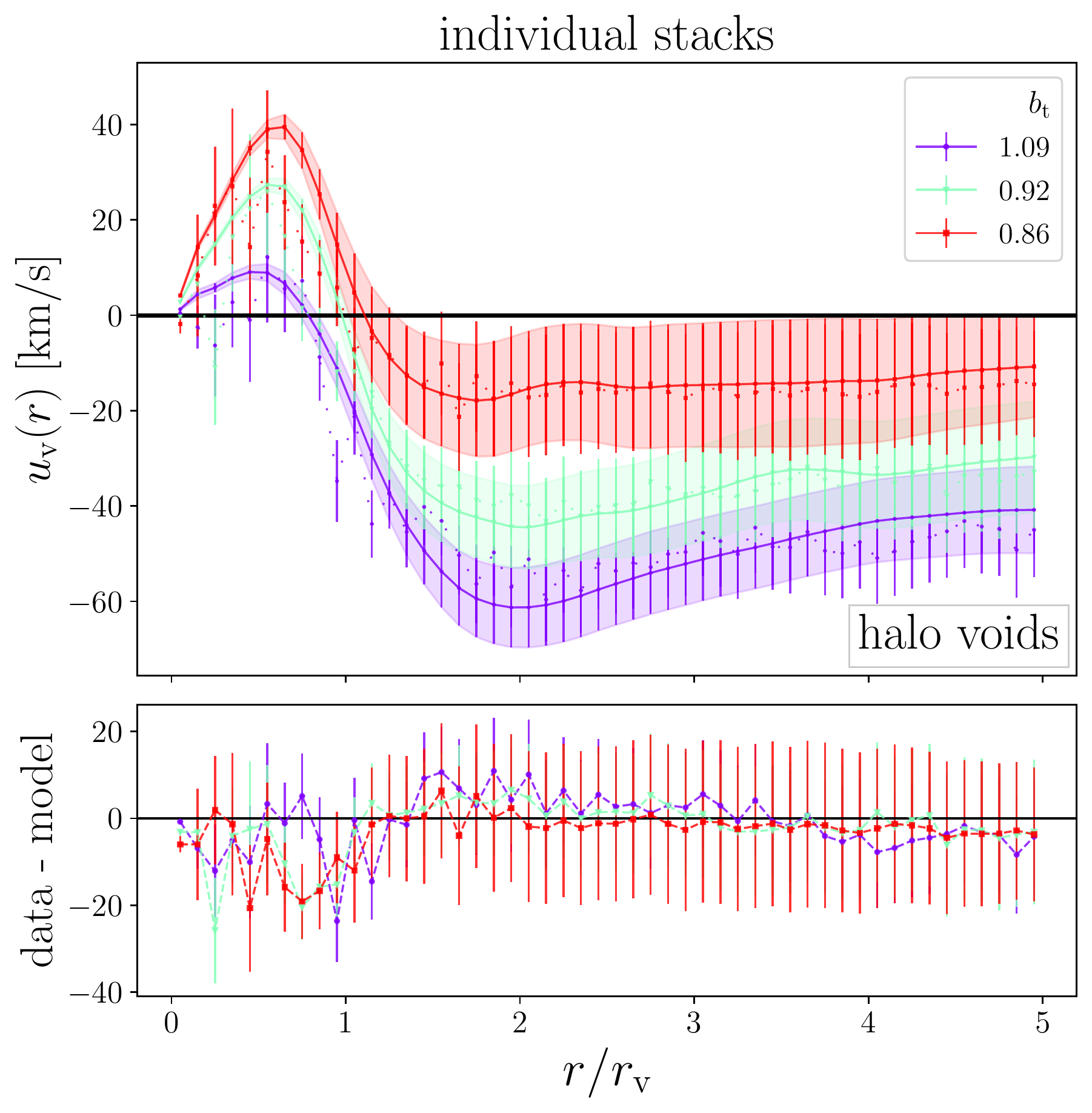}}

               \caption{Individual (top) and stacked (bottom) number density profiles of \emph{isolated} halo voids from the \UHR{} simulation on the left. Corresponding velocity profiles (dotted lines) and prediction of linear mass conservation (solid lines) via equation~(\ref{eq:velocity_relation}) on the right.}

               \label{fig_mass_conservation_uhr}

\end{figure}

Our tests on the linear continuity equation~(\ref{eq:velocity_relation}) in the previous subsections show that velocity profiles calculated with this model agree remarkably well with measured profiles, both for individual, as well as for stacked profiles of both CDM and halo voids. A comparison between \MR{} and \HR{} boxes has not revealed any limitations for the validity of the model, apart from the impact of tracer sparsity. Here we extend this resolution study by one more layer, making use of the \UHR{} simulation. It covers a box of merely $48\,\hMpc$ a side, but features a much higher resolution than \HR{}. We find a total of $346$ \emph{isolated} halo voids inside this simulation. This is too small of a sample to consider for realistic cosmological applications, but still very suited for testing linear dynamics at very small scales.

Figure~\ref{fig_mass_conservation_uhr} shows both individual and stacked density profiles of voids on the left. The latter clearly exhibit larger error bars and therefore more noise than the previously analysed profiles, owed to the much smaller sample size. Nevertheless, the characteristic features are consistent with our previous findings, except that here we are examining voids of only a few Mpc in size, with a maximum radius of $4.5\,\hMpc$. The right side of this figure presents the corresponding velocity profiles of the same voids, with individual stacks on the bottom. Most of these small voids experience infall from their environment, but there is one example of a clearly undercompensated void with $r_\void=4.15\,\hMpc$ (in green).

Based on their density profiles on the left side of figure~\ref{fig_mass_conservation_uhr}, we repeat the application of equation~(\ref{eq:velocity_relation}) to predict the velocity profiles on the right. From the individual profiles on the top we find that even the smallest depicted void with radius $1.38\,\hMpc$ (in red) is consistent with linear dynamics within the scatter. This is still in line with the individual void profiles in both \MR{} and \HR{} from section~\ref{subsec:lin_theory_individual_voids}. Despite their huge variety in shape, these individual profiles are all described remarkably well by the linear continuity equation. The stacked profiles on the bottom fully confirm this picture: even for the smallest bin in void radius between $1.0\,\hMpc$ and $2.5\,\hMpc$, linear theory profiles (based on individual fits) align with stacked velocity profiles within their error margins on all scales around the void center.

Such scales, reaching $1\,\hMpc$ and below, are typically considered as highly nonlinear in the field of large-scale structure. Nevertheless, we find no evidence for the onset of nonlinear dynamics in void environments of that size. It is possible that this is a general characteristic of voids and holds irrespective of scale. At the same time, voids dominate the volume fraction of the Universe, which implies that linear dynamics should prevail within the large-scale structure. It is conceivable that for practical reasons the tradition of studying the brightest galaxies (such as luminous red galaxies) and the most massive objects (such as galaxy clusters), which are located in the densest environments of the cosmos, may have concealed this insight from cosmologists so far.

\section{Conclusion\label{sec:conclusion}}

In this paper we have explored the interrelation between a variety of general void properties across a substantial dynamic range in mass and scale by analyzing the hydrodynamical simulation suite \Mag{}. This enabled us to reveal and inspect a number of universal characteristics of voids that persisted in all considered settings. Our main results can be summarized as follows:
\begin{itemize}
    \item Merging creates larger voids with a hierarchy of sub-voids. Stacking \emph{merged} voids leads to shallower central density profiles, due to sub-structure inside these voids, and slightly higher compensation walls (figures~\ref{fig_density_mr_CDM_halo_merging} and~\ref{fig_density_hr_CDM_halo_merging}). The smallest voids are not affected by this, because they host no sub-voids at a given tracer density. They are identical to \emph{isolated} voids. In contrast to \emph{isolated} voids, the size function of \emph{merged} voids converges on large scales, irrespective of resolution and tracer type (figure~\ref{fig_void_function_halo_CDM_merging_rad_ell_core}).
    
    \item The stacked number density profiles of halo voids exhibit the same characteristics as the matter density profiles of CDM voids, albeit with an enhanced amplitude caused by the bias of halos (figures~\ref{fig_density_mr_CDM_halo_merging} and~\ref{fig_density_hr_CDM_halo_merging}). Using their masses as weights in the density profile calculation even enhances this effect, as halo masses correlate with the density of their environment (figure~\ref{fig_massweight_mr_hr_halo}). However, the CDM density profiles of halo voids are very similar to the ones of CDM voids (figure~\ref{fig_matter_density_mr_hr_haloCDM}).
    
    \item The ellipticity of voids affects the shape of stacked density and velocity profiles via spherical shell averaging. More elliptical voids exhibit smoother density profiles with wider compensation walls and shallower cores (figure~\ref{fig_number_density_mr_halo_ellipticity_coreDens_compensation_halos}).
    
    \item The minimum and average density inside voids is anti-correlated with the height of their compensation wall: the shallowest voids feature the biggest walls and vice versa. Overcompensated voids, which live in overdense environments, are typically much smaller than undercompensated ones, which can be interpreted as sub-voids inside larger parent voids. These void properties are reflected in the general shape of their density and velocity profiles (figures~\ref{fig_number_density_mr_halo_ellipticity_coreDens_compensation_halos} and~\ref{fig_velocity_mr_halo_ellipticity_coreDens_compensation_halos}).
    
    \item The stacked velocity profiles of voids mirror their density structure: large and undercompensated voids are dominated by coherent outflow, while small and overcompensated ones are dominated by infall towards their compensation wall. Halos and CDM move at the same speed around voids, as expected from the equivalence principle in general relativity (figures~\ref{fig_velocity_mr_halo_CDM} and~\ref{fig_velocity_hr_halo_CDM}). Their radial velocity profiles accurately obey linear mass conservation. Residual deviations are caused by sampling artifacts arising from sparse tracer statistics. These become increasingly significant inside voids with extensions close to the resolution limit / mean tracer separation and depend on the type of estimator used for the velocity profile calculation (figures~\ref{fig_mass_conservation_stacked_density}--\ref{fig_stacks_density_velocity_mr_halos_massweight}).
    
    \item Mass cuts and subsampling affect different estimators for void profiles in different ways. Density profiles are stable against subsampling, while mass cuts have a similar effect as mass weighting. Global stacks for velocity profiles are more independent of tracer mass and subsampling than individual stacks (figures~\ref{fig_num_density_massbins_subsampling_hr} and~\ref{fig_velocity_massbins_subsamplings_hr}).
    
    \item All previous conclusions on stacked profiles remain valid even for individual voids, although their profiles are more affected by sampling variance and noise. Nevertheless, the interrelations between density and velocity profiles of individual voids obey linear mass conservation with an exquisite accuracy (figures~\ref{fig_individual_vel_relation_mr_CDM}--\ref{fig_individual_density_velocity_hr_halo}). This fact can be exploited to explain and improve the linear relationship between the corresponding average profiles of stacked voids, which plays an important role in cosmological applications.
    
    \item The apparent breakdown of linear dynamics inside the voids we analyze at moderate resolution is caused by sparse sampling of tracers close to their mean separation scale. We have confirmed this by comparing voids of a given size in simulations of increasing resolution and find no sign for the onset of nonlinear dynamics down to scales of order $1\,\hMpc$ (figure~\ref{fig_mass_conservation_uhr}). At a given resolution, velocity profiles are more severely affected by sparse sampling than density profiles, so the linearized continuity equation applied to densities is better suited for the modeling of velocities and observable RSD around voids than using more biased velocity profile templates measured in simulations.
\end{itemize}
These findings have a number of important implications for observational studies on cosmic voids and their use as cosmological probes. While it is not yet feasible to identify voids in the full three-dimensional matter distribution of the Universe, various tracers of the latter have already been considered for void finding, such as galaxies~\cite[e.g.,][]{Pan2012,Sutter2012a,Mao2017a,Hamaus2020,Aubert2022}, galaxy clusters~\cite{Pollina2019}, the Ly-$\alpha$ forest~\cite{Stark2015,Krolewski2018,Porqueres2019,Ravoux2022}, or the 21cm emission from neutral Hydrogen~\cite{White2017,Endo2020}. We cannot foresee a compelling reason as to why our conclusions above would not apply for the voids identified within any of these tracers.

This opens up a vast observational window to conduct cosmological experiments, allowing us to make use of voids ranging from a few (or possibly less) to hundreds of Mpc in diameter. As linear dynamics holds up in all voids irrespective of their size, void catalogs can not only be extended via enlarging survey volumes towards higher redshifts (as planned for a number of next-generation surveys, such as \emph{Euclid}~\cite{Hamaus2022,Contarini2022,Bonici2022}), but also by conducting deeper observations that provide denser tracer samples and hence smaller sub-voids (e.g., as planned for Roman~\cite{Spergel2015} and 4MOST~\cite{DeJong2019}), even at low redshift. This would open up the possibility to maximize the number of observable linear modes of the density field of large-scale structure that can be exploited for the purpose of cosmological inference, for example via the Alcock-Paczynski effect and RSD~\cite{Hamaus2016,Correa2021b,Hamaus2022}, far beyond the previously imposed limits. A similar conclusion has been reached in reference~\cite{Stopyra2021}, where voids with $r_\void>5\hMpc$ have been found to behave very linearly in the sense that they are well-described by the Zel’dovich approximation. Moreover, the upcoming data sets will contain hundreds of thousands of voids with numerous individual properties, providing a rich playground for the latest machine learning applications~\cite{Cousinou2019,Kreisch2022,Wang2022}.

\begin{acknowledgments}
We thank Angelo Caravano for useful comments. This research has been funded by the Deutsche Forschungsgemeinschaft (DFG, German Research Foundation) -- HA 8752/2-1 -- 669764. The authors acknowledge additional support from the Excellence Cluster ORIGINS, which is funded by the DFG under Germany's Excellence Strategy -- EXC-2094 -- 390783311. KD acknowledge funding for the COMPLEX project from the European Research Council (ERC) under the European Union’s Horizon 2020 research and innovation program grant agreement ERC-2019-AdG 882679. The calculations for the hydrodynamical simulations were carried out at the Leibniz Supercomputing Center (LRZ) under the project pr83li. We are especially grateful for the support by M. Petkova through the Computational Center for Particle and Astrophysics (C2PAP) and for the support by N. Hammer at LRZ when carrying out the Box0 simulation within the Extreme Scale-Out Phase on the new SuperMUC Haswell extension system.
\end{acknowledgments}

\bibliography{ms}

\bibliographystyle{JHEP}

\end{document}